\documentclass[]{aa}
\usepackage{tabularx}
\usepackage{amsmath}
\usepackage{natbib}
\usepackage{graphicx}
\usepackage{txfonts}

\usepackage{rotating}%
\usepackage{multirow}%

\bibpunct{(}{)}{;}{a}{}{,}

\newcommand{\msun}{~{\rm M}_\odot}

\newcommand{\mmin}{M_{\rm min}}
\newcommand{\mmax}{M_{\rm max}}
\newcommand{\qmin}{q_{\rm min}}
\newcommand{\qmax}{q_{\rm max}}

\newcommand{\binf}{\mathcal{B}}
\newcommand{\binfm}{\mathcal{B}_{M_1}(M_1)}
\newcommand{\binfall}{\mathcal{B}_{\rm all}}

\newcommand{\fq}{f_q(q)}
\newcommand{\hq}{h_q(q)}
\newcommand{\fqall}{f_{q,{\rm all}}(q)}
\newcommand{\fqm}{f_{q;M_1}(q)}

\begin{document}

\title{Exploring the consequences of pairing algorithms for binary stars} 

 \author{M.B.N. Kouwenhoven\inst{1,2}
          \and
          A.G.A. Brown\inst{3}
          \and
	  S.P. Goodwin\inst{1}
          \and
          S.F. Portegies Zwart\inst{2,4}
          \and
          L. Kaper\inst{2}
          }

 \offprints{M.B.N. Kouwenhoven \email{t.kouwenhoven@sheffield.ac.uk}}

 \institute{
   Department of Physics and Astronomy, University of Sheffield,
   Hicks Building, Hounsfield Road, S3~7RH, Sheffield, UK
   \\\email{t.kouwenhoven@sheffield.ac.uk; s.goodwin@sheffield.ac.uk}
   \and
   Astronomical Institute`Anton Pannekoek,
   University of Amsterdam,
   Kruislaan 403, 1098 SJ Amsterdam, The Netherlands
   \\\email{l.kaper@uva.nl}
   \and
   Leiden Observatory, University of Leiden,
   P.O. Box 9513, 2300 RA
   Leiden, The Netherlands \\\email{brown@strw.leidenuniv.nl}
   \and
   Section Computer Science, University of Amsterdam,
   Kruislaan 403, 1098 SJ Amsterdam, The Netherlands
   \\ \email{s.f.portegieszwart@uva.nl} }

 \date{Received ---; accepted ---}

\abstract{

Knowledge of the binary population in stellar groupings provides important information about the outcome of the star forming process in different environments \citep[e.g.,][and references therein]{blaauw1991}. Binarity is also a key ingredient in stellar population studies, and is a prerequisite to calibrate the binary evolution channels. 
In this paper we present an overview of several commonly used methods to pair individual stars into binary systems, which we refer to as pairing functions. These pairing functions are frequently used by observers and computational astronomers, either for their mathematical convenience, or because they roughly describe the expected outcome of the star forming process. We discuss the consequences of each pairing function for the interpretation of observations and numerical simulations. The binary fraction and mass ratio distribution generally depend strongly on the selection of the range in primary spectral type in a sample. The mass ratio distribution and binary fraction derived from a binarity survey among a mass-limited sample of targets is thus not representative for the population as a whole.
Neither theory nor observations indicate that random pairing of binary components from the mass distribution, the simplest pairing function, is realistic. It is more likely that companion stars are formed in a disk around a star, or that a pre-binary core fragments into two binary components.
The results of our analysis are important for (i) the interpretation of the observed mass ratio distribution and binary fraction for a sample of stars, (ii) a range of possible initial condition algorithms for star cluster simulations, and (iii) how to discriminate between the different star formation scenarios.


}

\maketitle

\section{Introduction}

Observations and simulations suggest that most stars form in binary systems \citep[e.g.][]{duquennoy1991,mason1998,goodwinkroupa2005,kobulnicky2007,kouwenhoven2005,kouwenhovenrecovery,goodwinpp2007}, and that a substantial fraction are part of a triple or higher-order system \citep[e.g.][]{tokovininsmekhov2002,correia2006,tokovinin2006,huyi2008}. Multiplicity is thus a fundamental property of the star forming process. Detailed knowledge of a young binary population can be employed to study the outcome of star formation, and consequently the star formation process itself.

Surveys for binarity have indicated that the properties of the binary population are a function of the spectral type of the primary star. Practically all O-type stars \citep{mason1998} and B/A-type stars \citep{shatsky2002,kobulnicky2007,kouwenhovenrecovery} are found in binary or multiple systems. \cite{abtlevy1976} report a multiplicity fraction of 55\% among F3--G2 stars, and in their CORAVEL spectroscopic study of F7--G9 stars \cite{duquennoy1991} find a binary fraction of $\sim 60\%$. The binary fraction among M-type stars is $30-40\%$ \citep{fischer1992,leinert1997,reid1997}. For late M-type stars and brown dwarfs the binary fraction decreases to $10-20\%$ \citep[e.g.,][]{gizis2003,close2003,bouy2003,burgasser2003,siegler2005,ahmic2007,maxted2008}.

In this paper we discuss in detail several methods of pairing individual stars into binary stars. We refer to the latter algorithms as ``pairing functions''. Several of these have a physical motivation, others are discussed because of their mathematical simplicity. All these pairing functions have in common that they are frequently used in literature. The main goal of this paper is to explain the consequences of adopting a particular pairing function when doing a numerical simulation, or when interpreting observations. A good understanding of the consequences of each pairing algorithm for the binary population is important for
\begin{itemize}
\item {\em The interpretation of observations}. Are the measured properties (e.g. mass ratio distribution and binary fraction) for a certain sample representative for the population as a whole? What is the role of selection effects, and how can the different pairing functions be distinguished?
\item {\em Initial conditions for simulations}. In N-body simulations, such as STARLAB \citep[e.g.,][]{ecology4} and NBODY6 \citep[e.g.,][]{aarseth1999}, a mass ratio distribution independent of primary mass is often adopted. What are the consequences of this approach?
\item {\em Star formation}. Which pairing functions are expected from the different star forming scenarios? What numerical simulations of clustered star formation \citep[e.g.,][]{bate2003,bate2009} predict? Do we expect random pairing from the initial mass function? Which observations are necessary to be able to distinguish between these scenarios?
\end{itemize}
This paper is organised as follows. In \S\,\ref{section:terminology} we introduce the terminology used and discuss our assumptions. In \S\,\ref{section:massdistribution} we discuss the mass distribution, and in \S\,\ref{section:massratiodistribution} the mass ratio distribution. We briefly describe the origin of the pairing between binary stars in \S\,\ref{section:origin}. The major part of this paper is  \S\,\ref{section:pairingfunction}, where we discuss the different binary populations resulting from the choice of the mass distribution and pairing function. The specific differences are discussed in detail in \S\,\ref{section:comparison_pairingfunctions}, and the dependence on the generating properties in \S\,\ref{section:dependence_generating_properties}. Observational complications and a strategy to recover the pairing function are described in  \S\,\ref{section:strategy}. Finally, we summarise our results in \S\,\ref{section:summaryandoutlook}.

\section{Method and terminology} \label{section:terminology}

We study the differences of the various methods of pairing individual stars into binary systems by analysing numerically simulated binary populations. For each binary system, we refer to the most massive component $M_1$ as the primary star, and the least massive component $M_2$ as the companion star. Our adopted definition is purely based on the {\em current mass} of the components, i.e., irrespective of their relative luminosity or initial mass. We define the mass ratio as $q \equiv M_2/M_1$, so that $0 < q \leq 1$. The total mass is denoted $M_T = M_1 + M_2$ for a binary system, and $M_T=M_S$ for a single star with mass $M_S$. In several cases, we construct binary systems from star forming cores of mass $M_C$ with star forming efficiency $\epsilon$, so that the total mass of the resulting objects is $M_T=\epsilon M_C$. 

In our simulations each star is given a mass. A subset of the stars is assigned a companion, the other stars remain single stars. The companion is given a mass according to a pairing algorithm. We refer to the algorithm that is used to combine individual stars into binary systems as the {\em pairing function}. The pairing function of a binary population may for example be {\em random pairing} of both companions from the mass distribution. In the particular case of random pairing, primary and companion are swapped, if necessary, so that the primary is the most massive star.

Depending on which pairing function is used, the mass of the primary star, and in several cases the companion star, is drawn from a mass distribution $f_M(M)$. We refer to $f_M(M)$ as the {\em generating} mass distribution. We denote the resulting mass distributions for primary stars, companion stars, systems, and single stars with $f_{M_1}(M)$, $f_{M_2}(M)$,  $f_{M_T}(M)$, and $f_{M_S}(M)$, respectively. Note that the distributions over primary and companion mass in a stellar grouping are never independent, $f_{M_1,M_2}(M_1,M_2) \neq f_{M_1}(M_1)f_{M_2}(M_2)$, as by definition $M_1 \geq M_2$. The resulting mass distribution for all single and primary stars is denoted with $f_{M_{1,S}}(M)$, and the mass distribution for all individual objects (i.e., singles, primaries and companions), is denoted with  $f_{\rm all}(M)$. The distribution $f_{\rm all}(M)$ that is present immediately after star formation is called the {\em initial mass function} (IMF). Note that $f_{\rm all}(M)$ is unequal to the {\em generating mass distribution} $f_M(M)$, except in the random pairing case (see \S\,\ref{section:pf_rp}).

For several pairing functions we generate companions by drawing the mass ratio for a binary from a (generating) mass ratio distribution $\fq$
Depending on the additional constraints specific for each pairing function, the resulting overall mass ratio distribution  $f_{q,all}(q)$ may or may not be equal to $\fq$. The resulting mass ratio distribution may or may not be a function of primary spectral type. We refer to the {\em specific} mass ratio distribution for all binaries with a primary of spectral type A or B, for example, with $f_{q,AB}(q)$. Throughout this paper we will mostly use the expression $\fqm$ for the specific mass ratio distribution for an ensemble of binaries with a limited primary mass range.

A common expression to quantify the multiplicity of a stellar population is the multiplicity fraction $\binf$ (which is often referred to as the {\em binary fraction}), defined as
\begin{equation} \label{equation:multiplicityfractions}
  \binf       =  \frac{ B+T+\dots    }{ S+B+T+\dots    } \\ 
\end{equation}
where $S$ is the number of single stars, $B$ the number of binaries, and $T$ the number of triple systems \citep[e.g.,][]{reipurth1993}. Throughout this paper we consider only single and binary stars, and do not consider higher-order systems, so that $\binf=B/(S+B)$. The number of systems is given by $N=S+B$ and the total number of (individual) stars is $S+2B = N (\binf + 1)$. 

Each pairing algorithm is provided with a (generating) binary fraction $\binf$, which describes the fraction of stars that is assigned a preliminary companion star. For most pairing functions this preliminary companion is accepted as such, so that the overall binary fraction $\binfall$ equal to the generating value $\binf$. For several pairing functions, however, additional constraints are set to the properties of the companion. For example, when a pairing function generates a Jupiter-mass companion around a solar-mass star, the ``primary'' is usually considered to be a ``single'' star. For such pairing functions, the resulting overall binary fraction $\binfall$ is smaller than $\binf$. We denote the {\em specific} binary fraction for the set of all single stars of spectral type A/B and binary systems with primary spectral type A/B as $\binf_{\rm AB}$.

During our analysis we make several assumptions for reasons of clarity; our models are simplifications of reality. Our results are not limited by these assumptions, and the models can easily be extended. Our main goal is to illustrate the implications of adopting a particular pairing function.

We assume that no triples or higher-order systems are present. Although observations have shown that a significant fraction ($\ga 15\%$) of the stars are part of a multiple system \citep{tokovininsmekhov2002,correia2006,tokovinin2006,eggleton2008,huyi2008}, the properties of higher-order systems are not well understood. Observational selection effects complicate the derivation of the properties of these systems significantly. Higher-order systems are often ignored in N-body simulations due to computational complications \citep[e.g.,][]{vandenberk2007}. A full understanding of star formation, however, ultimately requires a full knowledge of the formation and evolution of higher-order systems.

In our models the generating binary fraction for the population can be described with a single value $\binf$. For most pairing functions this results in a specific binary fraction $\binfm$ that is independent of primary mass $M_1$. However, in \S\,\ref{section:pairingfunction} we describe several cases where, as a result of the pairing properties, the binary fraction is a function of primary mass, even though this dependency is not included explicitly.

Selection effects play a major role in the interpretation of binary star observations. A detailed description of the selection effects, such as in \cite{kouwenhovenrecovery}, is necessary to derive the pairing function, the mass ratio distribution and the binary fraction from observations. A major bias is generally introduced by studying the binary population in a certain primary mass range; we describe this effect in detail for the different pairing functions. Throughout most of this paper we ignore the other selection effects.

\section{The mass distribution} \label{section:massdistribution}

\begin{figure}[tbp]
  \centering
  \includegraphics[width=0.5\textwidth,height=!]{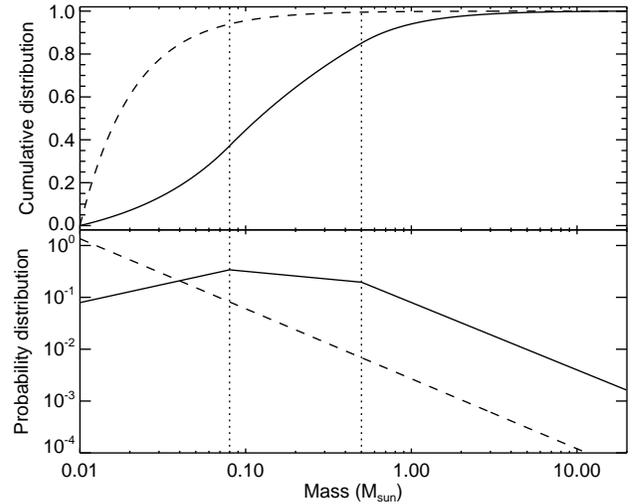}
  \caption{The cumulative mass distributions ({\em top}) and mass distributions
  ({\em bottom}) derived by Kroupa  (Eq.~\ref{equation:kroupaimf}; solid curve) 
  and Salpeter (Eq.~\ref{equation:salpeterimf}; dashed curve). 
  The masses at which the slope of the Kroupa
  mass distribution changes are indicated with the vertical dotted lines. 
 \label{figure:sos_massdistributions} }
\end{figure}

\begin{table}[tbp]
  \small
  \caption{The distribution of stars over main-sequence spectral type, for the Kroupa and Salpeter mass distributions. The spectral type and corresponding (approximate) mass range are listed in columns~1 and~2 (BD = brown dwarfs). For both the Kroupa and Salpeter mass distribution we list the fraction ${\mathcal F}_N$ of objects and the fraction ${\mathcal F}_M$ of the total mass in each mass range. We have only considered stars in the mass range $0.02\msun\leq M \leq 20\msun$.  \label{table:massfractions}} 
  \begin{tabular}{l r@{$-$}l r@{.}l r@{.}l r@{.}l r@{.}l}
    \hline
    \hline
    SpT                             & 
    \multicolumn{2}{c}{Mass range}  &
    \multicolumn{4}{c}{Kroupa}      & 
    \multicolumn{4}{c}{Salpeter}    \\
    \hline
                                    & 
    \multicolumn{2}{c}{$({\rm M}_\odot)$}  & 
    \multicolumn{2}{c}{${\mathcal F}_N$ (\%)}           & 
    \multicolumn{2}{c}{${\mathcal F}_M$ (\%)}           & 
    \multicolumn{2}{c}{${\mathcal F}_N$ (\%)}           & 
    \multicolumn{2}{c}{${\mathcal F}_M$ (\%)}           \\
    \hline
    B            & 3.0 & 20    &   1&45  &  24&76   &    0&10  &   8&66 \\
    A            & 1.5 & 3.0   &   2&28  &  13&05   &    0&18  &   5&31 \\
    F            & 1.0 & 1.5   &   2&72  &   9&15   &    0&22  &   3&86 \\
    G            & 0.8 & 1.0   &   2&22  &   5&52   &    0&18  &   2&30 \\
    K            & 0.5 & 0.8   &   7&45  &  12&92   &    0&61  &   5&45 \\
    M            & 0.08& 0.5   &  51&48  &  30&28   &   14&09  &  32&07 \\
    BD           & 0.02& 0.08  &  32&41  &   4&33   &   84&61  &  42&34 \\
    \hline
    All          & 0.02& 20    & 100&00  & 100&00   & 100&00   & 100&00 \\
    \hline
    \hline
  \end{tabular}

\end{table}

The mass distribution $f_M(M)$ defines the spectrum of masses in a stellar population, and is usually expressed as a single-component power law \citep{salpeter1955}, a multi-component power law \citep[e.g.,][]{kroupa2001} or a Gaussian distribution \citep[e.g.,][]{chabrier2003}. In our analysis we consider two mass distributions: the Kroupa mass distribution and the Salpeter mass distribution. The main difference between these is the presence or absence of a turnover in the low-mass regime. The mass distribution derived by \cite{kroupa2001} is given by
\begin{equation} \label{equation:kroupaimf}
  f_{\rm Kroupa}(M) \propto \left\{
  \begin{array}{llll}
    M^{-0.3}  & {\rm for \quad } 0.01 & \leq M/\msun & < 0.08 \\
    M^{-1.3}  & {\rm for \quad } 0.08 & \leq M/\msun & < 0.5   \\
    M^{-2.3}  & {\rm for \quad } 0.5  & \leq M/\msun & < \infty   \\
  \end{array}
  \right. .
\end{equation}
The classical Salpeter mass distribution \citep{salpeter1955} is given by
\begin{equation} \label{equation:salpeterimf}
f_{\rm Salpeter}(M) \propto M^{-2.35} \quad M \ga 1 \msun \,.
\end{equation}
The Salpeter mass distribution is derived for intermediate-mass stars in the Galactic field. Although it is known that the Salpeter mass distribution is incorrect for masses below $\sim 1\msun$, we use this mass distribution for comparison, to illustrate the effect of the slope of the mass distribution for low-mass stars. Note that, although, Eq.~(\ref{equation:kroupaimf}) introduces a turnover in Fig.~\ref{figure:sos_massdistributions}, which is logarithmically plotted, there is no real turnover in the mass distribution (see also Appendix~\ref{appendix:rp_powerlaw}). 

Table~\ref{table:massfractions} lists the fraction of stars of a given spectral type for both mass distributions. The corresponding probability distributions $f_M(M)$ and cumulative distributions $\int_0^M f_M(M')dM'$ are shown in Fig.~\ref{figure:sos_massdistributions}. Most young stellar populations in our Galaxy are described accurately with the Kroupa mass distribution \citep{bonnell2007}.

In our simulations we draw objects from the mass distribution in the range $0.02\msun \leq M \leq 20\msun$, i.e., stellar and substellar-mass objects. Objects less massive than $\mmin=0.02 \msun$ are considered to be planets, and form in a different way than stars or brown dwarfs \citep[see, e.g.,][]{pollack1996,kouwenhoven2007a}. The absolute maximum stellar mass is of order $M_{\rm max,abs} \approx 150 \msun$ \citep[][and references therein]{zinneckerreview2007}. The most massive star in a cluster may well depend on the total mass of the star cluster \citep{weidnerkroupa2006}. Stars more massive than $20 \msun$, however, are extremely rare, are very short-lived, and possibly even form by a different mechanism than most other stars \citep[e.g.,][]{zinneckerreview2007}.

The mass distribution of \cite{kroupa2001} is for {\em all} stars in a population, including single stars, primaries and companions. We note that choosing primaries from an IMF and then choosing secondaries from a mass ratio distribution (\S\,\ref{section:pf_pcp}) {\em will not} recover the original IMF.  Thus the primary mass distribution function cannot be exactly the same as the desired IMF \citep[see \S\,\ref{section:pf_fm};][]{malkovzinnecker2001,goodwinsplitup2008}. Technically, the mass distribution should therefore not be used to generate, for example, a primary mass distribution. For simplicity, however, we adopt the Kroupa mass distribution as the generating mass distribution for each pairing function. Ideally, one should iteratively determine the generating mass distribution by comparing the outcome of the pairing process with the Kroupa mass distribution. As this is computationally a very expensive exercise, we skip the iteration, and simply adopt the Kroupa mass distribution as the generating mass distribution. For a proper analysis of real observations, one should keep this issue in mind.

\section{The mass ratio distribution} \label{section:massratiodistribution}

The mass ratio distribution describes the distribution over mass ratio $q=M_2/M_1$ for a population of binary systems. The mass ratio distribution for binary systems has been studied thoroughly over the last decades \citep[see, e.g.,][]{zinnecker1984,hogeveen1992,mazeh2003,halbwachs2003}.  In this paper we discriminate between three different types of mass ratio distributions. The {\em generating} mass ratio distribution is an input distribution that is used by most pairing function algorithms to generate binaries, although some pairing functions (e.g., random pairing) do not require a generating mass ratio distribution. Note that the mass ratio distribution is not the same as the pairing function; the mass ratio distribution is a {\em property} of several (not all) pairing functions; see \S\,\ref{section:pairingfunction} for details. The {\em overall} mass ratio distribution is the mass ratio distribution resulting from the pairing mechanism, for all binaries in the population. The {\em specific} mass ratio distribution is that for a sample of stars with primaries in a given mass range. The latter is measured in observations, as a binarity survey is in most cases focused on a particular set of targets with given spectral types.

The (specific) mass ratio distribution is usually obtained by a fit to the observed mass ratio distribution of the sample. The observed distribution is often described with a simple power-law:
\begin{equation} \label{equation:q_powerlaw}
f_{\gamma_q}(q) \propto q^{\gamma_q} \quad \mbox{for} \quad q_0 < q \leq 1 \,,
\end{equation}
where the exponent $\gamma_q$ is fitted \citep[e.g.][]{hogeveen1992b,shatsky2002,kobulnicky2007,kouwenhovenrecovery}. Distributions with $\gamma_q =0$ are flat, while those with $\gamma_q < 0$ and $\gamma_q > 0$ are falling and rising with increasing $q$, respectively. Usually the adopted minimum value for the fit $q_0$ is the value of $q$ below which the observations become incomplete due to selection effects. For distributions with $\gamma_q \leq -1$, a minimum value $q_0 > 0$ is necessary such that $\fq$ can be normalised. Sometimes the necessity of $q_0>0$ is avoided by fitting a mass ratio distribution of the form
\begin{equation} \label{equation:1+q_powerlaw} f_{\Gamma_q}(q) \propto (1+q)^{\Gamma_q}
  \quad \mbox{for} \quad 0 < q \leq 1 \end{equation}
to the data \citep[e.g.][]{kuiper1935a,rensbergen2006}. This distribution is more commonly used to describe the mass ratio distribution of high-mass spectroscopic binaries, while the distribution in Eq.~(\ref{equation:q_powerlaw}) is often used for visual binaries. As Eq.~(\ref{equation:q_powerlaw}) and~(\ref{equation:1+q_powerlaw}) show a similar behaviour (both are either falling, flat, or rising), we will only consider the distribution of Eq.~(\ref{equation:q_powerlaw}). 

Alternatively, to allow for a mass ratio distribution with a peak in the range $0 < q \leq 1$, we also consider the Gaussian mass ratio distribution:
\begin{equation} \label{equation:q_gaussian}
f_{\rm Gauss}(q) \propto \exp \left\{ - \frac{(q-\mu_q)^2}{2\sigma_q} \right\} \quad \mbox{for} \quad 0 < q \leq 1 \,,
\end{equation}
where $\mu_q$ and $\sigma_q$ are free parameters, corresponding to the mean and standard deviation of a Gaussian distribution without the imposed limits on $q$. Models with $\mu_q < 0$ and $\mu_q > 1$ show a distribution $f_{\rm Gauss}(q)$ with an exclusively negative and positive slope, respectively. In the case where $\mu_q \ll 0$ and $\mu_q \gg 1$, the distribution may be approximated by a power-law (Eq.~\ref{equation:q_powerlaw}). Models with $\sigma_q \gg 0.5$ can be approximated with a flat mass ratio distribution. Note that the values of $\mu_q$ and $\sigma_q$ do not necessarily reflect physical properties. A value $\mu_q \ll 0$, for example, merely means that the mass ratio distribution in the interval $0 < q \leq 1$ can be described by Eq.~(\ref{equation:q_gaussian}) in this interval. A Gaussian mass ratio distributions was reported by, \cite{duquennoy1991} for solar-type stars in the solar neighbourhood, who find, based on the work on the initial mass function by \cite{kroupa1990}, $\mu_q=0.23$ and $\sigma_q=0.42$.

For our default model we adopt a (generating) mass ratio distribution of the form $\fq = 1$ with $0 < q \leq 1$, the flat mass ratio distribution.

\section{The origin of the pairing function} \label{section:origin}

We define the {\em pairing function} as the algorithm which is used to pair individual stars into binary systems. A well-known pairing function is random pairing from the mass distribution. Others include, for example, a fixed mass ratio distribution. The pairing of binary components in a stellar population results from the combined effect of star formation, stellar evolution, binary evolution, and dynamical interactions. By studying the pairing of binary stars, the contributions of the latter three effects can be evaluated, and an estimate for the primordial binary population can be obtained. This primordial pairing function allows us to constrain the process of star formation.

It is worth considering what pairing we might expect from the actual
star formation process, as opposed to the various theoretical
constructs we describe in this paper.

Often random pairing has been used to construct binary systems for various
models.  Random pairing has the obvious advantage that the chosen IMF is, by
design, automatically recovered.  However, there is no good theoretical
reason to suppose that the star formation process would produce a randomly
paired distribution. Furthermore, random pairing is ruled out observationally. The observed mass ratio distribution among intermediate mass stars  \citep{shatsky2002,kouwenhoven2005,kouwenhoven2007a,kouwenhovenrecovery} and brown dwarfs \citep{kraus2008} in the nearby OB association Scorpius-Centaurus have indicated that the binary components are not randomly paired from the mass distribution. The same result is found for the Cygnus OB2 association \citep{kobulnicky2007}, and also by Weidner, Kroupa \& Maschberger (in prep). Random pairing is further excluded by the large prevalence of massive binaries with a mass ratio close to unity, often referred to as the ``twin peak'' in the mass ratio distribution \citep[][see also \S\,\ref{section:constraintsfromobservations}]{lucy1979,tokovinin2000,garcia2001,pinsonneault2006,lucy2006,soderhjelm2007}.

Simulations have shown that it is impossible to form significant numbers of
binary systems from an initially single star distribution \citep{kroupa1995a},
therefore stars in binary systems must have predominantly formed in binary
systems.  Observations of pre-main sequence stars also suggest that they have a higher
multiplicity than field stars (at least for $>1\msun$), suggesting that
most stars form in binary (or higher-order) systems \citep[e.g.,][]{goodwinkroupa2005}.  
It is supposed that a primordial population with a multiplicity of almost 100\%
evolves into a field-like binary population through (a) the decay of
higher-order multiple systems \citep[e.g.,][]{goodwinkroupa2005}, and (b) the
dynamical destruction of binaries in binary-binary encounters in clusters
\citep[e.g.,][]{kroupa1995a,kroupa1995b}. Thus, the currently observed binary population is a
complex mixture of primordial binaries (i.e. in the same dynamical state as when
they formed), and dynamically evolved binaries (which may have different
characteristics, or even companions to their initial state).

Simulations of binary star formation have comprehensively failed to produce
systems that match observations, even when the dynamical evolution of the
initial states is accounted for \citep[see][and references therein]{goodwinpp2007}. However, hydrodynamic simulations of star formation suggest that
companions usually form by the fragmentation of massive, disc-like
circumstellar accretion regions around young stars (see \citealt{goodwinpp2007}
and references therein).

In such a scenario for companion formation it would be expected that the
secondary should have a roughly similar (i.e. within a factor of three or
four) mass to the primary, especially at low separations.  A massive enough
region to fragment is only present during the earliest (e.g. class 0/I)
phases of star formation before the star(s) have accreted the majority of
their natal core.  Thus, the secondary will be present whilst a large
reservoir of gas is also present around it. In the case of a star that will
eventually grow to be (for example) $5\msun$ the secondary will form
whilst the primary is only ${\mathcal O}(1\msun)$ and several solar masses of
gas are present in the circumstellar environment.  A secondary will
presumably form with an initial mass close to the opacity limit for
fragmentation, $\sim 10^{-2} \msun$.  However, it is difficult to imagine
a scenario in which the secondary fails to accrete at least some of the
circumstellar material, especially as the secondary will form with angular
momentum similar to that of the accretion region, and so will be more able
to accrete material (see \citealt{goodwinpp2007} and references therein; also
\citealt{delgado2005}). Therefore, the secondary mass is expected to
be a reasonably large fraction of the primary mass.  In particular, it
should be difficult for a companion to a B-star to remain at brown dwarf or
M-star mass due to the large amount of material available for accretion.
In particular, we would expect a rough correlation between separation and
mass ratio, with closer companions being generally more massive as is
observed \citep[e.g.][for field G-dwarfs]{mazeh1992alt}.

In addition, dynamical evolution in clusters will act to destroy the most
weakly bound systems (i.e. the widest and lowest-mass companions),
further biasing the mass ratio distributions away from low-$q$.

We would therefore argue that random pairing over the full mass range is the {\em last} type of pairing that would be expected from the star formation process (see, however, \S\,\ref{section:restrictedRP}).

\section{Analysis of frequently used pairing functions} \label{section:pairingfunction}

There are many ways to obtain a population of binary systems from a mass distribution $f_M(M)$. We analyse the most frequently used algorithms in the sections below. In general, the masses of the members of a stellar grouping are drawn from the mass distribution. A fraction of the stars is assigned a companion star (either from the mass distribution, or using a mass ratio distribution), or the mass is split into a primary and companion star. Four commonly used mechanisms are the following:
\begin{itemize}
\item[--] {\em Random pairing } (RP). The masses of both binary components are randomly drawn from the mass distribution $f_M(M)$. For each system, the most massive component is labelled ``primary star'', the other component ``companion star'' (see \S\,\ref{section:pf_rp}).
\item[--] {\em Primary-constrained random pairing} (PCRP). The primary mass $M_1$ is drawn from the mass distribution $f_M(M)$, and the companion mass $M_2$ is chosen from the same mass distribution, with the additional constraint that $M_2 \leq M_1$ (see \S\,\ref{section:pf_pcrp}).
\item[--] {\em Primary-constrained pairing} (PCP). The primary mass is drawn from the mass distribution $f_M(M)$. The companion mass is then determined by a mass ratio that is drawn from a distribution $\fq$, with $0 < q \leq 1$ (see below). 
\item[--] {\em Split-core pairing} (SCP). The total mass of the binary is drawn from the mass distribution $f_M(M)$. The mass ratio of the binary is then determined by a mass ratio distribution which is drawn from $\fq$ with $0 < q \leq 1$. Finally, the primary mass $M_1 = M (1+q)^{-1}$ and companion mass $M_2 = M (1+q^{-1})^{-1}$ are determined (see below).
\end{itemize}
In the case of PCP and SCP there is another complication, which occurs if $M_2=qM_1<M_{\rm 2,min}$. In this case, the mass ratio distribution generates a companion mass smaller than the permitted value $M_{\rm 2,min}$, for example the deuterium-burning limit, while such objects are usually not considered as companions in a binary system. There are three straightforward choices on how to handle such companions:
\begin{itemize}
\item[--] {\em Accept all companions} (PCP-I/SCP-I). All companions are accepted, regardless of their mass. The resulting mass ratio distribution obtained with this method is equal to the generating mass ratio distribution (see \S\,\ref{section:pf_pcpi} and \S\,\ref{section:pf_scpi}).
\item[--] {\em Reject low-mass companions} (PCP-II/SCP-II). If $M_{\rm 2,min}/M_{\rm 1,max} < \qmin$, a fraction of the companions has $M_2 < M_{\rm 2,min}$. All companions with $M_2<M_{\rm 2,min}$ are rejected, and the corresponding ``primaries'' are classified as single stars (see \S\,\ref{section:pf_pcpii}, \S\,\ref{section:pf_scpii}, and below). 
\item[--] {\em Redraw low-mass companions} (PCP-III/SCP-III). For all binaries with a companion mass $M_2 < M_{\rm 2,min}$ the mass ratio is redrawn from $\fq$. This procedure is repeated until $M_2 \geq M_{\rm 2,min}$. This method is equivalent to drawing a mass ratio from the distribution $\fq$ with limits $M_{\rm 2,min}/M_1 \leq q \leq 1$ (see \S\,\ref{section:pf_pcpiii},\S\,\ref{section:pf_scpiii}, and below).
\end{itemize} 
It is possible in pairing function SCP-II that the resulting single star mass is smaller than $\mmin$, and for SCP-III the splitting up is not possible if the binary system mass is smaller than $2\mmin$. In these cases there are three possibilities of dealing with this problem:
\begin{itemize}
\item[--] {\em Accept all singles/primaries} (SCP-IIa/SCP-IIIa). If a pairing mechanism produces a single or primary star less massive than $M_{\rm 2,min}$, it is accepted, and included in the model.
\item[--] {\em Reject low-mass singles/primaries} (SCP-IIb/SCP-IIIb). If a pairing mechanism produces a single or primary star less massive than $M_{\rm 2,min}$, it is rejected and removed from the model.
\item[--] {\em Do not split-up low-mass cores} (SCP-IIc/SCP-IIIc). Cores with a mass $M_T < 2\mmin$ are not split up; these become single stars of mass $M_S=M_T$.
\end{itemize} 
For reasons of simplicity and clarity, we use a generating mass distribution with a minimum value of $2\mmin$ for the SCP pairing functions, and adopt a minimum mass $\mmin$ for the companions. Implicitly, we therefore only consider the variants SCP-IIb and SCP-IIIb in this paper. From hereon, we use ``SCP-II'' and ``SCP-III'' to refer to SCP-IIb and SCP-IIIb, respectively.

\begin{figure*}[tbp]
  \centering
  \includegraphics[width=\textwidth,height=!]{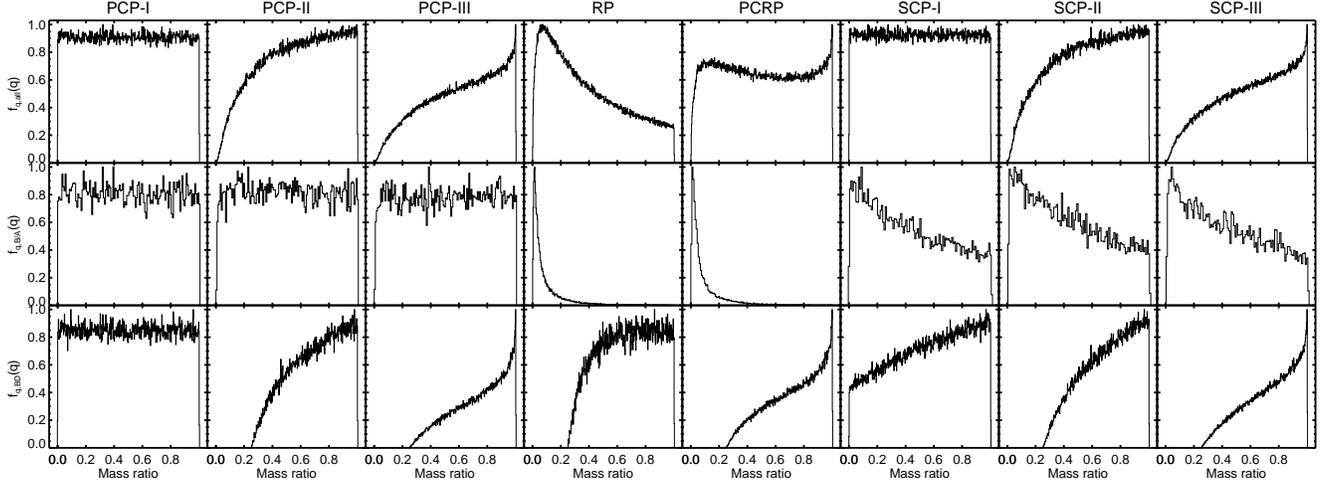}
  \caption{The (renormalised) mass ratio distributions resulting from the different pairing functions. From top to bottom, the panels show the overall mass ratio distribution (i.e., for all binaries), the mass ratio distribution for binaries with $1.5\msun \leq M_1 \leq 20\msun$, and for binaries with $0.02\msun \leq M_1 \leq 0.08\msun$. The models consist of $N=S+B=5\times 10^5$ particles and all have a generating binary fraction of $\binf=100\%$. For each model we adopt a Kroupa generating mass distribution in the mass range $0.02-20\msun$, and, when applicable, a flat generating mass ratio distribution $\fq =1$ ($0 < q \leq 1$). This figure illustrates that each pairing function results in a different overall or specific mass ratio distribution (see also Figs.~\ref{figure:resulting_massdistributions} and~\ref{figure:binaryfraction_versus_spectraltype}). 
  \label{figure:resulting_massratiodistributions} }
\end{figure*}

\begin{figure*}[tbp]
  \centering
  \includegraphics[width=\textwidth,height=!]{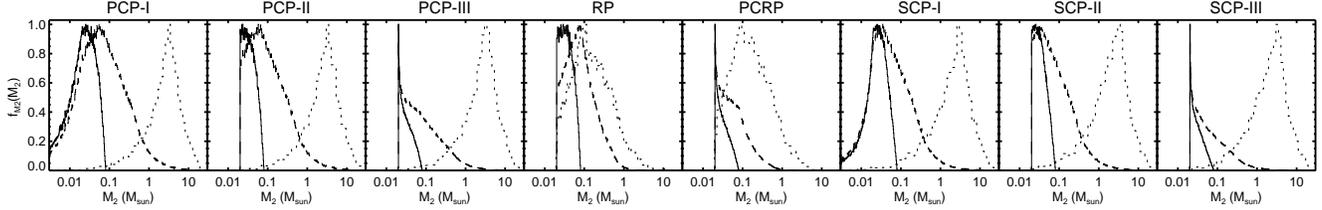}
  \caption{The (renormalised) companion mass distributions resulting from the different pairing functions, for the models shown in Fig.~\ref{figure:resulting_massratiodistributions}. The curves indicate the distribution for all binaries (dashed curves), for binaries with $1.5\msun \leq M_1 \leq 20\msun$ (dotted curves) and for binaries with $0.02\msun \leq M_1 \leq 0.08\msun$ (solid curves).
  \label{figure:resulting_companionmassdistributions} }
\end{figure*}

\begin{figure*}[tbp]
  \centering
    \includegraphics[width=\textwidth,height=!]{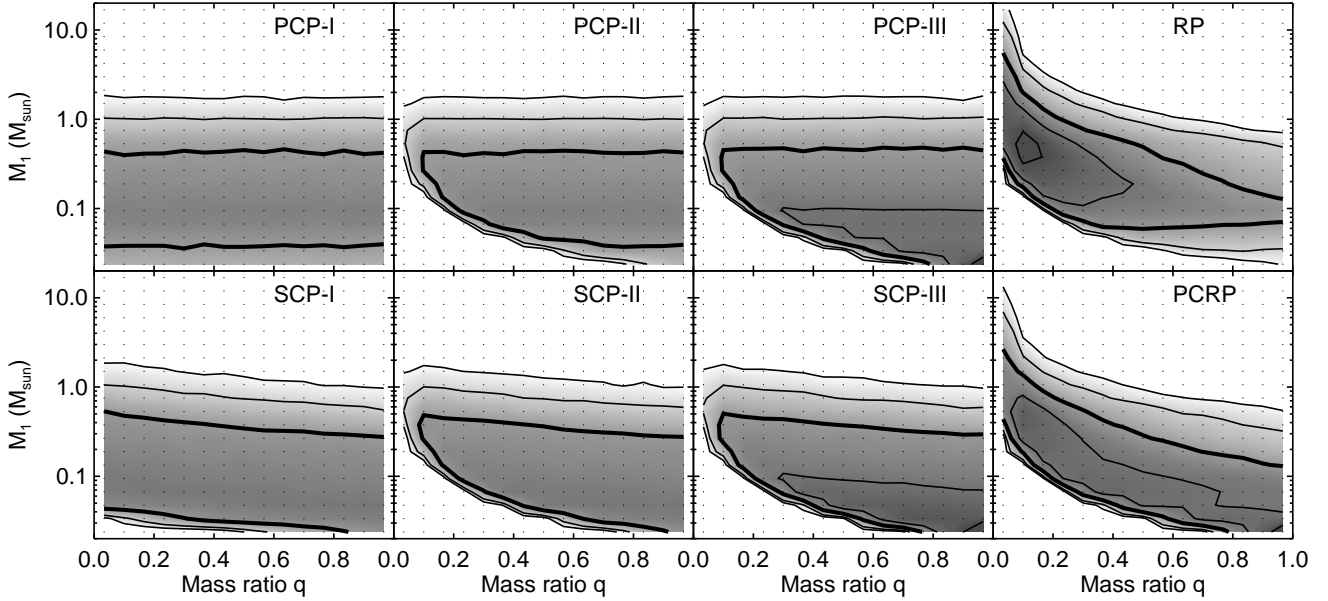}
  \caption{The distribution of binary systems over primary mass $M_1$ and mass
  ratio $q$ for the different pairing functions. 
All populations have a Kroupa generating mass distribution, a generating mass ratio distribution of the form $\fq = 1$ (when applicable), and a generating binary fraction $\binf$ of 100\%. From left to right, the panels show the results for pairing functions PCP-I, PCP-II, PCP-III and RP in the top row, and SCP-I, SCP-II, SCP-III and PCRP on the bottom row.   From light to dark, the contours bound values of 0.01\%, 0.02\%, 0.05\% (thick curve), 0.1\% and 0.2\% for $(q,\log M_1)$. Each model contains $10^6$ binaries. Each dot represents a bin in mass ratio and logarithmic primary mass. No selection effects have been applied.
  \label{figure:massratio_2d_fm} }
\end{figure*}

The above-mentioned pairing functions (RP, PCRP, and the three variations of PCP and SCP) are described in detail in the subsections below, while their differences are discussed in \S\,\ref{section:comparison_pairingfunctions}. Unless stated otherwise, we have adopted a Kroupa generating mass distribution, a flat generating mass ratio distribution (where applicable), and a generating binary fraction $\binf$ of 100\% (see \S\,\ref{section:dependence_generating_properties} for a discussion of these assumptions). 
The main differences between the (renormalised) resulting mass ratio distributions are show in Fig.~\ref{figure:resulting_massratiodistributions}, where the top panel represents the overall mass ratio distribution. The middle and bottom panels represent the specific mass ratio distributions $\fqm$ (the subscript $M_1$ indicates a {\em restricted} primary mass range) for binaries with high-mass primaries and low-mass primaries, respectively. Note that the derived mass ratio distribution for a sample of stars does not only depend strongly on the pairing function, but also on the targeted sample of stars. 
For the same reason, the (renormalised) companion mass distribution, shown in Fig.~\ref{figure:resulting_companionmassdistributions}, depends strongly on the primary mass range. A two-dimensional version of  Fig.~\ref{figure:resulting_massratiodistributions} is shown in Fig.~\ref{figure:massratio_2d_fm}. In the sections below we discuss in detail the pairing functions and the above-mentioned figures.

Note that the choices made in this paper do not imply that stellar populations indeed have these properties. It is not known how binary stars are formed, so that no robust predictions of their properties can be made. Different binary formation mechanisms may produce different mass ratio distributions, possibly varying with primary mass, period, or eccentricity \citep[e.g.][]{heggie1975,krumholz2007,zinneckerreview2007}. In addition, dynamical evolution after the formation process may alter the binary fraction and the mass and mass ratio distributions \citep[e.g.,][]{hills1975,heggie1975}, possibly as a function of environment \citep[see, e.g.,][and \S\,\ref{section:recoveringthepf}]{kroupa1999,preibisch2003,duchene2004,koehler2006,reipurth2007}. Other pairing functions suggested in literature include random pairing over a restricted mass range \citep[see also \S\,\ref{section:restrictedRP}]{kroupa1995a,kroupa1995b,kroupa1995c,kroupabouvierduchene,thies2007}, gravitationally-focused random pairing (Kouwenhoven et al, in prep.), ordered pairing (Oh et al, in prep.), pairing resulting from the dissolution of small-$N$ clusters \citep{clarke_smalln}, ``two-step'' pairing \citep{durisen2001}, binary formation from ring fragmentation \citep{hubber2005}, and numerous others.

\subsection{Pairing function RP (random pairing)} \label{section:pf_rp}

\begin{figure*}[tbp]
  \includegraphics[width=\textwidth,height=!]{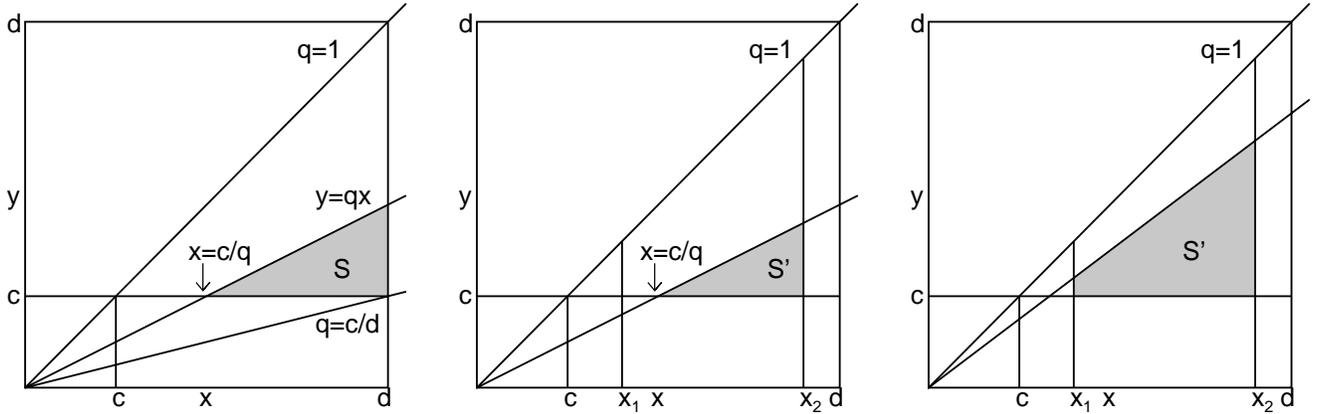} 
  \caption{The integration domains for pairing functions RP and PCRP. {\em Left}: the integration domain in the $(x,y)$ plane (where $x$ and $y$ are the primary and companion mass, respectively) for the determination of the cumulative distribution of $q=y/x$. The lower and upper limits on the generating mass distribution are given by $c$ and $d$, which implies that $c/d\leq q\leq 1$. The integration domain $S$ is given by: $c/q\leq x\leq d \wedge c\leq y\leq qx$. 
{\em Middle and right}: same, for the case that the primary mass $x$ is restricted by $x_1 < x < x_2$. If $q\leq c/x_1$ the integration domain $S'$ is given by $c/q\leq x\leq x_2 \wedge c\leq y\leq qx$ (middle panel), while for $q>c/x_1$, the domain $S'$ is defined by $x_1\leq x\leq x_2 \wedge c\leq y\leq qx$ (right panel).
    \label{fig:fqseldomain} }
\end{figure*}

\begin{figure*}[tbp]
  \centering
  \includegraphics[width=\textwidth,height=!]{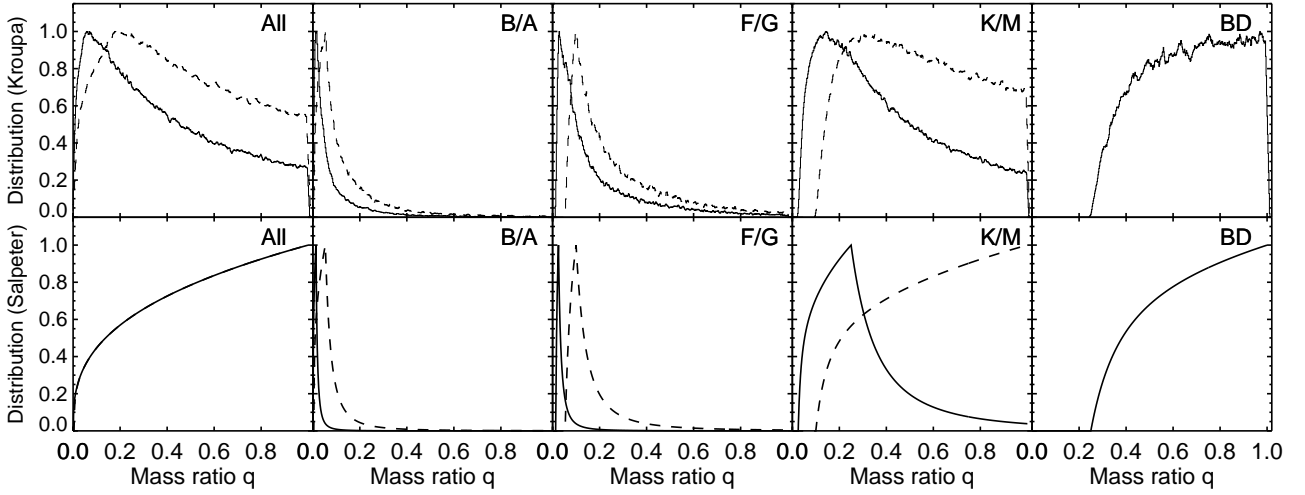}
  \caption{How does the specific mass ratio distribution $\fqm$ resulting from random pairing (RP) depend on the mass distribution and the selected primary mass range? This figure shows the (renormalised) resulting mass ratio distributions for the Kroupa mass distribution (top panels; Monte Carlo simulations) and the Salpeter mass distribution (bottom panels; analytical calculations) in the mass range $c \leq M_1 \leq 20\msun$. The lower limit of $f_M(M)$ is set to $c=0.02 \msun$ for the solid curves, and $c=0.08 \msun$ for the dashed curves; the difference between these models is thus the presence of brown dwarfs. In all panels the models with brown dwarfs have on average a lower mass ratio. The left-hand panel shows the {\em overall} mass ratio distribution (for all binaries in the population). The other panels show the {\em specific} mass ratio distribution for the binaries with a primary of the spectral type shown in the top-right corner of each panel (BD = systems with brown dwarf primaries). All distributions are normalized so that their maximum is unity. Each model contains $5\times 10^5$ binaries and has a binary fraction of 100\%. 
    \label{figure:rp_different_samples} }
\end{figure*}

\begin{figure*}[tbp]
  \centering
  \includegraphics[width=0.95\textwidth,height=!]{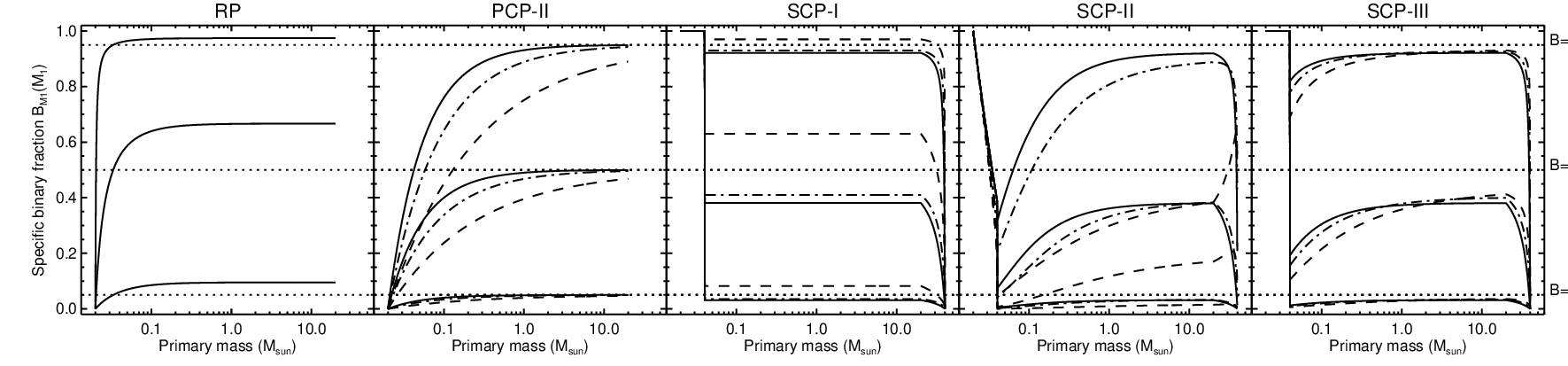} 
  \caption{The specific binary fraction $\binfm$ as a function of primary mass for pairing functions RP, PCP-II, SCP-I, SCP-II and SCP-III. Models with a generating binary fraction $\binf$ of (from bottom to top) 5\%, 50\% and 90\% are shown. The horizontal dotted lines in each panel represent the generating binary fraction. We adopt a Kroupa generating mass distribution with $c=0.02\msun$ and $d=20\msun$. For pairing functions PCP-II, SCP-I, SCP-II and SCP-III we show results for models with generating mass ratio distributions $\fq\propto q^\gamma$, with $\gamma=0$ (solid curves), $\gamma=-0.3$ (dash-dotted curves), and $\gamma=-0.6$ (dashed curves). 
    \label{figure:binaryfraction_versus_spectraltype} }
\end{figure*}

In the case of random pairing (RP), the primary and companion mass are both independently drawn from $f_M(M)$, and swapped, if necessary, so that the most massive star is the primary. As a result of this swapping, neither the resulting primary mass distribution $f_{M_1}(M_1)$, nor the companion mass distribution $f_{M_2}(M_2)$, nor the system mass distribution $f_{M_T}(M_T)$ is equal to the generating mass distribution $f_M(M)$; see, e,g., \cite{warner1961,tout1991,malkovzinnecker2001}. On the other hand, the mass distribution of {\em all} stars $f_{\rm all}(M)$, i.e., all singles, primaries and companions, is equal to $f_M(M)$. \cite{malkovzinnecker2001} derived general expressions for the distribution over primary star mass $M_1$, companion star mass $M_2$, and system mass $M_T=M_1+M_2$, respectively: 
\begin{eqnarray}
f_{M_1}(M_1) = & 2 f_M(M_1) \int_c^{M_1} f_M(M) \, dM  \\
f_{M_2}(M_2) = & 2 f_M(M_2) \int_{M_2}^d f_M(M) \, dM  \\
f_{M_T}(M_T) = & 2 \int_c^{M_T-c} f_M(M) \, f_M(M_T-M) \, dM  
\label{eq:rp_systemmass}
\end{eqnarray}

Below we calculate the mass ratio distributions resulting from random pairing. Let the generating mass distribution $f$ be defined on the interval $[c,d]$. For random pairing we draw two stars with masses $x$ and $y$ from $f$, and swap, if necessary, so that the primary is the most massive star. The resulting mass ratio distribution is bounded by $0 < c/d \leq q \leq 1$. To derive the overall mass ratio distribution we follow the appendix of \cite{piskunov1991} and derive $f_q(q)$ from its cumulative distribution function $F_q(q)=P(y/x\leq q)$:
\begin{equation}
  F_q(q)=\iint\limits_S f_\mathrm{rp}(x,y)dxdy=\int_{c/q}^d\int_c^{qx}
  2f(x)f(y)dydx\,,
  \label{eq:fqrpallcum}
\end{equation}
where the factor 2 accounts for the fact that the pairs of masses are swapped in order to ensure that $y \leq x$. The integration domain is shown in the left-hand panel of Fig.~\ref{fig:fqseldomain}. The probability density $f_q$ for $q$ (i.e., the overall mass ratio distribution) is then given by the derivative of Eq.~(\ref{eq:fqrpallcum}).

For observational reasons, surveys for binarity are often restricted to a certain range of primary spectral types. To derive the effects of the selection on primary mass, the derivation of $f_q(q)$ again proceeds via the cumulative distribution function, which is now given by:
\begin{equation}
  F_q(q)=P(y/x\leq q|x_1\leq x\leq x_2)\propto\iint\limits_{S'}f_\mathrm{rp}(x,y)dxdy\,,
\end{equation}
where the primary mass range is restricted to the range $[x_1,x_2]$. The integration domain $S'$ is as shown in Fig.~\ref{fig:fqseldomain}. The integration limits now depend on whether $q$ is larger or smaller than $c/x_1$. For $q\leq c/x_1$ the integration domain $S'$ is given by $c/q\leq x\leq x_2 \wedge c\leq y\leq qx$ (middle panel in Fig.~\ref{fig:fqseldomain}), while for $q>c/x_1$ $S'$ is defined by $x_1\leq x\leq x_2 \wedge c\leq y\leq qx$ (right panel in Fig.~\ref{fig:fqseldomain}). The minimum possible value of $q$ is $c/x_2$, so that $F(q)=0$ for $q<c/x_2$.
The general expression for the cumulative mass ratio distribution for a sample of binaries with a restricted primary mass range is given in Appendix~\ref{appendix:rp}. The dependence of the mass ratio distribution on the primary mass range for RP is shown in Fig.~\ref{figure:rp_different_samples} for the Kroupa (top) and Salpeter (bottom) generating mass distributions. Note that each mass ratio distribution is renormalised such that its maximum is unity. Note the differences between the overall mass ratio distributions $\fqall$ in the left-hand panels of Fig.~\ref{figure:rp_different_samples}. The Salpeter mass distribution results in a peak in the overall mass ratio distribution at $q\approx 1$. The Kroupa mass distribution, on the other hand, is on average much shallower, and therefore produces a peak at small $q$ (cf. the solid lines in the top panels of Fig.~\ref{fig:fqrpexample}).

In the special case where the mass distribution is of the form $f_M(M) \propto M^{-\alpha}$ with $\alpha \neq 1$, the overall mass ratio distribution is given by
\begin{equation}
  f_q(q) =
  \frac{\gamma}{(1-(c/d)^{\gamma})^2}\left(q^{\alpha-2}-(c/d)^{2\gamma}q^{-\alpha}\right)\,,
  \label{eq:fqsinglepow}
\end{equation}
where $\gamma=\alpha-1$. Realistic mass distributions cover a broad range of masses, i.e., $c \ll d$, for which the expression simplifies to:
\begin{equation}
  f_q(q)\approx
  \begin{cases}
    (\alpha-1)q^{\alpha-2} & \alpha>1 \\
    (1-\alpha)q^{-\alpha} & \alpha<1
  \end{cases}
  \label{eq:fqrpapprox}
\end{equation}
\citep{piskunov1991}. For distributions with $\alpha \approx 1$, the above expression is not a good approximation, and Eq.~(\ref{eq:fqsinglepow}) should be used. In the case of a power-law mass ratio distribution with $c\ll d$, the overall mass ratio distribution resulting from random pairing can thus be described by a simple power-law. Note that Eq.~(\ref{eq:fqrpapprox}) does not depend on $c$ or $d$; see also the bottom-left panel in Fig.~\ref{figure:rp_different_samples}. For a Salpeter mass distribution ($\alpha=2.35$), for example, the resulting overall mass ratio distribution is $\fqall \propto q^{0.35}$. The general expression for $f_q$ when the primary mass range is restricted to the range $[x_1,x_2]$ is given in Appendix~\ref{appendix:rp}. 
For a sample of binaries with a small primary mass range ($x_1 \approx x_2$), $f_q(q)$ is proportional to $q^{-\alpha}$ for $c/x_1\leq q\leq 1$ and zero otherwise:
\begin{equation}
  \fqm =
  \begin{cases}
    0 & q<c/x_1 \\
    \frac{\gamma}{1-(c/x_1)^{\gamma}}q^{-\alpha} & c/x_1\leq q\leq 1 \\
  \end{cases}\,.
  \label{eq:fqselonemass}
\end{equation}
Random pairing from a mass distribution that is approximately a single power-law (e.g. the Kroupa IMF) thus generally results in a three-segment mass ratio distribution, which is zero for $q \la c/x_2$ and exhibits a peak at $q \approx c/x_1$ (Eq.~\ref{eq:fqsel_rp_powerlaw}). As already noted by \cite{zinnecker1984}, a sample of binaries with high-mass primaries thus shows a peak in $\fq$ at low $q$, while a sample with low-mass primaries peaks at high $q$.

For pairing function RP the overall binary fraction $\binfall$ is equal to the generating binary fraction $\binf$. The {\em specific} binary fraction $\binfm$, however, depends on the surveyed primary mass range: the larger the primary mass, the higher the specific binary fraction (unless $\binf = 100\%$).
This can be understood as follows. For random pairing, the mass of the primary stars is drawn from the mass distribution. A fraction $\binf$ of the primaries is assigned a companion, and primary and companion are swapped, if necessary, so that the primary is the most massive star. This swapping leads to an increased number of binaries with a high-mass primary, and a decreased number of binaries with a low-mass primary, and hence a mass-dependent binary fraction.
The relation between specific binary fraction and primary mass for random pairing is given by
\begin{equation}
\binfm = \left( \frac{\binf^{-1} -1 }{2 F_M(M_1)} + 1 \right)^{-1},
\end{equation}
where $F_M(M_1)$ is the cumulative mass distribution evaluated at mass $M_1$ (see Appendix~\ref{appendix:rp_bf}). Fig.~\ref{figure:binaryfraction_versus_spectraltype} shows the binary fraction as a function of primary mass for several binary fractions, for the Salpeter mass distribution. Clearly, we have $\binfm \equiv 1$ when $\binf=100\%$, $\binfm \equiv 0$ when $\binf=0\%$, and $\binf_{M_1}(\mmin) = 0$. Systems with $M_1 > \langle M \rangle$ have $\binfm \geq \binf$, while systems with $M_1 < \langle M \rangle$ have $\binfm \leq \binf$ (see Appendix~\ref{appendix:rp_bf} for a full derivation). The magnitude of this difference depends on the shape of the mass distribution and the generating binary fraction. Note that the variation of binary fraction with primary mass is purely a result of the choice of pairing function; no explicit variation of binary fraction with primary mass is included in the simulations.

\subsubsection{Restricted random pairing (RRP)} \label{section:restrictedRP}

Restricted random pairing (RRP) is very similar to random pairing (RP)
as described in \S\,\ref{section:pf_rp}, with the difference that the
binary components are now drawn from a {\em limited} mass range. All
properties derived in \S\,\ref{section:pf_rp} are thus applicable to
the resulting binary sub-population resulting from RRP. However, the
nature of RRP implies the presence of one or more other
sub-populations that have a formed via another process. The other
sub-populations could have alternative pairing function, such as RRP
with different lower and upper mass limits, or a completely different
pairing function.

\cite{kroupa1995a,kroupa1995b,kroupa1995c} finds that observations of
binary systems are consistent with the population being born with
pairing function RRP in the stellar mass range, prior to the effects
of pre-main sequence eigenevolution. Further motivated by the
difference between the observed mass ratio distribution and semi-major
axis distribution of binary systems with a stellar primary and those
with a brown dwarf binary
\citep[e.g.,][]{bouy2003,burgasser2003,martin2003,close2003}, this
implies that the brown dwarf population has formed with a different
process \cite{kroupabouvierduchene,thies2007,thies2008}. Their
proposed model with stellar and substellar sub-populations is further
supported by the existence of the brown dwarf desert among solar-type
stars (see \S\,\ref{section:constraintsfromobservations}).

\subsection{Pairing function PCRP (primary-constrained random pairing)} \label{section:pf_pcrp}

For primary-constrained random pairing (PCRP), each primary mass $M_1$ is drawn from the mass distribution $f_M(M)$ with limits $c \leq M \leq d$. The companion mass $M_2$ is also drawn from the same mass distribution, but with the additional constraint that $M_2 \leq M_1$. The limits on the resulting mass ratio distribution are equivalent to those of random pairing $0 \leq c/d \leq q \leq 1$.
Writing the primary mass distribution as $f(x)$ and the re-normalised companion mass distribution as $f'(y)$, the expression for the joint probability distribution is $f_\mathrm{pcrp}(x,y) = f_{x}(x){f'}_{y}(y)$,
which is normalised to unity due to the re-normalisation of $f_{y}(y)$.
To derive the overall mass distribution $f_q$ one can proceed as for the RP case (see \S\,\ref{section:pf_rp}). The integration domain $S$ is again as shown in Fig.~\ref{fig:fqseldomain}, and 
\begin{equation}
  F(q)=P(y/x\leq q)=\int_{c/q}^d\int_c^{qx} f(x)f'(y)dydx\,.
  \label{eq:fqpcrpcumall}
\end{equation}
Note that the normalisation constant for $f'(y)$ depends on $x$. The expression for the sample with a restricted primary mass range $[x_1,x_2]$ now becomes:
\begin{equation}
  F_q(q) \propto
  \begin{cases}
    0 & q<c/x_2 \\[3pt]
    \int_{c/q}^{x_2}\int_c^{qx} f(x)f'(y)dydx & c/x_2\leq q<c/x_1 \\[3pt]
    \int_{x_1}^{x_2}\int_c^{qx} f(x)f'(y)dydx & c/x_1\leq q\leq 1
  \end{cases}\,,
  \label{eq:fqpcrpcum}
\end{equation}
where the integration domain $S'$ is as in Fig.~\ref{fig:fqseldomain}. The expression for the resulting $\fqall$ in the case of a single power-law generating mass distribution is given in Appendix~\ref{appendix:pcrp}. For PCRP the distribution $\fqall$ contains a term that diverges for $q \uparrow 1$, which can be seen in Fig.~\ref{figure:resulting_massratiodistributions} (see also Eq.~(\ref{eq:fqpcrp}) in Appendix~\ref{appendix:pcrp}).

For a sample with primary masses restricted to the range $[x_1,x_2]$, Eq.~(\ref{eq:fqpcrpcum}) has to be worked out. We will not explicitly show the results here. Most importantly, $f_q(q)$ is zero for $q<c/x_2$, and exhibits a peak at $q \approx c/x_1$. The distribution for a sample of binaries with high-mass primaries thus peaks at low $q$, and the distribution for low-mass binaries peaks at high $q$. If the primary mass range is small ($x_1 \approx x_2$), $f_q(q)$ can be approximated with Eq.~(\ref{eq:fqselonemass}). For a sample of stars with a very small primary mass range, the mass ratio distributions resulting from PCRP and RP thus give the same results. Differences between the two pairing functions become larger for realistic primary mass ranges.

For pairing function PCRP, the companion mass distribution
$f_{M_2}(M_2)$ {\em for a set of primaries of identical mass} is equal
to the generating mass distribution $f_M(M)$ in the mass range $\mmin
\leq M_2 \leq M_1$. The companion mass distribution can thus in
principle be used to derive the properties of the generating mass
distribution. For example, in a set of binaries with a primary mass of
$1\msun$, those with mass ratio $q<0.08$ have brown dwarf
companions. If the observations are of good enough quality to study
the mass ratio distribution below $q=0.08$, and it is known a-priori
that the pairing function is PCRP, the mass ratio distribution can be
used to constrain the slope of the mass distribution in the brown
dwarf regime.

The pairing algorithms RP and PCRP appear similar, but their
difference is for example seen in the primary mass distribution. For
RP there is a larger number of binary systems with high-mass
primaries, which can be understood as follows. Suppose a primary mass
$M_1$ of $5\msun$ is drawn from the mass distribution. For PCRP, the
companion mass $M_2$ is always smaller than the primary mass, while
for RP the companion mass can take any value permitted by the mass
distribution (i.e. also $M_2 > M_1$, after which the components are
switched). Another difference is that, unlike RP, for PCRP the binary
fraction is independent of primary spectral type: $\binfall = \binfm =
\binf$. For realistic mass distributions (e.g., Salpeter- or
Kroupa-like), the overall mass ratio distribution $\fqall$ of PCRP is
peaked towards high values of $q$, while that of RP is peaked towards
low values of $q$. The pairing functions RP and PCRP can be excluded
if more than a $1-2\%$ of the intermediate-mass stars are 'twins' ($q
\geq 0.8$).

\subsection{Pairing function PCP (primary-constrained pairing)} \label{section:pf_pcp}

In models with primary-constrained pairing (PCP), each binary system
is generated by drawing a primary mass $M_1$ from $f_M(M)$ in the
range $c \leq M \leq d$, and a mass ratio $q$ from the generating
distribution $\hq$. The companion mass is then calculated from $M_2 =
qM_1$. Due to the nature of this pairing mechanism it is possible that
the resulting companion is of very low mass, for example a planetary
mass if a very small mass ratio is drawn. Below we describe three
variants of pairing function PCP, each of which handles very low mass
companions in a different way: accepting all companions (PCP-I),
rejecting the very low-mass companions (PCP-II), and redrawing the
mass ratio if the companion mass is of very low mass (PCP-III).

\subsubsection{Pairing function PCP-I} \label{section:pf_pcpi}

PCP-I is the simplest variant of PCP: the primary mass is drawn from $f_M(M)$ and the mass ratio from $\hq$, and no further constraints are set. As a result, the {\em specific} mass ratio distribution $\fqm$ and overall mass ratio distribution $\fqall$ are equal to the generating mass ratio distribution $\hq$. Additionally, the specific binary fraction $\binfm$ and overall binary fraction $\binfall$ are equal to the generating binary fraction $\binf$. 

The companion mass for PCP-I can be arbitrarily small: $M_{\rm 2,min}=0$. Several companions may thus have masses significantly lower than the deuterium burning limit ($\sim 0.02\msun$). Even planetary companions are considered as companion ``stars'' for the pairing function PCP-I. However, if we do include planets, we make the implicit
assumption that the star formation process is scalable down to planetary masses.
This assumption is in contradiction with the theories that suggest that stars and brown dwarfs form by fragmentation \citep{goodwinpp2007,whitworth2007}, while planets form by
core-accretion \citep[see, e.g.,][]{pollack1996}. This is an important point to keep in mind when using PCP-I, i.e., when adopting a mass ratio distribution that is fully independent of primary spectral type.

\subsubsection{Pairing function PCP-II} \label{section:pf_pcpii}

For PCP-II, companions with $M_2 < c$, are rejected, and the corresponding primary stars are classified as single stars. There are two reasons why one may want to consider using PCP-II. First, one may wish to use this prescription if a minimum companion mass is expected from theory, for example the Jeans mass or the opacity limit for fragmentation \citep[e.g.,][]{hoyle1953,rees1976,low1976,silk1977a,silk1977b,silk1995,tohline1982,larson1969,larson1992,larson2005,masunaga2000}. The second, more observational approach may be to ``ignore'' the low-mass companions. Although this seems somewhat artificial, this method is often used in practice. Planets are usually not considered as companions (the Sun is a ``single star''), which implies a limit $c = 0.01-0.02\msun$.

Due to the rejection of low-mass companions the overall mass ratio distribution is zero for $0<q<c/d$. In the range $c/d\leq q \leq 1$, the expression for the overall mass ratio distribution is given by:
\begin{equation}
  \fqall=k\int_{c/q}^d f_\mathrm{pcp}(M_1,q)dM_1 = k\int_{c/q}^d
  \hq f_{M_1}(M_1) dM_1\,,
  \label{eq:fqpcp2expr}
\end{equation}
where $\hq $ is the generating mass ratio distribution and $k$ is a normalisation constant which ensures that $\int_0^1 \fqall dq=1$. The distribution $\fqall$ has a higher average mass ratio than the generating mass ratio distribution $\hq $ as a result of rejecting the low-mass companions. For a a sample with a restricted primary mass range, $c\leq x_1\leq M_1\leq x_2\leq d$, the expression for the specific mass ratio distribution is given in Appendix~\ref{appendix:pcp}, by Eq.~(\ref{eq:fqselpcp2expr}). The specific mass ratio distribution is given by a three-segment powerlaw, with slope changes at $q=c/x_2$ and $q=c/x_1$.
Note that for $q > c/x_1$, the distribution $\fqm$ is equal to the generating mass ratio distribution. For a sample of high-mass primaries, where $c/x_1 \leq 1$, we thus have $\fqm \approx \hq $. For a sample of low-mass primaries,  $c/x_2\leq q<c/x_1$ for most of the mass ratio range, and the corresponding term dominates. For these binaries, $\fqm$ differs significantly from $\hq $. 

The specific binary fraction for a sample of systems with primary mass $M_1$ is given by
\begin{equation} \label{equation:binaryfraction_pcp2}
  \binfm = \binf \int^1_{\qmin(M_1)} h_q(q) \,dq < \binf \,.
\end{equation}
Note that $\binfm$ is independent of $f_M(M)$ for PCP-II.
Fig.~\ref{figure:binaryfraction_versus_spectraltype} shows the specific binary fraction $\binfm$ as a function of $\binf$ and $\fq$. The specific binary fraction depends on the shape of the mass ratio distribution, and is independent of the mass distribution. The overall binary fraction $\binfall$ {\em after} rejection of the low-mass companion is smaller than the generating binary fraction $\binf$ (see Eq.~(\ref{equation:overall_bf_pcp}) in Appendix~\ref{appendix:pcp_bf}).
The binaries with high-mass primaries are hardly affected by the rejection algorithm. For these binaries the specific mass ratio distribution and specific binary fraction are practically equal to those for PCP-I: the specific binary fraction is equal to $\binf$, and the specific mass ratio distribution is equal to $\hq$.
For the very low-mass primaries, however, a large fraction of the companions is rejected, and therefore the specific binary fraction is low. The remaining companions of these stars have a mass comparable to that of their primary, and the resulting mass ratio distribution for the lowest-mass binaries is peaked to unity (see Fig.~\ref{figure:resulting_massratiodistributions}). 

As an example, consider a stellar population with $\binf=100\%$ and a generating mass ratio distribution $\fq =1$ and $M_{\rm 2,min}=0.02\msun$, systems with B-type primaries and systems with M-type primaries have a resulting specific binary fraction of 99\% and 87\%, respectively. If we also do not consider brown dwarfs as companions (so $M_{\rm 2,min}=0.08\msun$), then the specific binary fractions are 96\% and 51\%, respectively.

\subsubsection{Pairing function PCP-III} \label{section:pf_pcpiii}

For PCP-III the primary mass is drawn from $f_M(M)$ in the range $c \leq M_1 \leq d$, and the mass ratio is drawn from $\hq $. If the resulting companion star mass is smaller than $c$, the mass ratio is redrawn from $\hq $ until a companion with mass $M_2 \geq c$ is obtained. This is equivalent to renormalising $\hq$ in the range $\qmin(M_1) \leq q \leq 1$, where $\qmin(M_1) = c/M_1$. This effectively results in a mass-dependent generating mass ratio distribution ${h'}_q(q)$. The expression for the resulting overall mass ratio distribution $\fqall$ are then:
\begin{equation}
  \fqall =\int_{c/q}^d f_\mathrm{pcp}(M_1,q)dM_1 = \int_{c/q}^d
  {h'}_q(q)f_{M_1}(M_1) dM_1 \,.
\end{equation}
The specific mass ratio distribution for the restricted primary mass range $[x_1,x_2]$ is given in Appendix~\ref{appendix:pcp}.
A sample of binaries with high-mass primaries has $c/x_1 \ll 1$. For high-mass binaries $h'_q(q) \approx \hq$, and therefore $\fqm \approx \hq$.  For the very low-mass primaries, however, a large fraction of the companions is redrawn. Consequently, all binaries with a low-mass primary have a mass ratio close to unity. The resulting mass ratio distribution for the lowest-mass binaries is thus peaked to unity (see Fig.~\ref{figure:resulting_massratiodistributions}). 

As a result of the redrawing of the companions for pairing function PCP-III, the resulting overall binary fraction $\binfall$ equals the generating binary fraction $\binf$, and the specific binary fraction $\binfm$ equals $\binf$ for any primary mass range.

\subsection{Pairing function SCP (split-core pairing)} \label{section:pf_scp}

For split-core pairing  (SCP) one assumes that the system ``core'' mass $M_C$ is drawn from a core mass distribution $f_{M_C}(M_C)$ with $2 \epsilon c \leq M_C \leq 2 \epsilon d$, where $\epsilon$ is the star forming efficiency. Split-core pairing is frequently inferred from observations of dense cores in star forming regions, assuming that a fraction of the cores fragment into binaries \citep[see, e.g.,][]{goodwinsplitup2008,swift2008}. As a core collapses, it forms one or two stars with a total mass $M_T = \epsilon M_C$. The resulting minimum and maximum primary masses are thus $c$ and $2d$, respectively. The star forming efficiency may be a function of various parameters, for example the mass of the core. For simplicity in our analysis, however, we keep the star forming efficiency fixed to $\epsilon=1$ for all values of $M_C$. The total mass of each binary is thus $M_T=M_C$. Note that ``random fragmentation'' (random splitup of a clump into two stellar components) is very different from random pairing of two components from the IMF (see, e.g., Figs.~\ref{figure:resulting_massratiodistributions} and~\ref{figure:resulting_companionmassdistributions}).

For pairing function SCP, the binary total mass is thus drawn from a distribution $f_{M_C}(M_C)$. Note that, although we adopt $f_{M_C}(M_C)=f_{\rm Kroupa}(M)$ in this paper, there is no obvious prerequisite that $f_{M_C}(M_C)$ should be a standard IMF. The binary is split up according to a mass ratio that is drawn from a generating mass ratio distribution $\hq$. Given the core mass $M_C = M_1+M_2$ and the mass ratio $q$, the primary and companion mass are given by
\begin{equation} \label{equation:scpmasses}
  M_1 = \frac{M_C}{q+1} \quad \mbox{and} \quad M_2 = \frac{M_C}{q^{-1}+1} \,,
\end{equation}
respectively. As a result of this procedure, it may happen that a companion mass smaller than the minimum mass $c$ is drawn. Similar to pairing function PCP (\S\,\ref{section:pf_pcp}), there are three ways to address this issue: accepting the low-mass companions (SCP-I), rejecting the low-mass companions (SCP-II), and redrawing the mass ratio if the companion mass is too low (SCP-III). We discuss these three variants of SCP in the sections below.

\subsubsection{Pairing function SCP-I} \label{section:pf_scpi}

For pairing function SCP-I, all binary components resulting from the split-up mechanism are accepted, irrespective of their mass. Stars with substellar and planetary companions are thus also considered as ``binary stars''. 
Due to the nature of this pairing process, the {\em overall} mass ratio distribution is equal to the generating mass ratio distribution. The {\em specific} mass ratio distribution, however, is a function of spectral type. 
A full derivation of the specific mass ratio distribution is given in Appendix~\ref{appendix:scp}; see also \cite{clarke1996}. In most cases the primary mass range is contained within the range $[2c,d]$, which corresponds to ``case 7'' in Appendix~\ref{appendix:scp}.
Consider the special case of a single power-law mass distribution $f_{M_C}(M_C) \propto M_C^{-\alpha}$ ($\alpha\neq 1$) and a uniform mass ratio distribution $\hq =1$. Under these assumptions, the expression for the specific mass ratio distribution is $\fqm \propto (1+q)^{1-\alpha}$ for $0<q\leq 1$, if either the primary mass range is contained within the range $[2c,d]$ (case~7), or if $x_1=x_2$. Note that this expression is identical to that in Eq.~(\ref{equation:1+q_powerlaw}). The highest mass binaries thus have on average a low mass ratio, and the lowest mass binaries have on average a high mass ratio. Note that these trends are present, even though the generating mass ratio distribution produces mass ratios in the range $0<q\leq1$, irrespective of the core mass $M_T$.

Pairing function SCP-I naturally results in a mass-dependent binary fraction:
\begin{equation} \label{equation:specificbinaryfraction_scp1}
  \binf_{M_1}(M_1) = 
  \frac{    \binf f_{M_1}(M_1)  }{    \binf f_{M_1}(M_1) + (1-\binf) f_{M_C}(M_1)  } \,,
\end{equation}
where $f_{M_1}(M_1)$ is the primary mass distribution, $f_{M_C}(M_1)$ the generating (core) mass distribution evaluated at mass $M_1$, and $\binf$ the generating binary fraction; see Appendix~\ref{appendix:scp_bf} for the derivation. 
The mass-dependence of the binary fraction for SCP-I is illustrated in the middle panel of Fig.~\ref{figure:binaryfraction_versus_spectraltype}. In general, $\binf=100\%$ for $c \leq M_1<2c$, as $f_{M_C} = 0$ in this mass range. For $2c<M_1<d$ the specific binary fraction is more or less independent of $M_1$, while beyond $M_1=d$, it decreases down to zero at $M_1=2d$. The latter dependence is due to the fact that the cores are more massive than the primary stars they potentially form. As a result, the high-mass targets are dominated by cores that have not split up.
The overall binary fraction $\binfall$ is always equal to $\binf$.

\subsubsection{Pairing function SCP-II} \label{section:pf_scpii}

For SCP-II the companion is rejected if $M_2 < c$, and the primary star becomes single. The resulting primary mass range is then $c \leq M_1 \leq 2d-c$, and the companion mass is in the range $c \leq M_2 \leq d$.
The full derivation for the overall and specific mass ratio distributions is given in Appendix~\ref{appendix:scp}. In most realistic cases the primary mass range $[x_1,x_2]$ is fully enclosed in the mass range $2c<M_1<d$ (``case 1''). 
Unlike SCP-I, the resulting overall mass ratio distribution is unequal to the generating mass ratio distribution, but contains more high-$q$ binaries instead.

The specific binary fraction $\binfm$ resulting from SCP-II varies with primary mass $M_1$:
\begin{equation} \label{equation:scpii_binfm1}
  \binfm = 
  \frac{
    \binf_{100}(M_1)\binf f_{M_1}(M_1)
  }{
    \binf f_{M_1}(M_1) + (1-\binf) f_{M_C}(M_1)
  }
  \,,
\end{equation}
where $f_{M_1}$ is the primary mass distribution, $f_{M_C}$ is the generating (core) mass distribution, and $\binf$ is the generating binary fraction. The quantity $\binf_{100}(M_1)$ represents the value of $\binfm$ for a population with $\binf=100\%$. The full derivation of Eq.~(\ref{equation:scpii_binfm1}) is given in Appendix~\ref{appendix:scp_bf}.

Fig.~\ref{figure:binaryfraction_versus_spectraltype} shows the specific binary fraction $\binf(M_1)$ as a function of $\binf$ and $\hq$. 
The specific binary fraction decreases with decreasing primary mass, as, on average, more low-mass companions are rejected among lower-mass primaries. The majority of the newly formed single stars (due to rejection of low-mass companions) is thus of most low mass. For the very lowest-mass stars, however, the binary fraction increases to unity, as $\binf_{100}(M_1) \approx 1$ and $f_{M_C} =0$ for $M_1 \approx c$. The specific binary fraction then rises again to a maximum around $M_1 = d$, and then rapidly drops to zero at $M_1 = 2d-c$. As a result of the rejection of low-mass companions, the overall binary fraction is always smaller than $\binf$.

\subsubsection{Pairing function SCP-III} \label{section:pf_scpiii}

For SCP-III the mass ratio is redrawn when a companion with mass $M_2<c$ is produced by the splitting algorithm (similar to PCP-III). This effectively corresponds to a (mass-specific) re-normalised mass ratio distribution ${h'}_q(q)$ in the range $c/(M_C-c)\leq q\leq 1$. The resulting overall mass ratio distribution for SCP-III is given by:
\begin{equation}
  \fqall=k\int_{c(1+1/q)}^{2d} {h'}_q(q)f_{M_C}(M_C)dM_C\,,
\end{equation}
where the lower integration limit is set by the condition $M_2\geq c$ and $k$ is again a normalisation constant. The expressions for the specific distribution $\fqm$ for a restricted primary mass range are identical to those for the SCP-II case (see Appendix~\ref{appendix:scp}), except that ${h'}_q(q)$ replaces $h_q(q)$ everywhere. As a result of the renormalisation of the mass ratio distribution, the mass ratio distribution is a function of spectral type. The specific mass ratio distribution for the lowest-mass cores is strongly peaked to $q=1$. 

The overall binary fraction $\binfall$ equals $\binf$ for SCP-III. The specific binary fraction is given by Eq.~(\ref{equation:specificbinaryfraction_scp1}). Note however, that the primary mass distribution $f_{M_1}(M_1)$ resulting from SCP-III is different from that of SCP-III; see Appendix~\ref{appendix:scp_bf} for details.

\begin{figure*}[tbp]
  \centering
  \includegraphics[width=\textwidth,height=!]{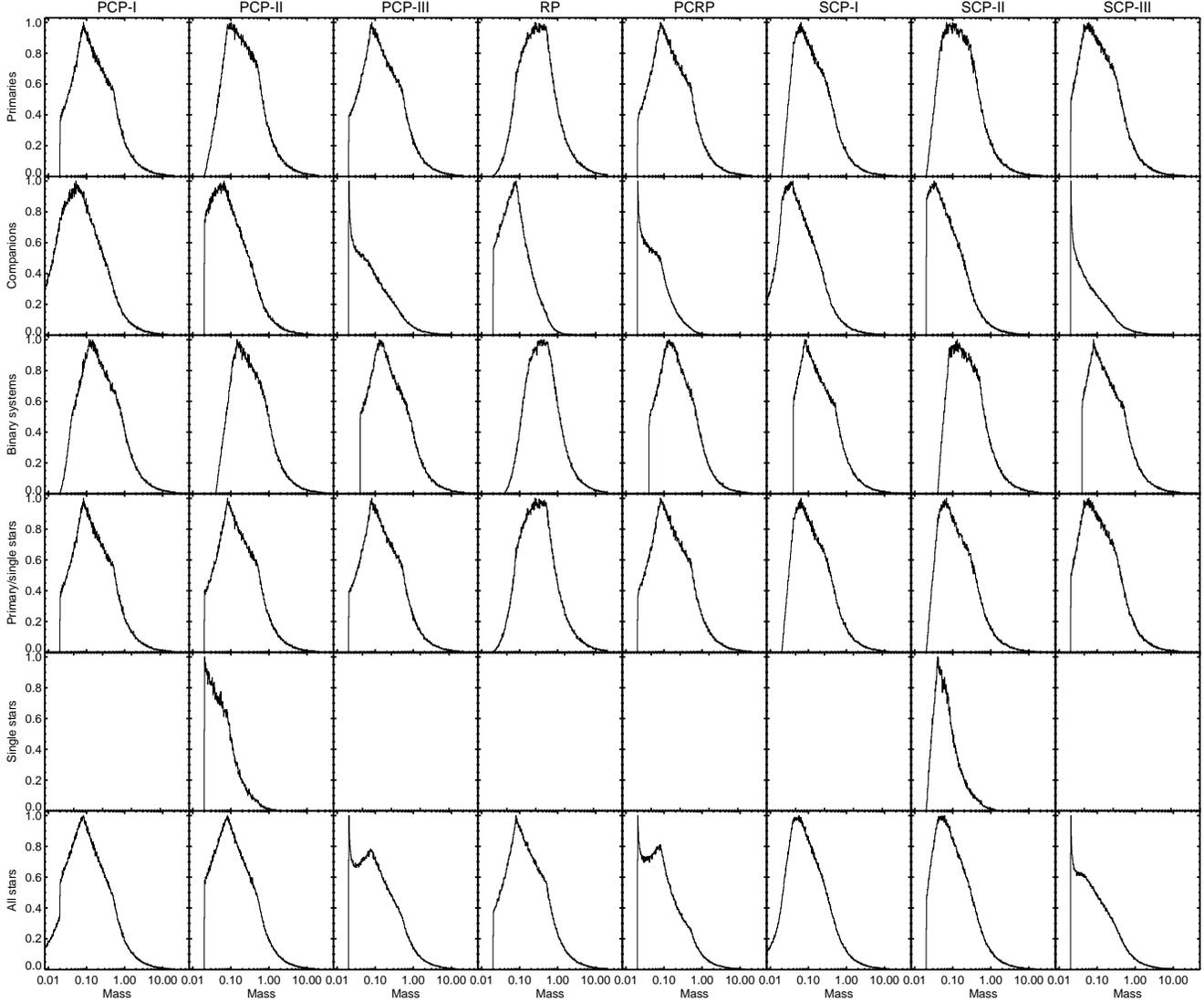}
  \caption{The mass distributions resulting from the different pairing functions, for the models shown in Fig.~\ref{figure:resulting_massratiodistributions}. From top to bottom, the panels show the primary mass distribution $f_{M_1}$, companion mass distribution $f_{M_2}$, the system mass distribution $f_{M_T}$, the mass distribution of the primaries and single stars combined $f_{M_{1,S}}$ (i.e., the targets in an observational binarity study), the single star mass distribution $f_{M_S}$, and the distribution of all stellar masses $f_{M_{\rm all}}$. Each distribution is normalised so that its maximum is unity. The vertical axis has a linear scale (rather than a logarithmic scale; cf. Fig.~\ref{figure:sos_massdistributions}) so that the differences are clearly shown. Each column corresponds to a different pairing function, which is indicated at the top. 
  \label{figure:resulting_massdistributions} }
\end{figure*}

\section{Differences between pairing functions} \label{section:comparison_pairingfunctions}

As discussed in the previous sections, each pairing function results in different properties of the binary population. In this section we provide an overview of the major differences and similarities. In general, each of the properties described below depends on the choice of the pairing function, a generating mass distribution $f_M(M)$, a generating mass ratio distribution $\fq$, and a generating binary fraction $\binf$. Our example models have a Kroupa generating mass distribution with $0.02 \leq M \leq 20 \msun$, a flat generating mass ratio distribution (if applicable), and a generating binary fraction of 100\%. The resulting mass distributions and mass ratio distributions for the different pairing functions are shown in Figs.~\ref{figure:resulting_massratiodistributions} and~\ref{figure:resulting_massdistributions}, respectively.

\subsection{The mass distributions} \label{section:pf_fm}

For pairing functions PCP-I, PCP-III and PCRP the primary mass distribution is identical to $f_M(M)$.  For pairing function PCP-II,  $f_{M_1}(M)$ is very similar to $f_M(M)$, but it contains less primaries of low mass due to rejection of very low-mass companions (which mostly occurs among low-mass primaries). The latter ``primaries'' are considered as single stars after removal of their companions. For pairing function RP the primary mass distribution is more massive than $f_M(M)$ due to swapping of the components that are drawn from $f_M(M)$. Pairing functions SCP-I, SCP-II and SCP-III result in smaller average primary masses than $f_M(M)$ due to the the core splitting.

For all pairing functions the companion mass distribution $f_{M_2}(M)$ is shifted to lower masses with respect to the generating mass distribution $f_M(M)$. Pairing functions PCP-III, PCRP and SCP-III result in a large number of low mass companions; the companion mass distribution is sharply peaked at $M_2 \approx c$; see Figs.~\ref{figure:resulting_companionmassdistributions} and~\ref{figure:resulting_massdistributions}. For a Kroupa generating mass distribution, the other pairing functions show a peak in the companion mass distribution around $M_2 \approx 0.05 \msun$. Pairing functions PCP-I and SCP-I may result in arbitrarily small companion masses, but all other pairing functions have $M_{\rm 2,min}=c$. All pairing functions have $M_{\rm 2,max}=d$. For each pairing function described in this paper, the companion mass distribution depends strongly on the selected primary mass range (see Fig.~\ref{figure:resulting_companionmassdistributions}), but is independent of $\binf$. Note, however, that \cite{metchev2008} find evidence for a universal companion mass distribution among stellar and substellar primaries.

We define the binary {\em system mass distribution} $f_{M_T}(M)$ as the distribution of masses $M_T = M_1 + M_2$ of all binary systems in a population. For pairing function SCP-I the system mass distribution equals the generating mass distribution: $f_{M_T}(M)=f_{M}(M)$. For all other pairing functions the expression for the system mass distribution is different from $f_M(M)$, $f_{M_1}(M)$ and $f_{M_2}(M)$. For random pairing, for example, the system mass distribution can be described as a convolution (Eq.~\ref{eq:rp_systemmass}).

Pairing functions with $\binf < 100\%$ result in a population of single stars. For PCP-II and SCP-II, even those with $\binf=100\%$ result in single stars. In our analysis we discriminate between the {\em single mass distribution} $f_{M_{S}}(M)$ and the {\em primary/single mass distribution} $f_{M_{1,S}}(M)$. The latter distribution includes both single stars and primaries and is important for observers. Target lists for binarity surveys are often defined by $f_{M_{1,S}}(M)$ for a certain mass range, as in practice it is often unknown whether a surveyed star has a companion star. 
For most pairing functions the single mass distribution $f_{M_S}(M)$ is equal to the generating mass distribution $f_M(M)$, and the fraction of single stars among the systems is given by $1-\binf$, where $\binf$ is the generating binary fraction. For PCP-II and SCP-II additional single stars are created due to the rejection of very low-mass companions. The generated single stars are mostly of low mass. 
Only for pairing functions PCP-I, PCP-II, PCP-III and PCRP, the primary/single star mass distribution $f_{M_{1,S}}(M)$ is equal to the generating mass distribution $f_M(M)$. The primary/single star mass distribution is biased to higher masses for RP, and to lower masses for SCP-I, SCP-II and SCP-III.

The bottom row of Fig.~\ref{figure:resulting_massdistributions} shows the mass distribution $f_{all}(M)$ of all stars, including primaries, companions, and single stars. This mass distribution, which includes all stars in the population, is referred to as the {\em mass function} of a stellar population. For a zero-age population, $f_{all}(M)$ is an {\em initial mass function} (IMF), e.g., the Kroupa IMF (Eq.~\ref{equation:kroupaimf}). Only for pairing function RP is the individual star mass distribution $f_{all}(M)$ equal to the generating mass distribution $f_M(M)$; the other pairing functions result in a distribution  $f_{all}(M)$ which is biased to lower values with respect to $f_M(M)$.

\subsection{The mass ratio distributions}   \label{section:pf_fqo}

The resulting overall mass ratio distribution $\fqall$ depends on the pairing function. The top panels in Fig.~\ref{figure:resulting_massratiodistributions} show the significant difference between the overall mass ratio distribution for the different pairing functions. The overall mass ratio distribution $\fqall$ is equal to the generating mass ratio distribution $\fq$ for pairing functions PCP-I and SCP-I. For pairing functions PCP-II, PCP-III, SCP-II and SCP-III the overall mass ratio distribution is biased to higher values of $q$ with respect to $\fq$. For these pairing functions binaries with very low companion masses, and thus often very low mass ratios, are either rejected or redrawn, resulting in systematically higher values of $q$. The overall mass ratio distribution of pairing functions RP and PCRP are purely a result of the mass distribution $f_M(M)$ and depend strongly on its lower and upper limits.

Only for pairing function PCP-I is the mass ratio distribution independent of spectral type, and thus equal to the overall and generating mass ratio distributions. For all other pairing functions the specific mass ratio distribution is a function of the primary mass. The middle and bottom panels in Fig.~\ref{figure:resulting_massratiodistributions} show for each pairing function the {\em specific} mass ratio distribution for target samples of different spectral types. For each of these pairing functions, high-mass binaries have on average a lower mass ratio than low-mass binaries. The lowest-mass binaries have a mass ratio distribution peaked to $q=1$. 

Figs.~\ref{figure:resulting_massratiodistributions} and~\ref{figure:rp_different_samples} illustrate the strong dependence of the specific mass ratio distribution on the targeted sample in the survey. Fig.~\ref{figure:massratio_2d_fm} shows a generalised version of these figures. Each panel shows the two-dimensional distribution $f(q,M_1)$ for the different pairing functions. Care should thus be taken when extrapolating the results to the population as a whole. The interpretation of the observations is further complicated by the instrument bias and observational errors, an effect we will discuss in \S\,\ref{section:strategy}. An overview of the mass ratio distribution changes is presented in Table~\ref{table:pf_massratiodistribution}.

\begin{table}[tbp]
  \caption{The mass ratio distribution resulting from different pairing functions, for different samples, as compared to the generating mass ratio distribution $\hq$. The columns show the pairing function, the overall mass ratio distribution, the specific mass ratio distribution for high-mass stars, the specific mass ratio distribution for low-mass stars. The symbols the table indicate whether the mass ratio distribution is equal to ($=$), almost equal to ($\approx$), or biased to low mass ratios ($\downarrow$) or high mass ratios ($\uparrow$), with respect to $\hq$. As the distribution $\hq$ is undefined for RP and PCRP, the properties of $\fqm$ with respect to $\fqall$ are indicated.
    \label{table:pf_massratiodistribution} }
  \begin{tabular}{lccc}
    \hline
    \hline
    Pairing & $\fqall$ & $\fq_{M_1 \approx \mmax}$ & $\fq_{M_1 \approx \mmin}$ \\
    \hline
    RP         & $-$           & $\downarrow$    & $\uparrow$    \\
    PCRP       & $-$           & $\downarrow$    & $\uparrow$    \\
    \hline
    PCP-I      & $=$           & $=$             & $=$           \\
    PCP-II     & $\downarrow$  & $\approx$       & $\uparrow$     \\
    PCP-III    & $\uparrow$    & $\approx$       & $\uparrow$   \\
    \hline
    SCP-I      & $=$           & $\downarrow$    & $\uparrow$    \\
    SCP-II     & $\uparrow$    & $\downarrow$    & $\uparrow$   \\
    SCP-III    & $\uparrow$    & $\downarrow$    & $\uparrow$    \\
    \hline
    \hline
  \end{tabular}
\end{table}

\subsection{The binary fractions} \label{section:pf_binf}

The overall binary fraction $\binfall$ is equal to the generating binary fraction $\binf$ for most pairing functions. Only for PCP-II and SCP-II the overall binary fraction is smaller than $\binf$ because of the rejection of low-mass companions. 
Table~\ref{table:pf_binaryfraction} provides an overview of the changes in the binary fraction as a function of spectral type and primary mass range. Most pairing functions result in a mass-dependent binary fraction. If the generating binary fraction is smaller than 100\%, all pairing functions except PCP-I, PCP-III and PCRP result in a specific binary fraction that depends on primary mass (see, e.g., Fig.~\ref{figure:binaryfraction_versus_spectraltype}). On the other hand, if $\binf=100\%$, only pairing functions PCP-II and SCP-II result in a varying $\binfm$, due to the rejection of low-mass companions.

For a sample of binaries with high-mass primaries, $\binfm$ is approximately equal to $\binf$ for pairing functions PCP-I, PCP-II, PCP-III and PCRP. For RP, the specific binary fraction for high-mass binaries is larger than $\binf$ (unless $\binf=100\%$), while for SCP-I, SCP-II and SCP-III the binary fraction for high-mass stars is smaller than $\binf$ (unless $\binf=100$). For low-mass binaries, $\binfm$ is equal to $\binf$ for pairing functions PCP-I, PCP-III, and PCRP. For RP, the specific binary fraction for low-mass stars is only equal to the generating binary fraction if the latter is $\binf = 0\%$ or $\binf = 100\%$, and smaller in the other cases. For PCP-II the specific binary fraction for low-mass stars is smaller than $\binf$. For SCP-I and SCP-III the binary fraction for low-mass stars is larger than $\binf$. For SCP-II the specific binary fraction for low-mass stars may be larger or smaller than $\binf$, depending on the properties of $f_M(M)$ and $\fq$ and the value of $\binf$.

\begin{table}[tbp]
  \caption{The specific binary fraction $\binfm$ as compared to the generating binary fraction $\binf$, for different the pairing functions. The columns show the pairing function, the overall binary fraction $\binfall$, the specific binary fraction for high-mass stars $\binf_{M_1 \approx \mmax}$ and for low-mass stars $\binf_{M_1 \approx \mmin}$. The last column shows whether the resulting binary fraction is equal to $\binf$ for any primary mass range. See also Fig.~\ref{figure:binaryfraction_versus_spectraltype} for several examples.
    \label{table:pf_binaryfraction} }
  \begin{tabular}{lllll}
    \hline
    \hline
    Pairing & $\binfall$ & $\binf_{M_1 \approx \mmax}$ & $\binf_{M_1 \approx \mmin}$ & $\binfm = \binf$? \\
    \hline
    RP         & $=$  & $>$        & $<$         & no, unless $\binf=100\%$  \\
    PCRP       & $=$  & $=$        & $=$         & yes  \\
    \hline
    PCP-I      & $=$  & $=$        & $=$         & yes  \\
    PCP-II     & $<$  & $\approx$  & $<$         & no   \\
    PCP-III    & $=$  & $=$        & $=$         & yes \\
    \hline
    SCP-I      & $=$  & $<,\approx,>$  & $<,\approx,>$ & no, unless $\binf=100\%$  \\
    SCP-II     & $<$  & $<,\approx,>$  & $<,\approx,>$ & no            \\
    SCP-III    & $=$  & $<,\approx,>$  & $<,\approx,>$ & no, unless $\binf=100\%$   \\
    \hline
    \hline
  \end{tabular}
\end{table}

\section{Dependence on generating properties} \label{section:dependence_generating_properties}

\begin{table*}
  \begin{tabular}{| l | p{0.2cm}p{0.2cm}p{0.2cm}p{0.2cm} p{0.2cm}p{0.2cm}p{0.2cm}p{0.2cm} | p{0.2cm}p{0.2cm}p{0.2cm}p{0.2cm} p{0.2cm}p{0.2cm}p{0.2cm}p{0.2cm} | p{0.2cm}p{0.2cm}p{0.2cm}p{0.2cm} p{0.2cm}p{0.2cm}p{0.2cm}p{0.2cm} |}
    \hline    
    Generating prop. & \multicolumn{8}{c}{$f_M(M)$}  & \multicolumn{8}{|c|}{$f_q(q)$}  & \multicolumn{8}{c|}{$\binf$} \\ 
    \hline
    \multirow{0}{*}{Pairing function}   & 
    \multirow{0}{*}{ \rotatebox{-90}{\mbox{\!\!\!\!RP}}  }  & 
    \multirow{0}{*}{ \rotatebox{-90}{\mbox{\!\!\!\!PCRP}}  }  & 
    \multirow{0}{*}{ \rotatebox{-90}{\mbox{\!\!\!\!PCP-I}}  }  & 
    \multirow{0}{*}{ \rotatebox{-90}{\mbox{\!\!\!\!PCP-II}}  }  & 
    \multirow{0}{*}{ \rotatebox{-90}{\mbox{\!\!\!\!PCP-III}}  }  & 
    \multirow{0}{*}{ \rotatebox{-90}{\mbox{\!\!\!\!SCP-I}}  }  & 
    \multirow{0}{*}{ \rotatebox{-90}{\mbox{\!\!\!\!SCP-II}}  }  & 
    \multirow{0}{*}{ \rotatebox{-90}{\mbox{\!\!\!\!SCP-III}}  }  & 
    \multirow{0}{*}{ \rotatebox{-90}{\mbox{\!\!\!\!RP}}  }  & 
    \multirow{0}{*}{ \rotatebox{-90}{\mbox{\!\!\!\!PCRP}}  }  & 
    \multirow{0}{*}{ \rotatebox{-90}{\mbox{\!\!\!\!PCP-I}}  }  & 
    \multirow{0}{*}{ \rotatebox{-90}{\mbox\!\!\!\!{PCP-II}}  }  & 
    \multirow{0}{*}{ \rotatebox{-90}{\mbox{\!\!\!\!PCP-III}}  }  & 
    \multirow{0}{*}{ \rotatebox{-90}{\mbox{\!\!\!\!SCP-I}}  }  & 
    \multirow{0}{*}{ \rotatebox{-90}{\mbox{\!\!\!\!SCP-II}}  }  & 
    \multirow{0}{*}{ \rotatebox{-90}{\mbox{\!\!\!\!SCP-III}}  }  & 
    \multirow{0}{*}{ \rotatebox{-90}{\mbox{\!\!\!\!RP}}  }  & 
    \multirow{0}{*}{ \rotatebox{-90}{\mbox{\!\!\!\!PCRP}}  }  & 
    \multirow{0}{*}{ \rotatebox{-90}{\mbox{\!\!\!\!PCP-I}}  }  & 
    \multirow{0}{*}{ \rotatebox{-90}{\mbox{\!\!\!\!PCP-II}}  }  & 
    \multirow{0}{*}{ \rotatebox{-90}{\mbox{\!\!\!\!PCP-III}}  }  & 
    \multirow{0}{*}{ \rotatebox{-90}{\mbox{\!\!\!\!SCP-I}}  }  & 
    \multirow{0}{*}{ \rotatebox{-90}{\mbox{\!\!\!\!SCP-II}}  }  & 
    \multirow{0}{*}{ \rotatebox{-90}{\mbox{\!\!\!\!SCP-III}}  }  \\
    & \multicolumn{8}{c}{}  & \multicolumn{8}{|c|}{}  & \multicolumn{8}{c|}{} \\ 
    & \multicolumn{8}{c}{}  & \multicolumn{8}{|c|}{}  & \multicolumn{8}{c|}{} \\    
    \hline
    $f_{M_1}$          & $\surd$  & $\surd$  & $\surd$  & $\surd$  & $\surd$  & $\surd$  & $\surd$  & $\surd$    & --       & --       & $\times$ & $\surd$  & $\times$ & $\surd$  & $\surd$  & $\surd$    & $\times$ & $\times$ & $\times$ & $\times$ & $\times$ & $\times$ & $\times$ & $\times$ \\
    $f_{M_2}$          & $\surd$  & $\surd$  & $\surd$  & $\surd$  & $\surd$  & $\surd$  & $\surd$  & $\surd$    & --       & --       & $\surd$  & $\surd$  & $\surd$  & $\surd$  & $\surd$  & $\surd$    & $\times$ & $\times$ & $\times$ & $\times$ & $\times$ & $\times$ & $\times$ & $\times$ \\
    $f_{M_T}$          & $\surd$  & $\surd$  & $\surd$  & $\surd$  & $\surd$  & $\surd$  & $\surd$  & $\surd$    & --       & --       & $\surd$  & $\surd$  & $\surd$  & $\times$ & $\surd$  & $\times$   & $\surd$  & $\surd$  & $\surd$  & $\surd$  & $\surd$  & $\times$ & $\surd$  & $\times$ \\
    $f_{M_{1,S}}$      & $\surd$  & $\surd$  & $\surd$  & $\surd$  & $\surd$  & $\surd$  & $\surd$  & $\surd$    & --       & --       & $\times$ & $\times$ & $\times$ & $\surd$  & $\surd$  & $\surd$    & $\surd$  & $\times$ & $\times$ & $\times$ & $\times$ & $\surd$  & $\surd$  & $\surd$  \\
    $f_{M_S}$          & $\surd$  & $\surd$  & $\surd$  & $\surd$  & $\surd$  & $\surd$  & $\surd$  & $\surd$    & --       & --       & $\times$ & $\surd$  & $\times$ & $\times$ & $\surd$  & $\times$   & $\times$ & $\times$ & $\times$ & $\surd$  & $\times$ & $\times$ & $\surd$  & $\times$ \\
    $f_{M_{\rm all}}$ (``IMF'')  & $\surd$  & $\surd$  & $\surd$  & $\surd$  & $\surd$  & $\surd$  & $\surd$  & $\surd$    & --       & --       & $\surd$  & $\surd$  & $\surd$  & $\surd$  & $\surd$  & $\surd$    & $\times$ & $\surd$  & $\surd$  & $\surd$  & $\surd$  & $\surd$  & $\surd$  & $\surd$  \\
    \hline
    $\fqall$           & $\surd$  & $\surd$  & $\times$ & $\surd$  & $\surd$  & $\times$ & $\surd$  & $\surd$    & --       & --       & $\surd$  & $\surd$  & $\surd$  & $\surd$  & $\surd$  & $\surd$    & $\times$ & $\times$ & $\times$ & $\times$ & $\times$ & $\times$ & $\times$ & $\times$ \\
    $\fqm$             & $\surd$  & $\surd$  & $\times$ & $\times$ & $\times$ & $\surd$  & $\surd$  & $\surd$    & --       & --       & $\surd$  & $\surd$  & $\surd$  & $\surd$  & $\surd$  & $\surd$    & $\times$ & $\times$ & $\times$ & $\times$ & $\times$ & $\times$ & $\times$ & $\times$ \\
    \hline
    $\binfall$         & $\times$ & $\times$ & $\times$ & $\surd$  & $\times$ & $\times$ & $\surd$  & $\times$   & --       & --       & $\times$ & $\surd$  & $\times$ & $\times$ & $\surd$  & $\times$   & $\surd$  & $\surd$  & $\surd$  & $\surd$  & $\surd$  & $\surd$  & $\surd$  & $\surd$  \\
    $\binfm$           & $\surd$  & $\times$ & $\times$ & $\times$ & $\times$ & $\surd$  & $\surd$  & $\surd$    & --       & --       & $\times$ & $\surd$  & $\times$ & $\surd$  & $\surd$  & $\surd$    & $\surd$  & $\surd$  & $\surd$  & $\surd$  & $\surd$  & $\surd$  & $\surd$  & $\surd$  \\
    \hline
  \end{tabular}
  \caption{Do the properties of a population resulting from a certain pairing function depend on its generating properties? The three generating properties $f_M(M)$, $\fq$ and $\binf$ are listed in the top row. For each of the ten quantities listed in the left-most column, we list whether or not they depend on the choice of $f_M(M)$, $\fq$ and $\binf$. We indicate dependence and independence with the symbols $\surd$ and $\times$, respectively. The results for $\fqm$ and $\binfm$ are valid only for a population of binaries with identical primary mass $M_1$. The distribution over total system mass $f_{M_T}$ includes both binary systems and single stars. 
  \label{table:dependencies}
}
\end{table*}

In this section we discuss how the properties of a binary population
depend on the attributes for the pairing functions: the generating
mass distribution $f_M(M)$, the generating mass ratio distribution
$\fq$, and the generating binary fraction $\binf$. For the most
important properties of a population (with respect to binarity), we
list in Table~\ref{table:dependencies} whether or not they depend on
$f_M(M)$, $\fq$, or $\binf$, for each of the eight pairing functions
described in this paper. Note that the system mass distribution
$f_{M_T}$ includes both single stars and binary systems. The specific
mass ratio distribution $\fqm$ and specific binary fraction $\binfm$
in Table~\ref{table:dependencies} only refer to samples where all
binaries have identical primary mass $M_1$. For a sample with a finite
primary mass, range, the results for the specific mass ratio
distribution and specific binary fraction are mostly identical to
those of $\fqall$ and $\binfall$,
respectively. Table~\ref{table:dependencies} also illustrates which
properties of the population can be used to recover $f_M(M)$, $\fq$,
and $\binf$. For example, for RP, the specific binary fraction
$\binfm$ provides information on the generating mass distribution
(e.g., the IMF).

{\em The generating mass distribution}. For obvious reasons, all mass
distributions listed in Table~\ref{table:dependencies} depend on
$f_M(M)$. The mass ratio distributions for RP and PCRP are defined by,
and depend strongly on the properties of $f_M(M)$ and $\binf$. For
PCP-I/II/III, $\fqm$ does not depend on $f_M(M)$, as both $M_1$ and
$q$ are drawn independently from their generating
distributions. Obtaining the overall mass ratio distribution $\fqall$
requires integration over the primary mass distribution; as for PCP-II
and PCP-III $\fqm$ varies with $M_1$, so does $\fqall$. The overall
binary fraction $\binfall$ is independent of $f_M(M)$, except for
PCP-II and SCP-II, for which low-mass companions are rejected. The
specific binary fraction $\binfm$, does not depend on $f_M(M)$ for
PCRP and PCP-I/II/III due to the independent drawing of $M_1$ and
$M_2$ (or $q$), while it does vary with $f_M(M)$ for SCP-I/II/III as
$M_1$ and $q$ are not drawn independently, and for RP as $M_1$ and
$M_2$ are not drawn independently (due to the swapping of the
components; see \S\,\ref{section:pf_rp}).

{\em The generating mass ratio distribution}. The
generating mass ratio distribution $\fq$ is undefined for RP and
PCRP. For the other pairing functions, the dependence of $\fqm$ and
$\fqall$ on $\fq$ is obvious. As companion masses are derived from
$q$, the distributions $f_{M_2}$ and $f_{M_{\rm all}}$ depend on the choice of
$\fq$. Note that that for SCP-II, all parameters vary with $\fq$, and
for PCP-II most parameters (except the primary/single mass
distribution) vary with $\fq$. The properties of the single stars do
not depend on $\fq$, except for PCP-II and SCP-II, where additional
single stars are created due to the rejection of low-mass companions.

{\em The generating binary fraction}. The dependence of
$\binfall$ and $\binfm$ on the generating binary fraction $\binf$ is
trivial. The distributions that do not involve single stars, such as
$f_{M_1}$, $f_{M_2}$, $\fqall$ and $\fqm$, by definition never depend on the choice
of $\binf$. The mass distribution of all stars $f_{M_{\rm all}}$ (the
``IMF'') depends on the choice of $\binf$ for all pairing functions,
except for RP.

\section{Interpretation of observations} \label{section:strategy}

Binarity and multiplicity provide important information about the
outcome of the star forming process in different environments
\citep{blaauw1961}. In this paper we explore this issue by making
the assumption that binary stars are formed through a simple "pairing
function". In reality the distribution of stars in
$f_{M_1,M_2}(M_1,M_2)$ is the result of complex physics involving the
collapse of a molecular cloud into stars and stellar systems with
disks (which can themselves fragment), followed by dynamical evolution
of the protocluster (see \S\,\ref{section:origin}). The resulting
"pairing function" may thus not be describable in terms of the simple
probability distributions given in this paper
(\S\,\ref{section:pairingfunction}). However, even if we proceed from
our assumption that the pairing of binary stars involves the random
selection of a primary mass followed by the secondary (RP, PCRP,
PCP-I/II/III) or the random splitting of cores (SCP-I/II/III), the
interpretation of the observations is not trivial:
\begin{itemize}
\item There is a large space of possible models for the formation and
  evolution of a binary population. In this paper we describe eight
  pairing mechanisms. For each pairing function there are a large
  number of possibilities for the generating mass distributions
  $f_M(M)$, mass ratio distributions $\fq$, and binary fractions
  $\binf$. In addition, we have no a priori restriction on the
  plausible formation mechanisms (see, however,
  \S\,\ref{section:origin}).
\item The observations of a binary population are generally
  limited. The surveys are incomplete, and are affected by selection
  effects and observational biases, and often only a limited set of
  binary population parameters is measured.
\end{itemize}
When the inverse problem of obtaining the binary formation mechanism
from the data is so poorly constrained, it is not possible to find the
``best model'' for binary formation from a fit to the data. The only
thing one can realistically do is to exclude models that are not
capable of reproducing the data and accept that all other models offer
plausible binary formation prescriptions.

\subsection{Constraints from observations} \label{section:constraintsfromobservations}

Below we list several properties that have been identified for various binary populations over the last decades. These provide important information on the primordial pairing function, and the formation and evolution of binary populations.

{\em The observed mass-dependent binary fraction}.
The observed binary fraction is known to increase with increasing primary spectral type; see, e.g., \cite{sterzik2004,koehler2006,lada2006,bouy2006} for an overview. For early-type (O/B/A) stars the binary fraction approaches 100\% \citep[e.g.,][]{abt1990,mason1998,shatsky2002,kobulnicky2007,kouwenhovenrecovery}. The binary fraction decreases to $50-60\%$ for F/G-type stars \citep{abtlevy1976,duquennoy1991}. For M-type stars the binary fraction is $30-40\%$ \citep{fischer1992,leinert1997,reid1997}, and for late M-type stars and brown dwarfs the binary fraction decreases to $10-20\%$ \citep{gizis2003,close2003,bouy2003,burgasser2003,siegler2005,maxted2008}. Note that this correlation between mass and binary fraction is also predicted by the hydrodynamical/sink particle simulations of \cite{bate2009}.
Assuming that this observational trend is not induced by selection effects, is
inconsistent with pairing functions PCP-I, PCP-III and SCP-I, for
which the binary fraction is independent of primary mass. Furthermore,
observations have ruled out pairing functions RP and PCRP in various
stellar groupings (see \S\,\ref{section:origin}). Pairing functions
PCP-II, SCP-II and SCP-III remain options to describe the binary
population, as for these pairing functions the binary fraction increases
with increasing stellar mass, and the average mass ratio decreases
with increasing mass. However, in this paper we merely describe
simplistic (but frequently used) pairing functions. A deeper analysis,
including a study of more complicated pairing functions, combined with
further observations, is necessary for a full description of the
pairing function in the different stellar populations.

{\em Twin binaries}. 
 Observationally, there is a large prevalence of massive binaries with a mass ratio close to unity, often referred to as the ``twin peak'' in the mass ratio distribution \citep{lucy1979,tokovinin2000,pinsonneault2006,lucy2006,soderhjelm2007}.
High-mass twin binaries are extremely rare for RP and PCRP. For pairing functions PCP and SCP, high-mass twin binaries only occur frequently when this is explicitly put into the generating mass ratio distribution. A high prevalence of low-mass twin binaries, on the other hand, naturally results from {\em all} pairing functions except for PCP-I. In general, peaks in the mass ratio distribution can occur for any mass ratio (see, e.g., Fig.~\ref{figure:rp_different_samples}). The location of the peak depends on the pairing function and the primary mass range, and, if applicable, the mass ratio distribution. In general, the peaks occur at low-$q$ for a sample of high-mass binaries, and at high-$q$ for a sample of low-mass stars. Pairing functions RP and PCRP are thus excluded, while pairing functions PCP and SCP can only result in massive twin binaries if the corresponding generating mass ratio distribution is strongly peaked to $q=1$.

{\em The brown dwarf desert}.  The brown dwarf desert is defined as a
deficit (not necessarily a total absence) of brown dwarf companions,
either relative to the frequency of stellar companions, or relative to
the frequency of planetary companions
\citep{mccarthy2004,grether2006}. Theories have been developed
that explain the existence of the brown dwarf desert using migration
\citep{armitage2002} or ejection \citep{reipurth2001} of brown
darfs. The latter scenario ``embryo ejection'' is most popular, and
predicts ejectino of brown dwarfs soon after their formation. In this
scenario, brown dwarfs could be considered as failed
stars. \cite{kouwenhoven2007a}, however, show that the scarcity of
brown dwarf companions among intermediate-mass stars can also be explained by
an extrapolation of the mass ratio distribution into the brown dwarf
regime; PCP-I/II/III are thus not excluded by the presence of the
brown dwarf desert.

{\em The (initial) mass distribution}.  The initial or present-day
mass distribution $f_{\rm all}(M)$ of a stellar population sets strong
constraints on the star formation process, and is an important feature
of each pairing function. The distribution is often derived after its
members are securely identified
\citep[e.g.,][]{kroupa2001,preibisch2002,harayama2008,stolte2008}. The
{\em measured} mass distribution is often the distribution of
single/primary masses $f_{M_{1,S}}(M)$, as it is not known which
members are single and which are binary, which results in a measured
mass distribution that is biased to higher masses with respect to the
overall mass distribution $f_{\rm all}(M)$, which, if measured just
after star formation, is the IMF. As stellar masses are often derived
from measured luminosities, the presence of unresolved binaries and
crowding may further bias the measured (initial or present-day) mass
distribution to higher masses
\citep[e.g.,][]{vanbeveren1982,maiz2008}.
Over the last decade, considerable effort has been put into studying
possible environmental dependences of the IMF \citep[see, e.g.,][for
an overview and examples]{elmegreen2007,kroupa2008}. The IMF of a
population is presumably a result of the form of the initial core mass
function, {\em and} the primordial pairing function \citep[i.e., how these
cores fragment into multiple systems, see][]{goodwinsplitup2008}. 
An environmental dependence of the primordial pairing function (e.g.,
mass ratio distribution, binary fraction) implies a different outcome
of the star formation process with environment, it also almost
certainly implies an environmental dependence of the IMF (unless the
core mass function changes in such as way as to mask this change).

\subsection{Recovering the pairing function} \label{section:recoveringthepf}

The pairing function, $f_M(M)$, $f_q(q)$ and $\binf$ can in principle
be derived from observations of binary systems, provided that the
observations cover a large part of the parameter space $\{M_1,q\}$;
see \S\,\ref{section:dependence_generating_properties}. A significant
complication, however, is introduced by selection effects, in
particular by detection limits that prevent the detection of faint
companion stars.  As an example, a twin peak for high mass binaries
would rule out all of the models presented in this paper if $f_q(q)$
is assumed to be flat for PCP and SCP (see
\S\,\ref{section:constraintsfromobservations}).  On the other hand, a
twin peak at the low mass end only rules out PCP-I ({\em if} the
generating $\fq$ is flat). So how should we proceed when interpreting
the observational data?  If we assume for the moment that we somehow
know that one of the eight pairing mechanisms discussed in this paper
occurs in Nature, we can advise the following:
\begin{itemize}
\item Make sure the survey of the binary population is complete, i.e.,
  that all primary masses are sampled. As is clear from
  Fig.~\ref{figure:resulting_massratiodistributions}, looking only at
  the low mass stars does not allow differentiating PCP-II, PCP-III,
  RP, PCRP, and SCP-I/II/III (all show a twin peak).
\item Examine not only the overall distribution of a certain parameter
  (such as $q$ or the binary fraction) but also study how it varies
  with primary mass. Again, from
  Fig.~\ref{figure:resulting_massratiodistributions} it can be seen
  that when considering only a single row of panels it is not possible
  to easily differentiate the pairing mechanisms. However, when looking
  at the overall {\em and} specific mass ratio distributions, the
  differences do become clear.
\item Examine the combined behaviour of each parameter (the mass ratio
distribution and the binary fraction) as a function of primary
mass. The combination of these parameters constrains the possible pairing
mechanisms significantly further.
\item List all mechanisms capable of reproducing the observations as
  possible solutions to the inverse problem. Do not try to give a
  single answer if this is not warranted by the data.
\end{itemize}
In reality the number of possible models is of course much larger,
especially if we start from arbitrary probability distributions that
are not constrained by an understanding of the physics of binary
formation. This is illustrated by the simple example in
\S\,\ref{section:example} which shows that allowing a power-law
distribution for $\fq$ means that only RP and PCRP can be excluded based
on an observed flat distribution of $q$. 

The only practical way of excluding models of binary formation based on observations is to treat the inverse problem with Monte Carlo methods where the observations are predicted from the model and compared to the real observations.  In this method selection and observational biases should be included \citep[see, e.g.,][]{kouwenhoventhesis,kouwenhovenrecovery}.
Starting from models based on probability distributions for a set of parameters may not be the most useful way of constraining the formation mechanism for binaries as this leaves a lot of freedom. It is more fruitful (but also more difficult) to start from actual physical models of binary formation and see if these are capable of reproducing the observations.

A further complication occurs when one wants to recover the {\em
primordial pairing function}, i.e., the pairing function that is
present just after star formation, as the pairing function of a
stellar population evolves over time as a result of several
processes. During the first stages of star formation, the newly formed
proto-binaries are affected by pre-main sequence eigenevolution
\citep{kroupa1995a,kroupa1995b,kroupa1995c} due to interaction with
the remaining gas in the circumbinary disk. Dynamical interactions can
result in ionisation of binaries, the formation of new binary systems,
and exchange interactions, and thus alters the pairing function of a
stellar population \citep[see,
e.g.,][]{kroupa1999,kroupaaarseth2001,preibisch2003,duchene2004,koehler2006,reipurth2007}. The
pairing function also changes due to stellar evolution, which can
change the mass of one or both of the components of a binary system,
and in some cases in a merger \citep[e.g.,][]{sills2002,gaburov2008},
or in the ejection of one of the components during a supernova event
\citep[e.g.,][]{blaauw1961}. The primordial pairing function can be
constrained using the technique of inverse dynamical population
synthesis \citep[e.g.,][]{kroupa1995b,kroupa1995c}, in which the
outcome of $N$-body simulations is compared with present-day binary
population in a Monte Carlo way.

Finally we stress that any interpretation of observations of a binary population in terms of the formation of binaries should start by stating the assumptions one makes in order to restrict the number of solutions to explore. That is, in the context of what class of binary formation mechanisms are the observations interpreted?

\subsection{An example -- an observed flat mass ratio distribution } \label{section:example}

\begin{table}[tbp]
  \caption{Suppose that a stellar population has an {\em observed} mass ratio distribution $f_{q,{\rm obs}}(q) =1$ for $\qmin < q \leq 1$. What is the pairing function and the generating mass distribution $\fq$? This table lists the exponent $\gamma_q$ of the generating mass distribution $\fq \propto q^{\gamma_q}$ that is most compatible with the observations. We assume that $\qmin=0.1$, and binaries with $q<0.1$ cannot be detected. We ignore the other selection effects. The numbers in the three columns represent the most compatible values of $\gamma_q$ for the overall mass ratio distribution $\fq$, and for the specific mass ratio distribution of binaries with A/B primaries and brown dwarf primaries, respectively. 
  \label{table:examples}}
  \begin{tabular}{lccc}
    \hline 
    \hline
    Pairing function & $\gamma_q$ (all stars) & $\gamma_q$ (AB stars) & $\gamma_q$ (BDs) \\
    \hline
    Observed          & 0.00 & 0.00    & 0.00 \\
    \hline
    PCP-I             & 0.00 & 0.00    & 0.00 \\
    PCP-II            & 0.35 & 0.00    & 1.30 \\
    PCP-III           & 0.55 & 0.00    & 1.80 \\
    RP                & excluded  & excluded     & excluded  \\
    PCRP              & excluded  & excluded     & excluded  \\
    SCP-I             & 0.00 & $-0.40$ & 0.10 \\
    SCP-II            & 0.30 & $-0.40$ & 1.50 \\
    SCP-III           & 0.55 & $-0.40$ & 1.75 \\
    \hline
    \hline 
  \end{tabular}
\end{table}

Most pairing functions result in a mass ratio distribution that varies with primary spectral type. For this reason one has to be cautious when interpreting the observations of a sample of binaries. Given the observed dataset, what is the pairing function, and what is the {\em generating} mass ratio distribution $\fq$? The answer partially depends on the generating mass distribution, which we assume to be the Kroupa mass distribution for now. More importantly, the answer depends on the properties of the surveyed targets. In this example we analyse three cases: a sample where {\em all} binaries are studied, a sample of spectral type A/B targets, and a sample of brown dwarf targets. Using Monte-Carlo techniques we determine which pairing function is consistent with an observed flat mass ratio distribution $f_{q,{\rm obs}}(q)=1$ for each subset, and, if applicable, which generating mass ratio distribution. We assume that the generating mass ratio distribution has the form $\fq \propto q^{\gamma_q}$. We further assume that the observations are complete in the range $\qmin <q \leq 1$ with $\qmin=0.1$, and that no binaries with $q<\qmin$ are observed due to incompleteness.

Table~\ref{table:examples} shows the best-fitting values of $\gamma_q$ for each pairing function. For each for the three samples, pairing functions RP and PCRP are excluded with high confidence; these are unable to reproduce the observed flat mass ratio distribution. The best-fitting value for pairing function PCP-I is $\gamma_q=0$ for all samples. This is not surprising, as for $\fqm = \fq$ for this pairing function. For PCP-II and PCP-III the derived $\gamma_q$ for high-mass stars equals the observed value, but the other two samples contain more binaries with high mass ratios. For pairing functions SCP, the best-fitting intrinsic value of $\gamma_q$ is smaller than the observed value of $\gamma_q$ for high-mass binaries, but larger for low-mass binaries. This example illustrates that the inferred intrinsic pairing properties may be significantly different from the observed pairing properties, depending on the pairing function and the selected sample of binaries.

In practice, parameter distributions are often represented with a functional form. Suppose, for example, that we {\em assume} that the generating mass ratio distribution has the form $\fq\propto q^{\gamma_q}$. If we use this functional form for our model population, and compare simulated observations with the true observations for different values of $\gamma_q$, we will likely find a best-fitting $\gamma_q$. This does not necessarily mean that the generating mass ratio distribution has indeed the form $\fq\propto q^{\gamma_q}$. In this example we have {\em added another assumption}, i.e., that the mass ratio distribution has the form $\fq\propto q^{\gamma_q}$.

\section{Summary and discussion} \label{section:summaryandoutlook}

We have described several methods of pairing individual stars into
binary systems. We refer to these algorithms as {\em pairing
functions}. Each pairing function is characterized by a generating
mass distribution $f_M(M)$ and a generating binary fraction $\binf$,
and most additionally by a generating mass ratio distribution $\fq$. Each
pairing function results in a significantly different binary
population. Depending on the pairing function and the mass range of
the binaries studied, the {\em resulting} binary population may or may
not have a mass ratio distribution or binary fraction that is equal
to $\fq$ or $\binf$, respectively. The binary fraction and mass ratio
distribution generally depend strongly on the number of substellar objects in
the population, and on the properties of the surveyed sample.

Eight pairing mechanisms are discussed in detail. For random pairing
(RP) both components are randomly drawn from the mass distribution
$f_M(M)$. For primary-constrained random pairing (PCRP), both
components are drawn from $f_M(M)$, with the constraint that the
companion is less massive than the primary. For primary-constrained
pairing (PCP-I, PCP-II, and PCP-III), the primary is drawn from
$f_M(M)$, and the companion mass is determined using a mass ratio
distribution $\fq$. For split-core pairing (SCP-I, SCP-II, and
SCP-III), the core mass is drawn from $f_M(M)$, and the masses of the
binary components are determined by the mass ratio $\fq$, which splits
up the core into two stars. The difference between the variants of
pairing functions PCP and SCP lies in the treatment of low-mass
companions (see \S\,\ref{section:pairingfunction}). Seven pairing
functions naturally result in a specific mass ratio distribution that depends on
primary spectral type, and five naturally result in a mass-dependent binary
fraction. Seven out of eight pairing functions always produce a twin peak for low-mass binaries, while none result in a twin peak for high-mass binaries, unless the generating mass ratio distribution is strongly peaked to $q=1$.

The differences between pairing functions are important for (i) the
interpretation of observations, (ii) initial conditions of numerical
simulations, and (iii) understanding the outcome of star formation:

\noindent
(i) {\em The interpretation of observations} The choice of the
observational sample may mislead the observer in deriving the overall
properties of a stellar population, as most pairing functions have a
mass-dependent binary fraction and mass ratio distribution. If the
binary fraction or mass ratio distribution of two samples (e.g.,
systems with B-type primaries and those with M-type primaries) are
different, this does {\em not} necessarily mean that the underlying
pairing function is different.  
A significant further complication is introduced by observational selection effects, which artificially decrease the binary fraction and increase the average mass ratio. The only practical way to account for these is by using a Monte Carlo approach, and to compare simulated observations of a model population with the results of the binarity survey, taking into account all sampling and selection effects (\S\,\ref{section:strategy}).

\noindent
(ii) {\em Initial conditions for numerical simulations} A choice for
the pairing function has to be made when generating initial conditions
for simulations of star cluster simulations with binaries. The
simplest choice is random pairing (RP), although this pairing function
is excluded from observations and not expected from star
formation. When modeling star clusters, one has to be aware that most
pairing mechanisms result in mass-dependent properties, such as mass
ratio distribution and binary fraction. The choice of the pairing
function affects the outcome of the simulations, such as the dynamical
evolution of star clusters, mass segregation, and the number of
contact binaries and supernovae.

\noindent
(iii) {\em Star formation}.  Different star formation scenarios result
in different mechanisms of pairing stars into binary systems. After
star formation, the pairing function is altered by dynamical
interactions and stellar evolution. Random pairing, however, is not
predicted by star formation models, and is excluded by observations
(see \S\,\ref{section:origin}). The binary fraction and mass ratio
distribution can be used to discriminate between the different pairing
functions. Although the pairing functions described in this paper
are common in literature, we do not suggest that one of these pairing
functions indeed describes the natural outcome of the star forming
process. It is, for example, possible that the pairing properties are
a function of primary mass or core mass. If this is the case, it may
indicate different formation processes for different masses. Due to
the lack of observations, this has not been studied in detail, apart
from the extreme ends of the mass distribution (very massive stars and
brown dwarfs). \cite{kroupabouvierduchene} and \cite{thies2007}, for
example, find that the observed IMF and binary population can be
explained by separate formation mechanisms for stars and brown dwarfs
(see \S\,\ref{section:restrictedRP}). Nevertheless, the pairing
functions described in this paper are useful tools to describe the
outcome of star formation simulations. In order to constrain the
primordial binary population from observations, one does not only have
to take into account the selection effects, but also the change in the
pairing function that has occured due to the effects of stellar and
dynamical evolution.
Over the last decade, considerable effort has been put into studying
possible environmental dependences of the IMF.  The IMF of a
population is presumably a result of the form of the initial core mass
function, {\em and} the primordial pairing function \citep[i.e., how
these cores fragment into multiple systems,
see][]{goodwinsplitup2008}.  An environmental dependence of the
primordial pairing function (e.g., mass ratio distribution and binary
fraction) implies a different outcome of the star formation process
and almost certainly an environmental dependence of the IMF (unless
the core mass function changes in such as way as to exactly mask this
change); see \S\,\ref{section:dependence_generating_properties}.

Each pairing function, as well as each subset of stars, results in a
different mass ratio distribution and binary fraction. It is therefore
of great importance to carefully study selection effects in
observations, and to clearly state the pairing mechanism used in
simulations, in order to make statements about the star formation
process. The pairing functions described in this paper are likely too
simplistic to describe a realistic stellar population. However, they
are frequently used to describe observations and simulations. The next
step forward is to {\em fully} characterize the binary population of
several young stellar groupings; only in this way the star formation
process can be recovered.

\begin{acknowledgements}
We would like to thank the referee, Rainer K\"{o}hler, for detailed comments and suggestions which helped to improve this paper significantly. We are also thankful to Pavel Kroupa for valuable comments and suggestions. M.K. was supported by PPARC/STFC under grant number PP/D002036/1 and by NWO under project number 614.041.006. M.K. and S.G. acknowledge a Royal Society International Joint Project grant between Sheffield and Bonn. This research was supported by the Netherlands Research School for Astronomy (NOVA) and by the Leids Kerkhoven Bosscha Fonds (LKBF).
\end{acknowledgements}

\bibliographystyle{aa}
\bibliography{bibliography}

\Online

\begin{appendix}

  \section{The pairing function and mass ratio distributions} \label{appendix}

In this appendix we discuss for each of the pairing mechanisms described in the
main paper how to calculate the mass ratio distribution function $f_q(q)$. The
masses of primary ($M_1$) and secondary ($M_2$) are drawn from the same
generating mass distribution $f_M(M)$, or alternatively, the primary is drawn from 
$f_M(M)$ while
the secondary is drawn from a generating mass ratio distribution. The 
generating mass distribution is treated
throughout this appendix as a probability density:
\begin{equation}
  P(M\leq t)=F_M(t)=\int_{-\infty}^t f_M(x)dx\,,
  \label{eq:imfdef}
\end{equation}
and
\begin{equation}
  \int_{-\infty}^\infty f_M(M)dM=1\,.
\end{equation}

In the subsequent section we describe for each of the pairing functions first
how to calculate $f_q(q)$ without specifying the generating mass distribution and then we work out the
resulting mass-ratio distributions for the single power-law $f_M(M)$ which is given
here as:
\begin{equation}
  f(M)=aM^{-\alpha} \quad\mathrm{with}\quad
  a=\frac{1-\alpha}{d^{1-\alpha}-c^{1-\alpha}} = 
  \frac{\gamma c^\gamma}{1-(c/d)^{\gamma}}\,,
  \label{eq:singlepow}
\end{equation}
where $\alpha\neq1$ and $\gamma=\alpha-1$ and $c$ and $d$ are the lower and
upper limits on the mass distribution.

\subsection{Random pairing} \label{appendix:rp}

The simplest choice for a pairing function is that of `random pairing' (RP). In
this case both component masses are drawn independently from $f_M(M)$ and swapped, 
if necessary, to ensure that $M_2\leq M_1$. In
this case the joint distribution function $f_\mathrm{rp}(M_1,M_2)$ for the
component masses is given by:
\begin{equation}
  f_\mathrm{rp}(M_1,M_2)=2f_{M_1}(M_1)f_{M_2}(M_2)\,,
  \label{eq:rpjoint}
\end{equation}
where the factor 2 accounts for the fact that the pairs of masses are swapped in
order to ensure that $M_2\leq M_1$. The masses are restricted by a lower limit
$c$ and and upper limit $d$ which leads to the domain of
$f_\mathrm{rp}(M_1,M_2)$ being defined as $c\leq M_1\leq d$ and $c\leq M_2 \leq
M_1$.

\subsubsection{Mass ratio distributions for RP}

First, general expressions for the mass ratio distributions are derived before
working out specific examples. To derive the mass ratio distribution we follow
the appendix of \cite{piskunov1991} and derive $f_q(q)$ from its cumulative
distribution function:
\begin{eqnarray}
  P(y/x\leq q)=F_q(q) & = &\iint\limits_S f_\mathrm{rp}(x,y)dxdy \nonumber \\
  & = &\int_{c/q}^d\int_c^{qx} 2f(x)f(y)dydx\,,
  \label{eq:fqrpallcuM_app}
\end{eqnarray}
where for ease of notation we use $M_1=x$ and $M_2=y$. The integration domain $S$
is shown in the left-hand panel of Fig.~\ref{fig:fqseldomain}. The probability density $f_q$ is
then given by:
\begin{equation}
  f_q(q) = \frac{d}{dq}F_q(q)\,.
  \label{eq:fqrpall_app}
\end{equation}
In observational surveys for binarity one is often restricted to a certain range
of primary spectral types for observational reasons. To derive the effects of
the selection on primary mass, the derivation of $f_q(q)$ again proceeds via the
cumulative distribution function, which is now given by:
\begin{equation}
  F_q(q)=P(y/x\leq q|x_1\leq x\leq x_2)\propto\iint\limits_{S'}f_\mathrm{rp}(x,y)dxdy\,,
\end{equation}
where the primary mass range is restricted to $x_1\leq M_1\leq x_2$ and the
integration domain $S'$ is as shown in Fig.~\ref{fig:fqseldomain}. The
integration domain limits now depend on whether $q$ is larger or smaller than
$c/x_1$. For $q\leq c/x_1$ the integration domain $S'$ is given by: $c/q\leq
x\leq x_2 \wedge c\leq y\leq qx$ (middle panel in Fig.~\ref{fig:fqseldomain}),
while for $q>c/x_1$ $S'$ is defined by $x_1\leq x\leq x_2 \wedge c\leq y\leq qx$
(right-hand panel in Fig.~\ref{fig:fqseldomain}). The minimum possible value of $q$
is $c/x_2$ in this case, implying that $F(q)=0$ for $q<c/x_2$.
Thus the expression for the mass ratio distribution now becomes:
\begin{equation}
  F_q(q) =\left\{
  \begin{array}{ll}
    0 & q<\frac{c}{x_2} \\[5pt]
    k\int_{c/q}^{x_2}\int_c^{qx} 2f(x)f(y)dydx & \frac{c}{x_2}\leq q<\frac{c}{x_1} \\[5pt]
    k\int_{x_1}^{x_2}\int_c^{qx} 2f(x)f(y)dydx & \frac{c}{x_1}\leq q\leq 1
  \end{array}\right.\,,
  \label{eq:fqrpcuM_app}
\end{equation}
and the probability density $f_q$ is again derived according to equation
(\ref{eq:fqrpall_app}). The normalisation constant $k$ can be calculated from the
the condition $F_q(1)=1$.

\begin{figure*}
  \begin{center}
    \includegraphics[width=0.48\textwidth]{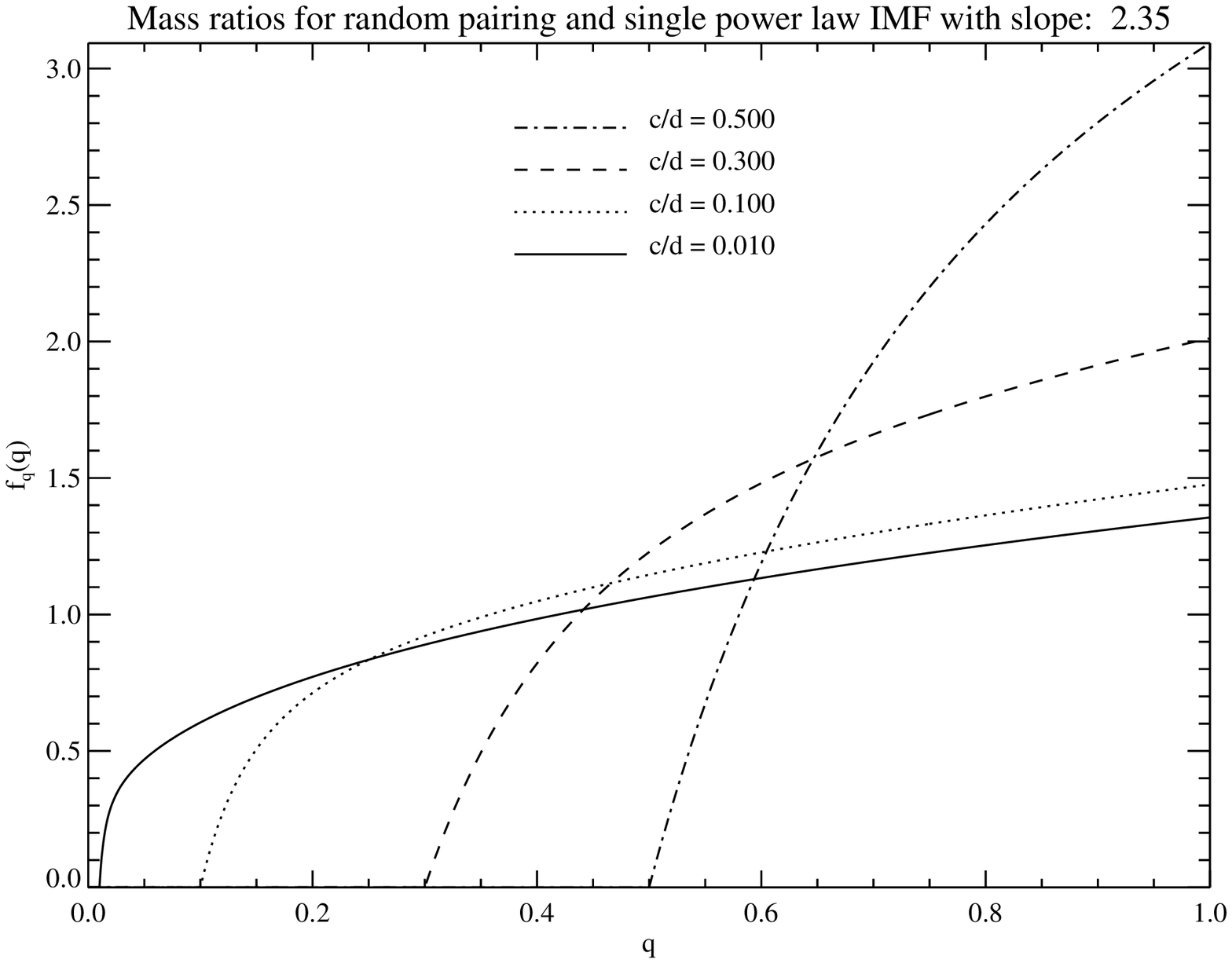}\hfil
    \includegraphics[width=0.48\textwidth]{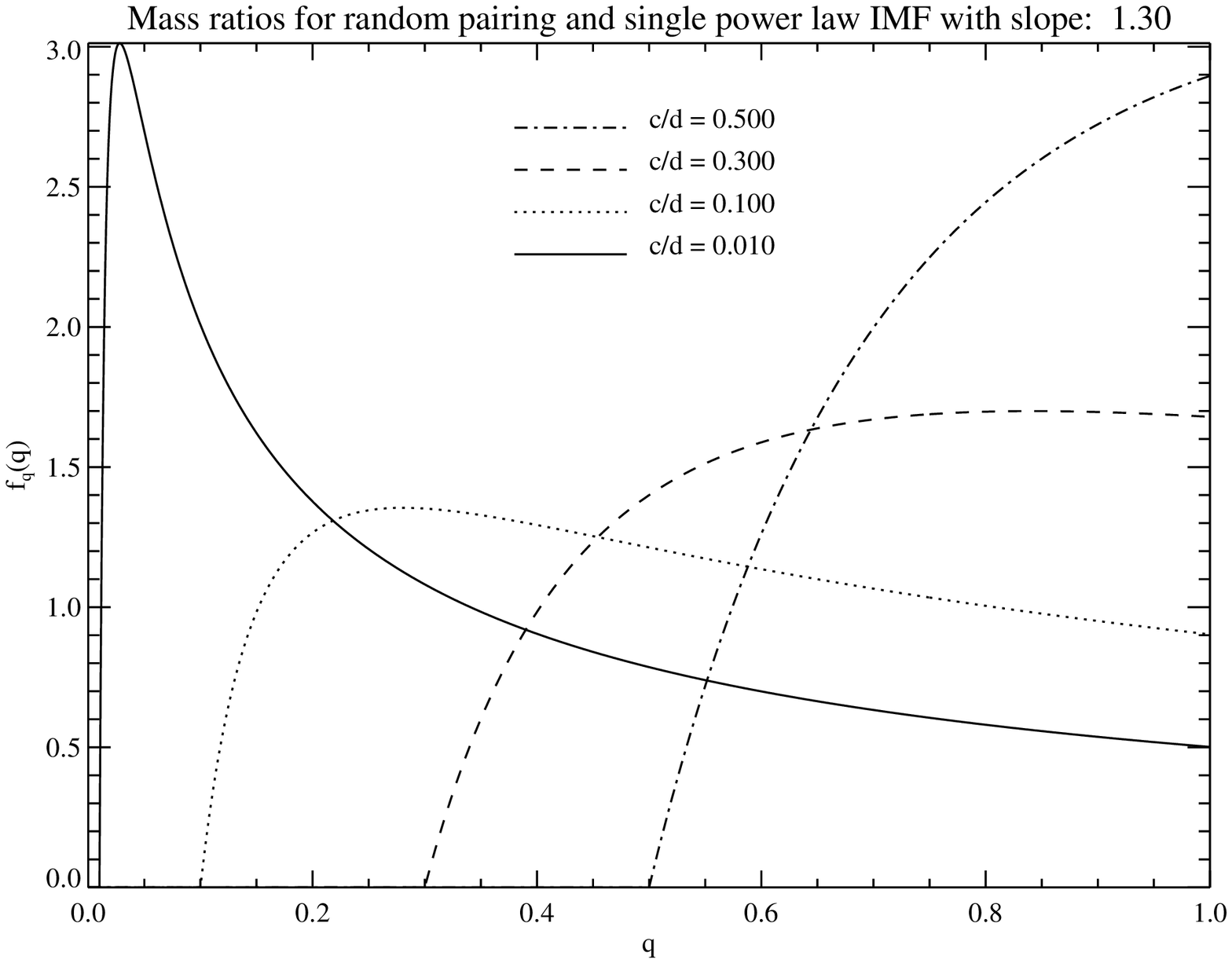}\\
    \includegraphics[width=0.48\textwidth]{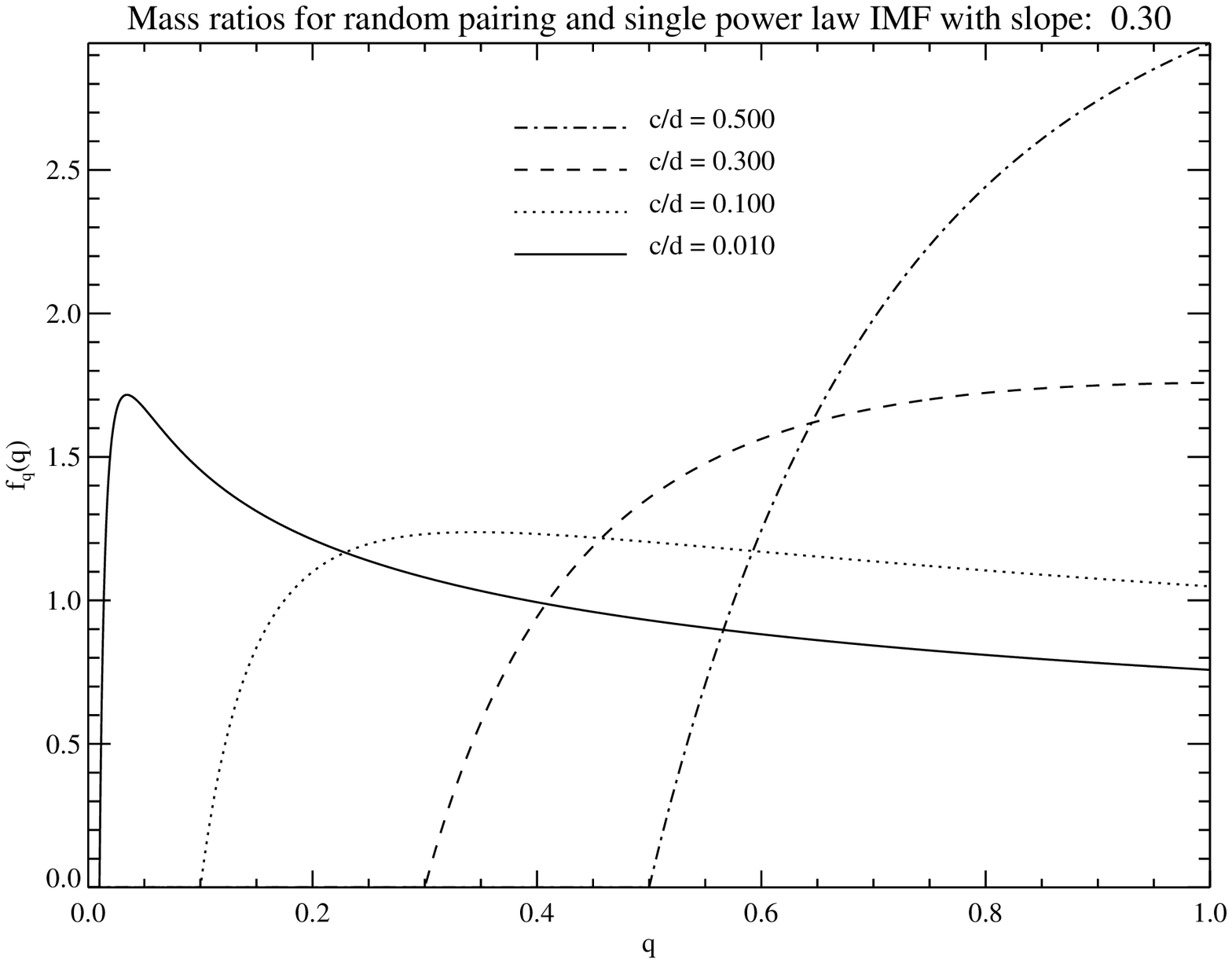}\hfil
    \includegraphics[width=0.48\textwidth]{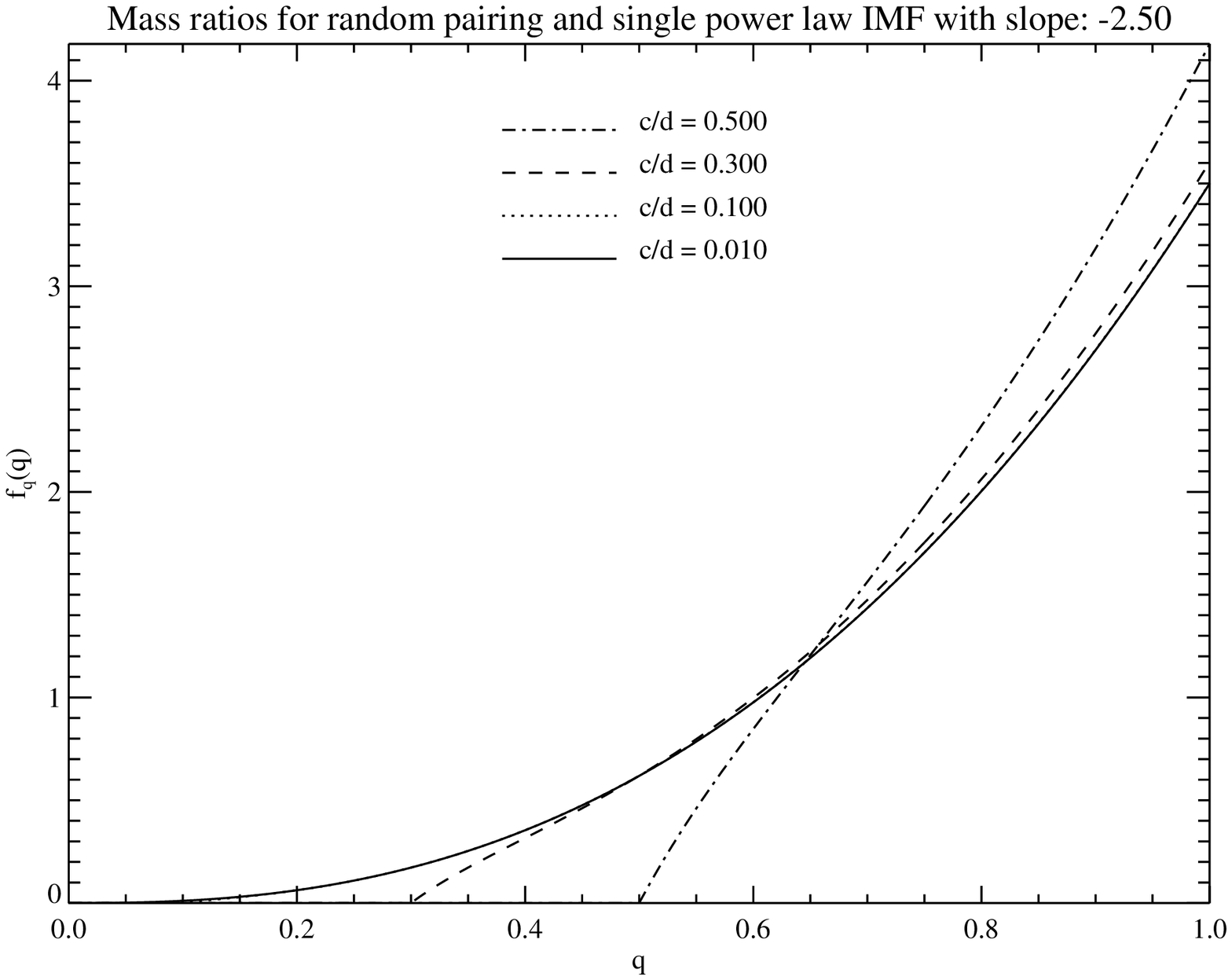}
  \end{center}
  \caption{Mass ratio distributions $f_q(q)$ for random pairing from the single
  power-law $f_M(M)$. The curves are shown for $\alpha=2.35,\,1.30,\,0.30,\,-2.50$ and
  four ratios of the lower to the upper mass limit of $f_M(M)$:
  $c/d=0.01,\,0.1,\,0.3,\,0.5$.\label{fig:fqrpexample}}
\end{figure*}

\begin{figure*}
  \begin{center}
    \includegraphics[width=0.48\textwidth]{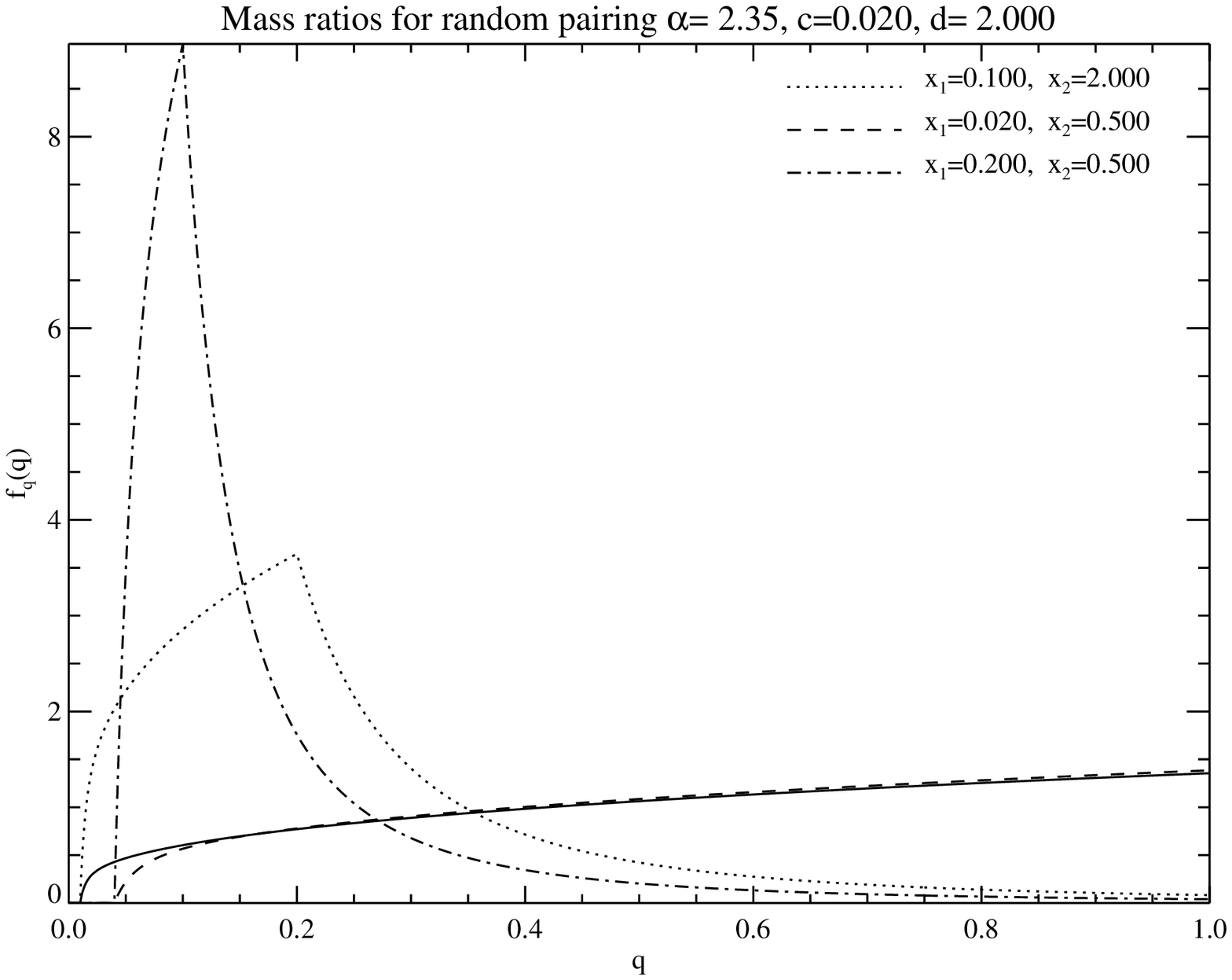}\hfil
    \includegraphics[width=0.48\textwidth]{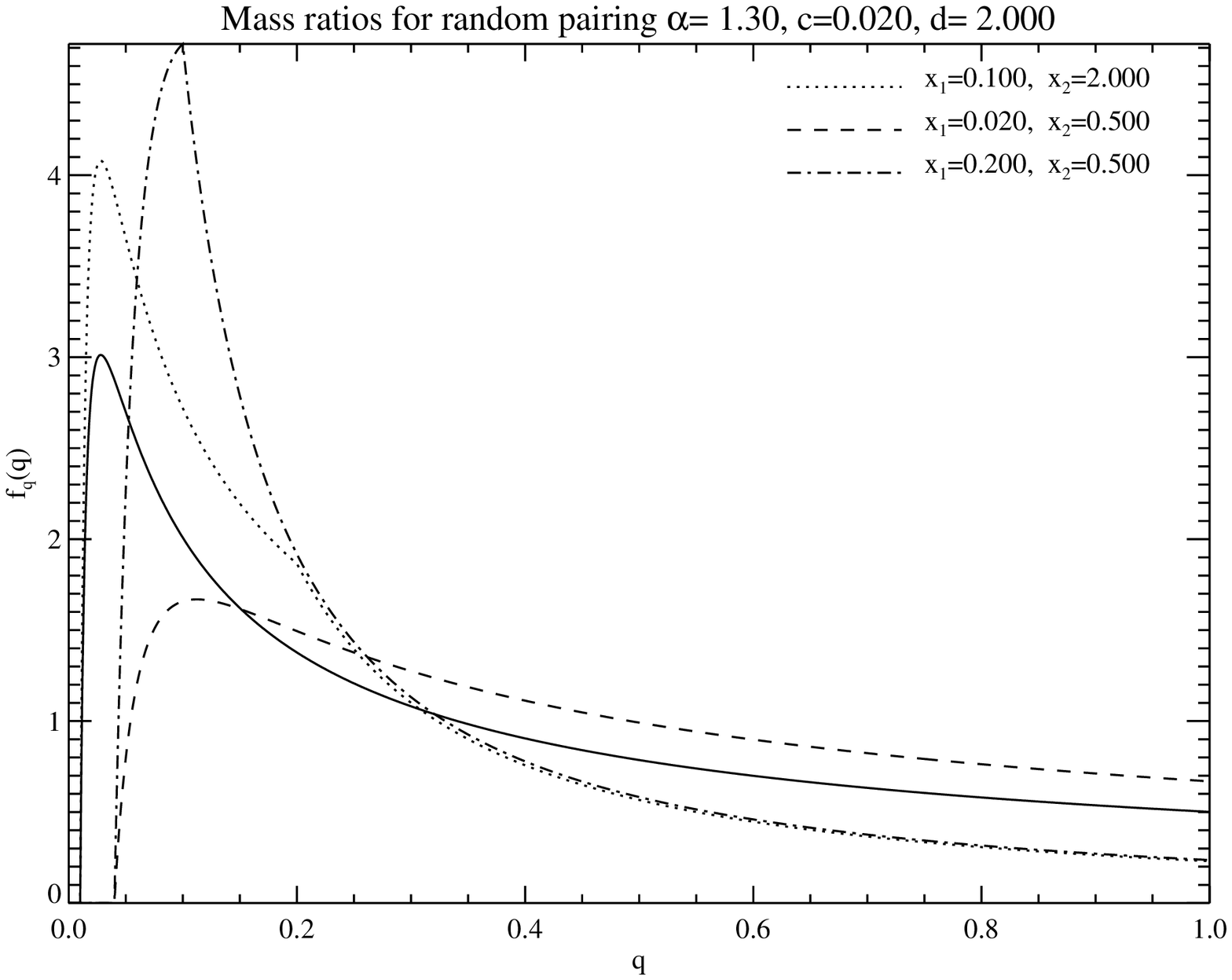}\\
    \includegraphics[width=0.48\textwidth]{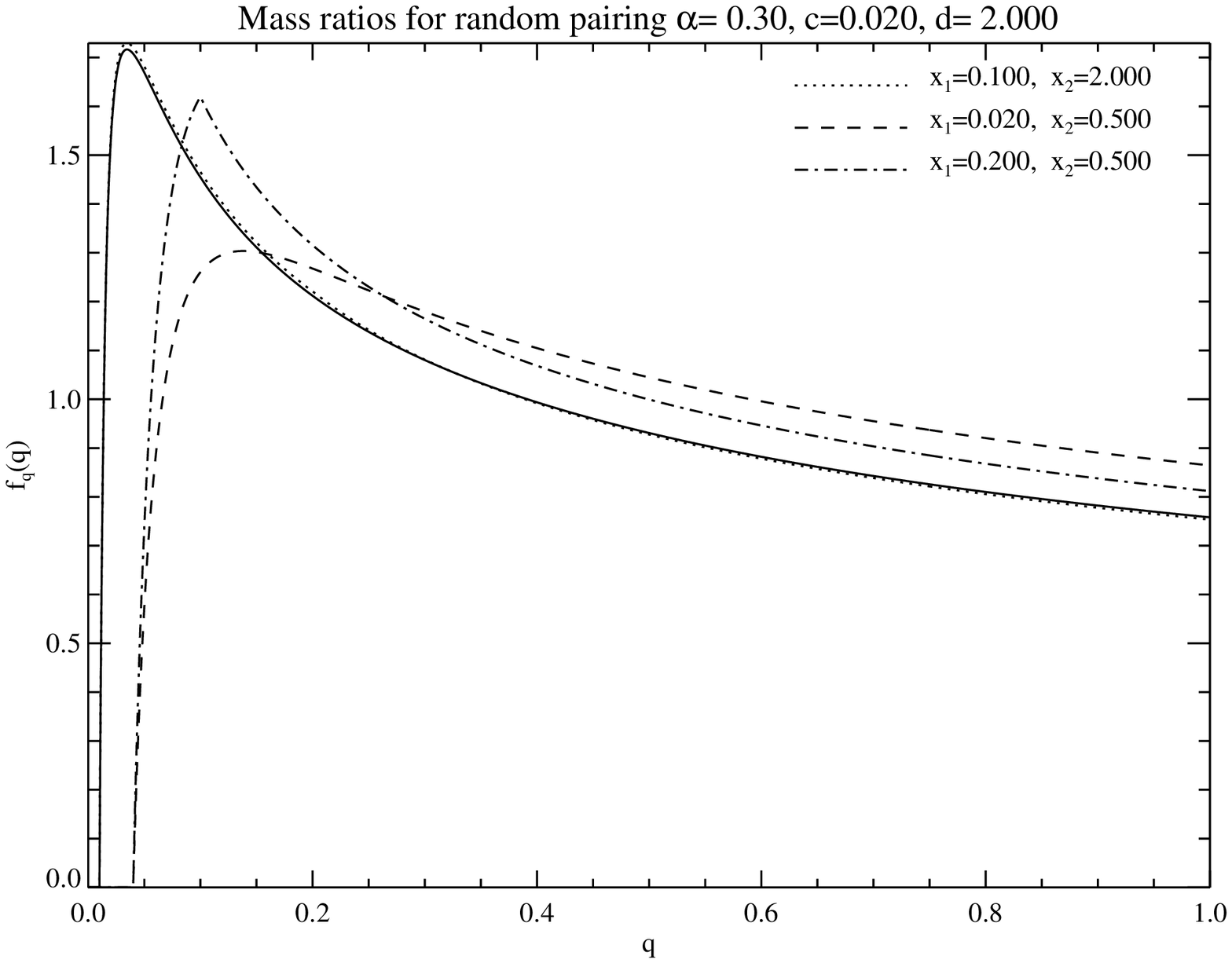}\hfil
    \includegraphics[width=0.48\textwidth]{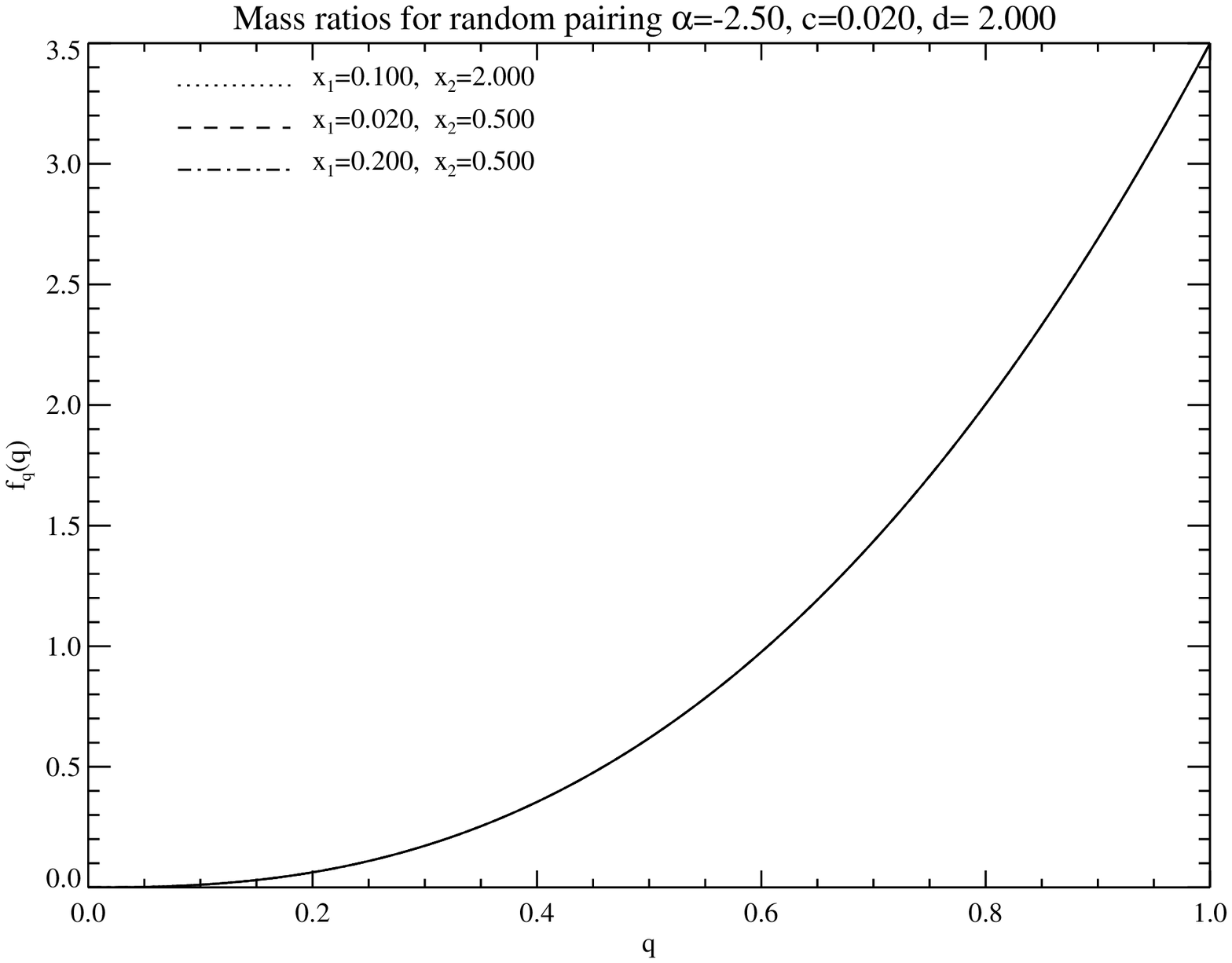}
  \end{center}
  \caption{Mass ratio distributions $f_q(q)$ for random pairing from a single
  power-law $f_M(M)$. The solid curve shows the complete mass ratio distribution (for
  all binaries in the population). The other curves show what happens to the
  observed $f_q(q)$ if the primary mass $M_1$ is restricted to $x_1\leq M_1\leq
  x_2$. The curves are shown for $\alpha=2.35,\,1.30,\,0.30,\,-2.50$. The value
  of $c/d$ is $0.01$ and $x_1$ and $x_2$ are listed in the
  panels.\label{fig:fqselrpexample}}
\end{figure*}

\subsubsection{Binary fractions for RP} \label{appendix:rp_bf}

For RP, the overall binary fraction $\binfall$ is equal to the generating binary fraction $\binf$. The specific binary fraction $\binfm$, however, is generally a function of primary mass. For a given primary mass, the number of single stars $S(M_1)$ with mass $M_1$ and the number of binary stars $B(M_1)$ with primary mass $M_1$ is given by:
\begin{equation}
S_{M_1} = S f_M(M_1) dM_1 \,,
\end{equation}
where $S$ is the total number of single stars in the system, and
\begin{equation}
B_{M_1} =  2B f_M(M_1) dM_1 \int_c^{M_1} f_M(M')dM' \,,
\end{equation}
where $B$ is the total number of binary stars in the system and $c$ is the minimum stellar mass. The specific binary fraction $\binfm = B_{M_1}/(S_{M_1}+B_{M_1})$ is then given by:
\begin{equation}
\binfm = \frac{ 2 B\, F_M(M_1) }{ 2 B\, F_M(M_1) + S  } = \left( \frac{\binf^{-1} -1 }{2 F_M(M_1)} + 1 \right)^{-1} \,,
\end{equation}
where $F_M(M_1)$ is the cumulative mass distribution (i.e., the primitive of $f_M$) evaluated at mass $M_1$.

An example of the mass-dependent binary fraction resulting from RP is shown in Fig.~\ref{figure:binaryfraction_versus_spectraltype}. Clearly, we have $\binfm \equiv 1$ when $S=0$, and $\binfm \equiv 0$ when $B=0$. Also, $\binf_{M_1}(\mmax) = 2B/(2B+S) \geq \binf$ and $\binf_{M_1}(\mmin) = 0$. The mass at which the specific binary fraction equals the generating binary fraction can be found by solving $\binf = \binfm$, which gives $F_M(M_1) = 0.5$. In other words, for pairing function RP the overall binary fraction can be found at the median stellar mass $\langle M \rangle$. For larger primary masses, $\binfm$ is larger, and for smaller primary masses, $\binfm$ is smaller.

\subsubsection{Single power-law mass distribution} \label{appendix:rp_powerlaw}

From Eqs.~(\ref{eq:fqrpallcuM_app}), (\ref{eq:fqrpall_app}), and (\ref{eq:fqrpcuM_app})
mass ratio distributions for specific choices of the generating mass
distribution can be derived. The
single power-law defined in Eq.(\ref{eq:singlepow}) is considered here.
Without restrictions on the primary mass range, the cumulative distribution for
$q$ is given by:
\begin{eqnarray*}
  F_q(q) &=&\int_{c/q}^d\int_c^{qx} 2a^2x^{-\alpha}y^{-\alpha}dydx \\
  & = & \frac{a^2}{\gamma^2c^{2\gamma}}\left(
  (c/d)^{2\gamma}q^{-\gamma}+q^\gamma - 2(c/d)^\gamma\right)\,.
\end{eqnarray*}
The probability density for $q$ is then obtained as the derivative with respect
to $q$ of $F_q$ which after some algebraic manipulation leads to:
\begin{equation}
  f_q(q) =
  \frac{\gamma}{(1-(c/d)^\gamma)^2}\left(q^{\alpha-2}-(c/d)^{2\gamma}q^{-\alpha}\right)\,.
\end{equation}
For $c/d\ll 1$ one obtains:
\begin{equation}
  f_q(q)\approx\left\{
  \begin{array}{ll}
    (\alpha-1)q^{\alpha-2} & \alpha>1 \\[5pt]
    (1-\alpha)q^{-\alpha} & \alpha<1
  \end{array}\right.\,.
  \label{eq:fqrpapprox_app}
\end{equation}
These approximations are poor when $\alpha \approx 1$.

Fig.~\ref{fig:fqrpexample} shows a number of examples of the resulting mass
ratio distributions for different values of $c/d$ and the power-law slope
$\alpha$. The values for $\alpha$ represent the Salpeter mass distribution ($2.35$), the
slopes at the lower mass end for the \cite{kroupa2001} mass distribution ($1.3$ and $0.3$),
and a possible slope at the very low mass end of the mass distribution, where the number of
stars increases with $m$. The latter value may occur in a multi-part power-law
mass distribution with a real turnover at the low mass end. For $\alpha>1$ the mass distribution decreases
with $\log m$ and for $\alpha<1$ it increases with $\log m$. However this does
not represent a real turnover in the mass distribution, the number of stars still increases as
the mass goes down as long as $\alpha>0$. Note how the peak in the $q$
distribution changes as the values of $\alpha$ and $c/d$ are changed. For very
small values of $c/d$ one can see from the approximation (\ref{eq:fqrpapprox_app})
that $f_q$ will be flat for $\alpha\approx 0$ and $\alpha\approx 2$ and that it
will peak at low values of $q$ for $0<\alpha<2$. For values of $\alpha$ larger
than 2 the number of low-mass stars is so dominant that high values of $q$ are
favoured (i.e., both $M_1$ and $M_2$ are likely to be small). Conversely, for
$\alpha<0$ the rise of the number of stars with $m$ again favours high values of
$q$. For $0<\alpha<2$ the ratio of probabilities to obtain low or high-mass
stars is such that drawing two equal mass star is unlikely thus favouring low
values of $q$.

To find the expression for $f_q$ when the primary mass range is restricted
Eq.~(\ref{eq:fqrpcuM_app}) has to be worked out for the single power-law mass distribution.
For $0< q< c/x_2$ $F_q(q)=0$, while for $c/x_2\leq q< c/x_1$ $F_q(q)$ is
given by:
\begin{eqnarray*}
  F_q(q) & \propto \int_{c/q}^{x_2}\int_c^{qx} x^{-\alpha}y^{-\alpha}dydx =
  \frac{1}{1-\alpha}\int_{c/q}^{x_2}
  x^{-\alpha}(q^{1-\alpha}x^{1-\alpha}-c^{1-\alpha}) dx \\[5pt]
  & \propto \frac{1}{\gamma^2}\left(
  \frac{1}{2}(q^{-\gamma}x_2^{-2\gamma}+c^{-2\gamma}q^{\gamma}) -
  c^{-\gamma}x_2^{-\gamma}\right) \,. \\
\end{eqnarray*}
For $c/x_1\leq q\leq 1$ the cumulative distribution for $q$ is given by:
\begin{eqnarray*}
  F_q(q) & \propto & \int_{x_1}^{x_2}\int_c^{qx} x^{-\alpha}y^{-\alpha}dydx
  \\[5pt]
  & \propto & \frac{1}{\gamma^2}\left(
  \frac{1}{2}q^{-\gamma}(x_2^{-2\gamma}-x_1^{-2\gamma}) -
  c^{-\gamma}(x_2^{-\gamma}-x_1^{-\gamma})\right) \,. \\
\end{eqnarray*}
The normalisation constant $k$ for the probability density of $q$ can now be
found by substituting $q=1$ in the last expression for $F_q(q)$:
\begin{equation}
  k=\gamma^2\left(
  \frac{1}{2}(x_2^{-2\gamma}-x_1^{-2\gamma}) -
  c^{-\gamma}(x_2^{-\gamma}-x_1^{-\gamma})\right)^{-1}
  \label{eq:fqselnorm}
\end{equation}
Now the expressions for $f_q(q)$ can be derived by taking the derivative of the
integrals above:
\begin{equation}
  f_q(q) = \left\{
  \begin{array}{ll}
    0 & q<\frac{c}{x_2} \\[5pt]
    \frac{k}{2\gamma c^{2\gamma}}\left(q^{\alpha-2}-\left(\frac{c}{x_2}\right)^{2\gamma}
    q^{-\alpha}\right)
    & \frac{c}{x_2}\leq q<\frac{c}{x_1} \\[5pt]
    \frac{k}{2\gamma
    x_1^{2\gamma}}\left(1-\left(\frac{x_1}{x_2}\right)^{2\gamma}\right)q^{-\alpha}
    & \frac{c}{x_1}\leq q\leq 1
  \end{array}\right.\,,
  \label{eq:fqsel_rp_powerlaw}
\end{equation}
If $x_1=x_2$ $f_q(q)$ will be proportional to $q^{-\alpha}$ for $c/x_1\leq
q\leq 1$ and zero otherwise:
\begin{equation}
  f_q(q) =\left\{
  \begin{array}{ll}
    0 & q<\frac{c}{x_1} \\[5pt]
    \frac{1-\alpha}{1-(c/x_1)^{1-\alpha}}q^{-\alpha} & \frac{c}{x_1}\leq q\leq 1
  \end{array}\right.\,.
  \label{eq:fqselrponemass}
\end{equation}
Fig.~\ref{fig:fqselrpexample} shows examples of the behaviour of $f_q(q)$ when
the primary mass range is restricted. The solid lines show the distribution of
$q$ for a complete binary sample. The dotted lines show what happens if there is
only a lower limit ($>c$) on the primary mass range. For $\alpha>1$ the high
mass ratios are preferentially removed because low-mass primaries are removed.
Conversely if there is only an upper limit on $M_1$ ($<d$) only low mass ratios
are removed and the resulting $f_q(q)$ is given by the dashed lines.  The
dot-dashed lines show a generic case with $c<x_1<x_2<d$. The latter case for
very narrow primary mass ranges $f_q(q)$ will converge to $f_q(q)\propto
q^{-\alpha}$. For $0<\alpha<1$ the effect of mass selection is to remove the low
values of $q$, thus flattening the distribution and moving the peak. For
$\alpha<0$ the mass selection does not have much effect. This figure illustrates
that the interpretation of mass ratio distributions in terms of random pairing
is not straightforward unless the generating mass function is well known and the
observations are indeed complete over a known primary mass range.

The expression for the specific binary fraction $\binfm$ resulting from random pairing 
is listed in Table~\ref{table:binfracm_powerlaw}, and, for the Salpeter generating mass distribution ($\alpha=2.35$), shown in Fig.~\ref{figure:binaryfraction_versus_spectraltype}.

\subsection{Primary-constrained random pairing} \label{appendix:pcrp}

In this case (PCRP) the primary and secondary are again drawn independently from
the generating mass distribution, however for the secondary the condition $M_2\leq M_1$ is
imposed before drawing the secondary mass. That is the probability density
$f_{M_2}(M_2)$ is re-normalised to the interval $[c,M_1]$ (recall that $c$ is
the lower mass limit on the mass distribution).

\subsubsection{Mass ratio distributions for PCRP}

Writing the re-normalised secondary mass distribution as
${f'}_{M_2}(M_2)=f'(y)$, the expression for the joint probability distribution
$f_\mathrm{pcrp}(M_1,M_2)$ is:
\begin{equation}
  f_\mathrm{pcrp}(M_1,M_2) = f_{M_1}(M_1){f'}_{M_2}(M_2)\,,
\end{equation}
which is normalised to 1 due to the re-normalisation of ${f'}_{M_2}(M_2)$.

To derive $f_q$ one can proceed as for the RP case. The integration domain is as
shown in the left-hand panel of Fig.~\ref{fig:fqseldomain} and the expression for $F(q)$ can be written
as:
\begin{equation}
  F(q)=\int_{c/q}^d\int_c^{qx} f(x)f'(y)dydx\,.
\end{equation}
Note that the normalisation constant for $f'(y)$ depends on $x$. The expression
for the restricted primary mass range becomes:
\begin{equation}
  F_q(q) =\left\{
  \begin{array}{ll}
    0 & q<\frac{c}{x_2} \\[5pt]
    k\int_{c/q}^{x_2}\int_c^{qx} f(x)f'(y)dydx & \frac{c}{x_2}\leq
    q<\frac{c}{x_1} \\[5pt]
    k\int_{x_1}^{x_2}\int_c^{qx} f(x)f'(y)dydx & \frac{c}{x_1}\leq q\leq 1
  \end{array}\right.\,,
  \label{eq:fqpcrpcuM_app}
\end{equation}
where the integration domains for $c/x_2\leq q <c/x_1$ and $c/x_1 \leq q \leq 1$ are shown in the middle and right-hand panels of Fig.~\ref{fig:fqseldomain}, respectively.

\begin{figure*}
  \begin{center}
    \includegraphics[width=0.48\textwidth]{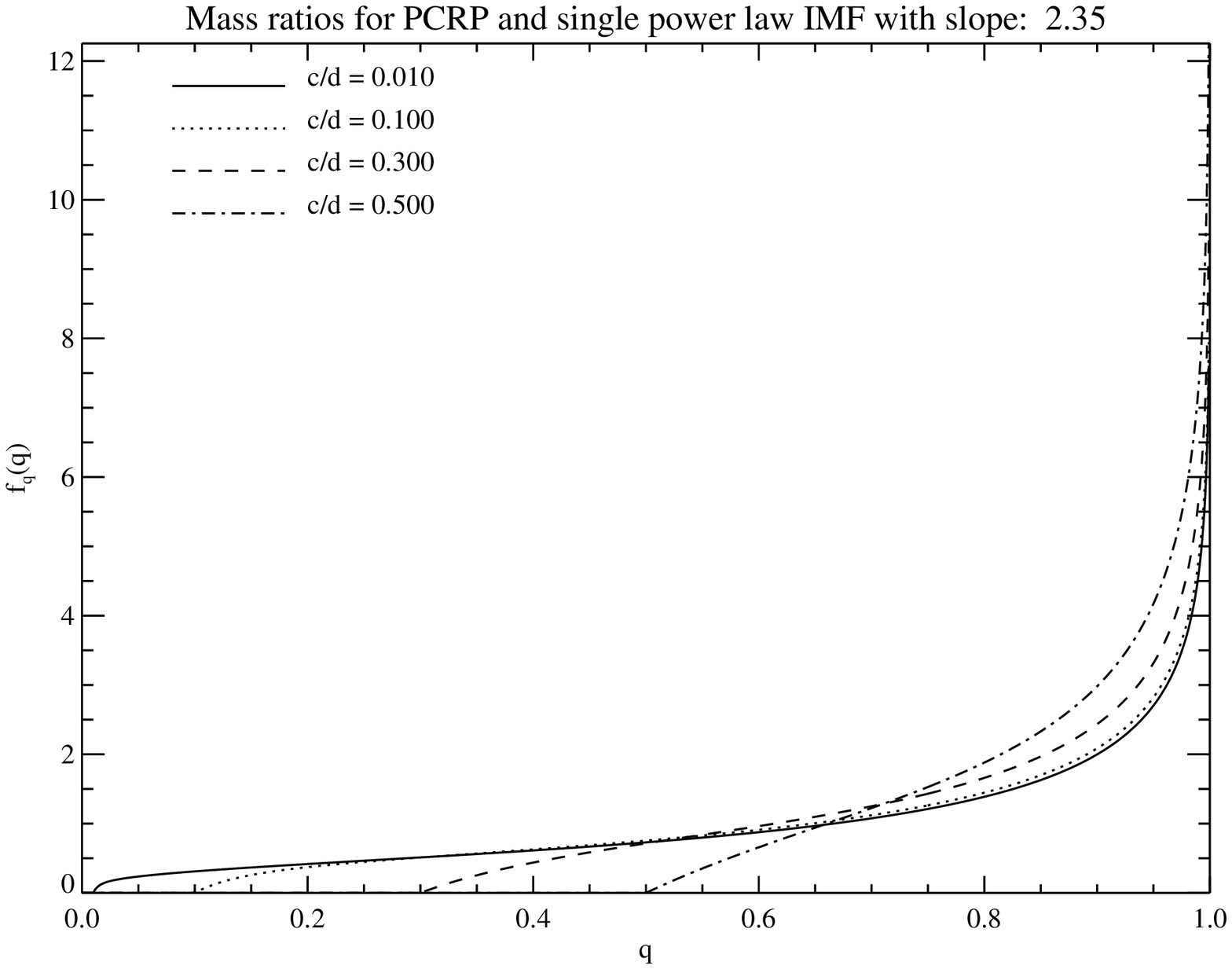}\hfil
    \includegraphics[width=0.48\textwidth]{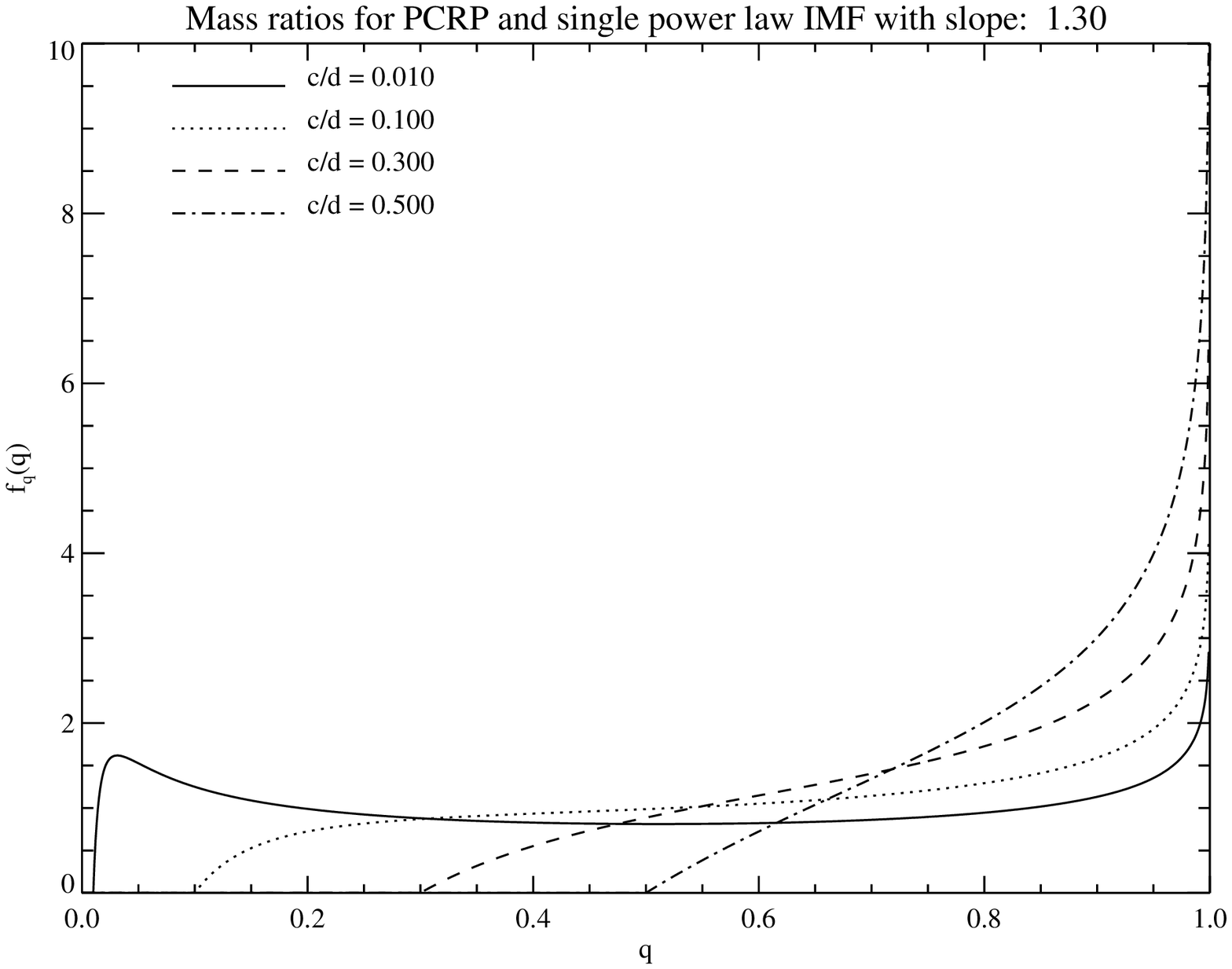}\\
    \includegraphics[width=0.48\textwidth]{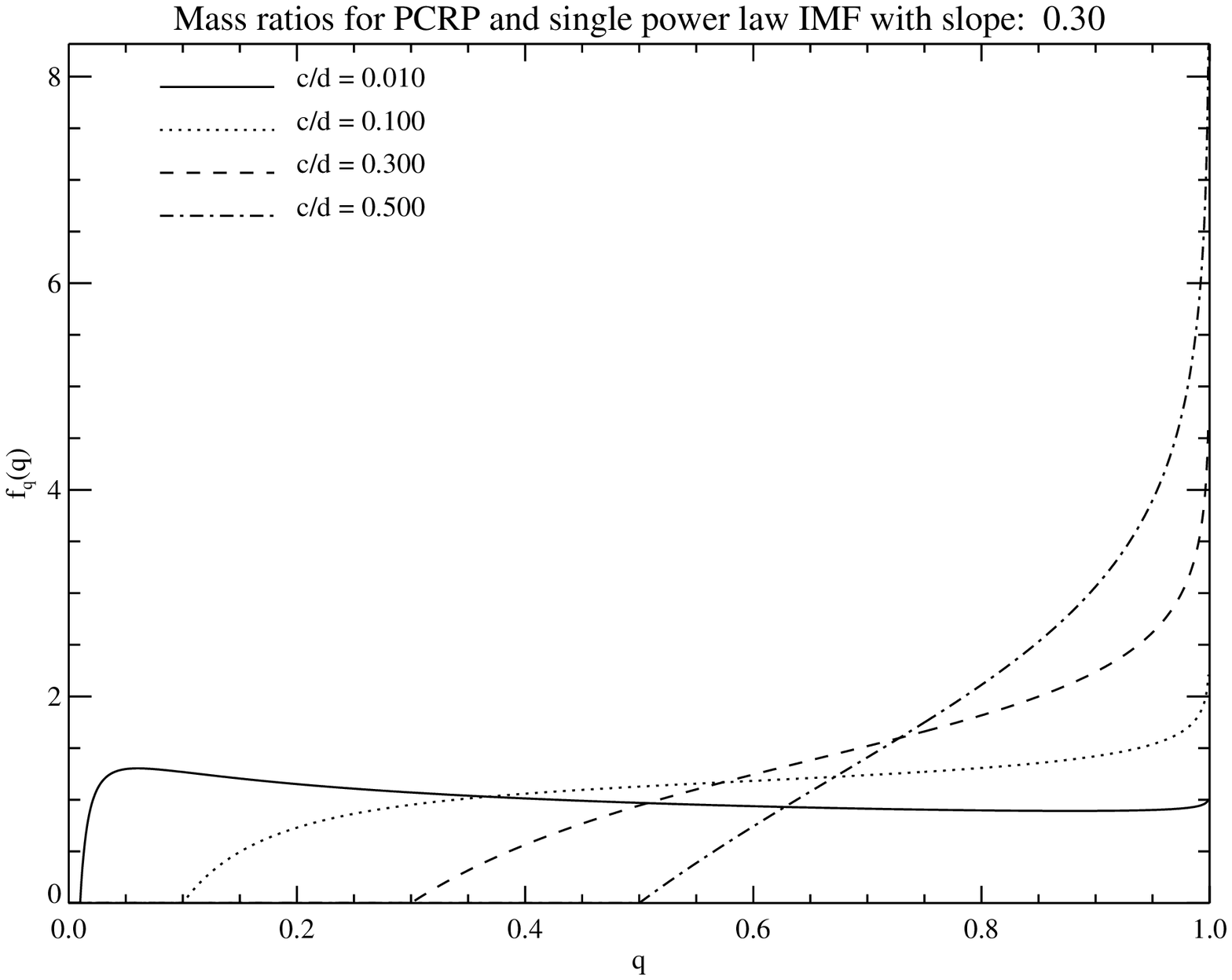}\hfil
    \includegraphics[width=0.48\textwidth]{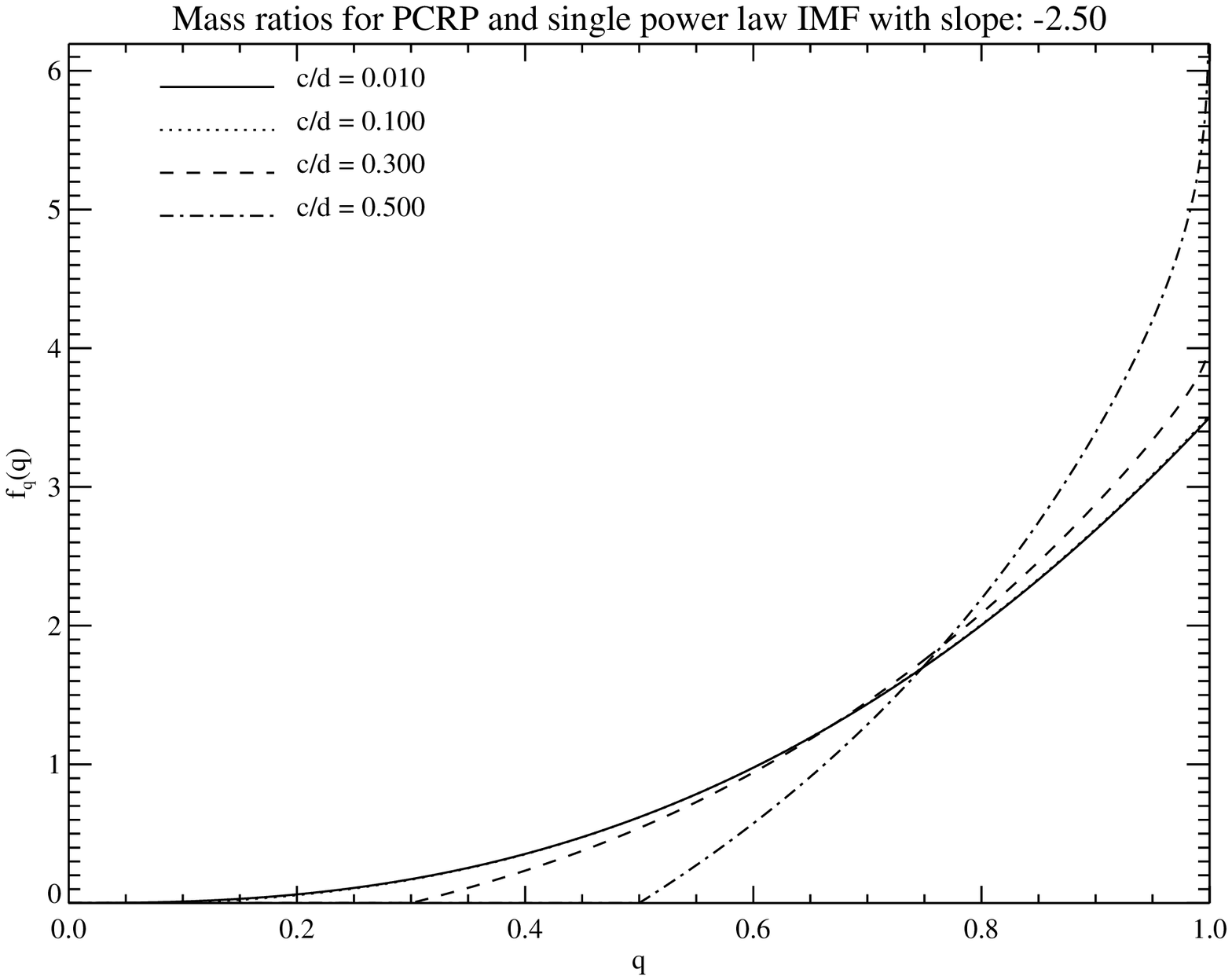}
  \end{center}
  \caption{Mass ratio distributions $f_q(q)$ for PCRP
    from a single power-law mass distribution. The curves are shown for
  $\alpha=2.35,\,1.30,\,0.30,\,-2.50$ and four ratios of the lower to the upper
  mass limit of the mass distribution: $c/d=0.01,\,0.1,\,0.3,\,0.5$.
  \label{fig:fqpcrpexample}}
\end{figure*}

\subsubsection{Binary fractions for PCRP}

For PCRP, the overall binary fraction $\binfall$ equals $\binf$, and specific binary fraction $\binfm$ is equal to the generating binary fraction $\binf$ for any mass $M_1$.

\subsubsection{Single power-law mass distribution}

For the single power-law mass distribution, with lower and upper mass limits $c$ and $d$, the
joint distribution of $x=M_1$ and $y=M_2$ is:
\begin{equation}
  f_\mathrm{pcrp}(x,y)=a_1x^{-\alpha}a_2(x)y^{-\alpha}\,,
\end{equation}
where the normalisation constant for $f(y)$ depends on the primary mass $x$:
\begin{equation}
  a_1=\frac{\gamma c^\gamma}{1-(c/d)^\gamma} \quad\mathrm{and}\quad
  a_2(x)=\frac{\gamma c^\gamma}{1-(c/x)^\gamma}\,.
\end{equation}
A distinction has to be made between the cases $\alpha<1$ and $\alpha>1$ as will
become clear below.

$F_q(q)$ is given by:
\begin{eqnarray*}
  F_q(q) & = &\int_{c/q}^d\int_c^{qx}\frac{\gamma^2
  c^{2\gamma}x^{-\alpha}}{(1-(c/d)^\gamma)(1-(c/x)^\gamma)}y^{-\alpha}dydx \\[5pt]
  & = & a_1\int_{c/q}^d x^{-\gamma-1}\left[
  \frac{q^{-\gamma}(c/x)^\gamma - 1}
  {(c/x)^\gamma-1}\right]
  dx\,.
\end{eqnarray*}
Using the substitution $z=c/x$ (which implies $dx=-(c/z^2)dz$) the integral can
be written in a more convenient form and its solution can be written as linear
combination of the terms $z^\gamma$ and $\ln(1-z^\gamma)$ for $\alpha>1$, while
for $\alpha<1$ the terms $z^\gamma$ and $\ln(z^\gamma-1)$ are involved.

\begin{figure*}
  \begin{center}
    \includegraphics[width=0.48\textwidth]{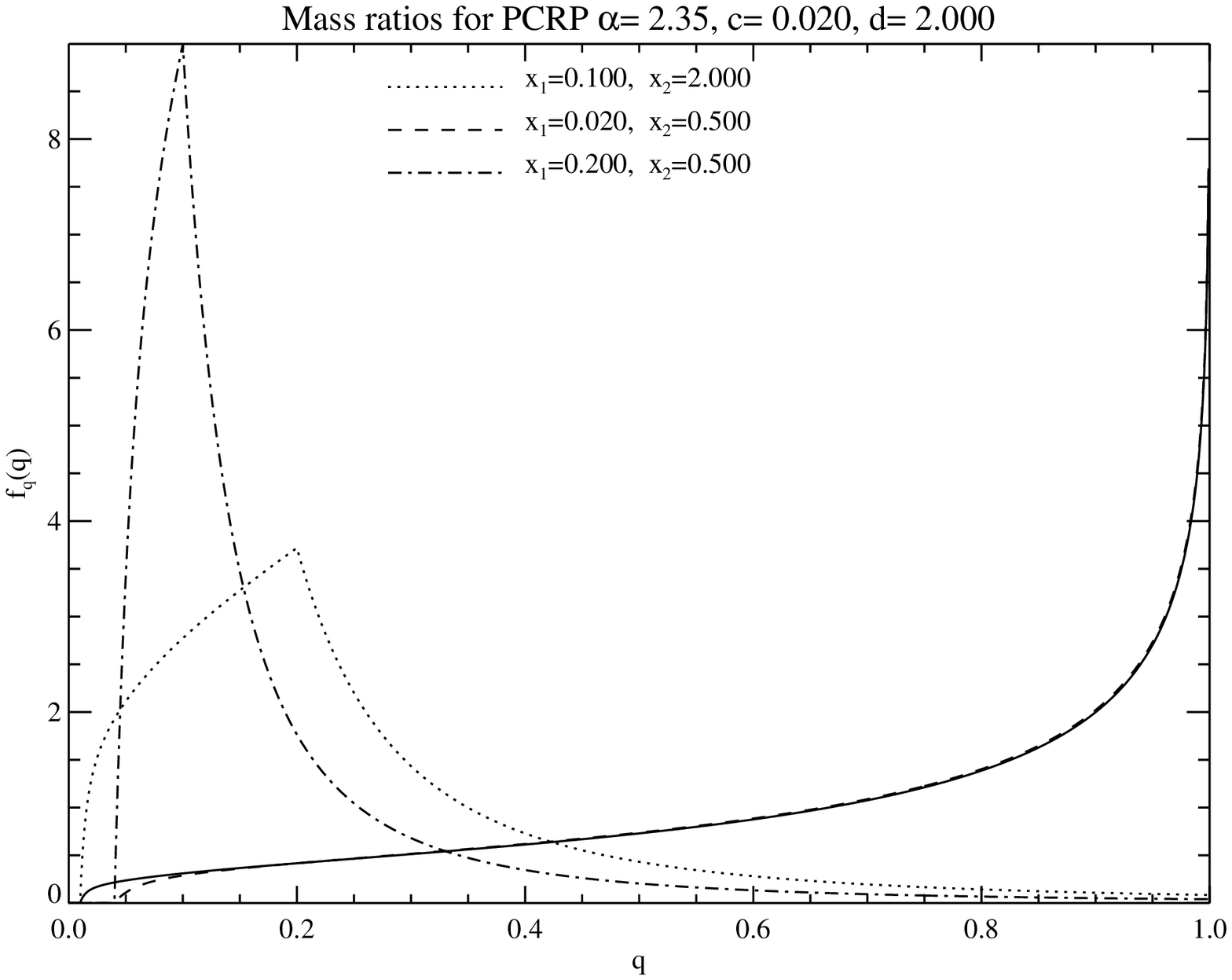}\hfil
    \includegraphics[width=0.48\textwidth]{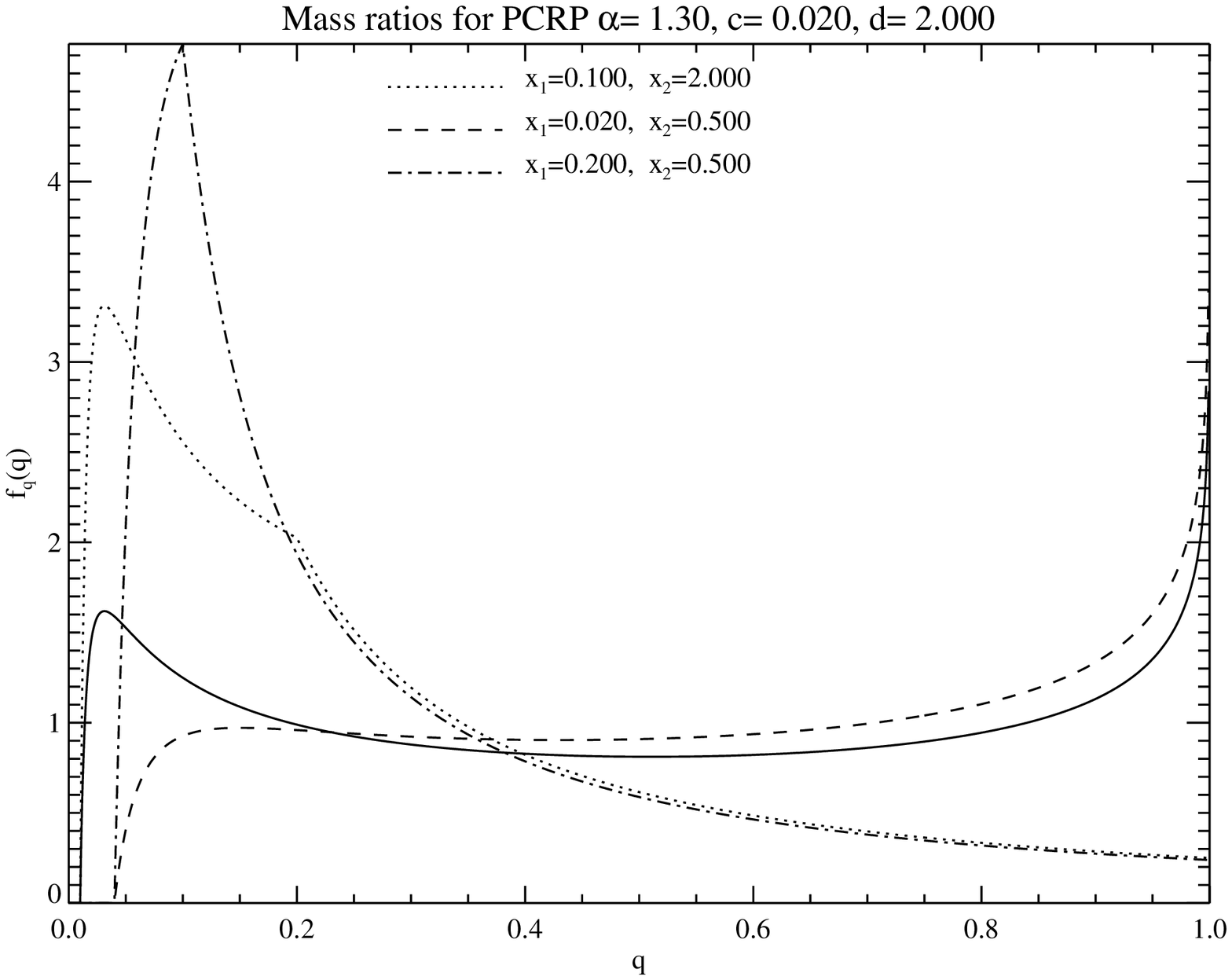}\\
    \includegraphics[width=0.48\textwidth]{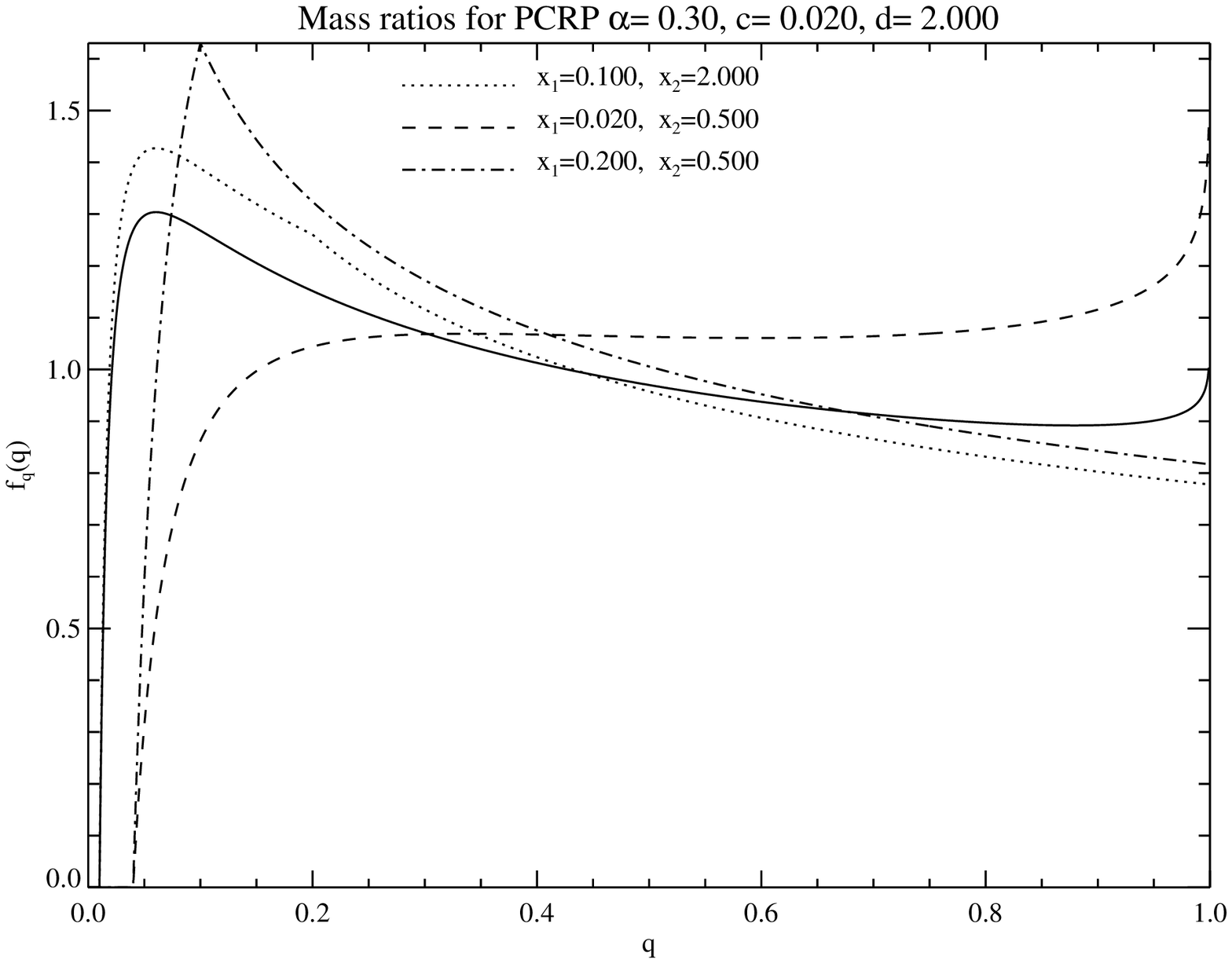}\hfil
    \includegraphics[width=0.48\textwidth]{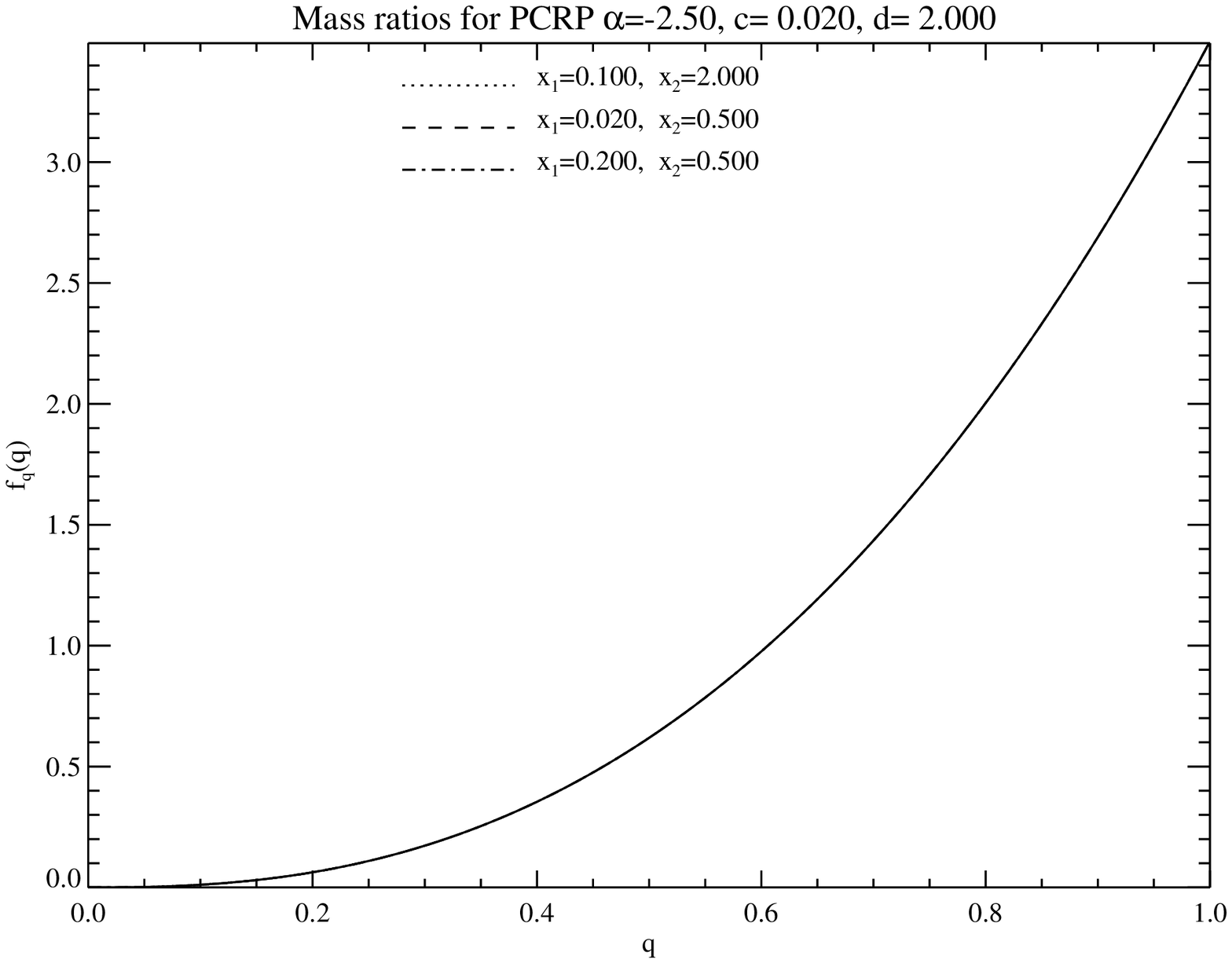}
  \end{center}
  \caption{Mass ratio distributions $f_q(q)$ for PCRP
    from a single power-law mass distribution. The solid curves show the complete mass
  ratio distribution (for all binaries in the population). The other curves show
  what happens to the observed $f_q(q)$ if the primary mass $M_1$ is restricted
  to $x_1\leq M_1\leq x_2$. The value of $c/d$ is $0.01$ and $x_1$ and $x_2$
  are listed in the panels.\label{fig:fqselpcrpexample}}
\end{figure*}

The resulting expression for $f_q(q)$ is:
\begin{equation}
  f_q(q)=\left\{
  \begin{array}{ll}
    \frac{\gamma q^{-\gamma-1}}{1-(c/d)^\gamma}\left[
    \ln\left(\frac{(c/d)^\gamma-1}{q^\gamma-1}\right)+(c/d)^\gamma-q^\gamma\right] &
    \alpha<1 \\[5pt]
    \frac{\gamma q^{-\gamma-1}}{1-(c/d)^\gamma}\left[
    \ln\left(\frac{1-(c/d)^\gamma}{1-q^\gamma}\right)+(c/d)^\gamma-q^\gamma\right] &
    \alpha>1
  \end{array}\right.\,.
  \label{eq:fqpcrp}
\end{equation}
This expression diverges as $q\rightarrow 1$ which is due to the low-mass end of
the primary mass distribution. As $M_1$ approaches $c$ the values of $q$ will
increasingly all be close to 1. The rate at which $f_q$ diverges depends on the
value of $\alpha$. For $\alpha>0$ the mass distribution peaks at the low-mass end thus causing
a rapid divergence.

Fig.~\ref{fig:fqpcrpexample} shows four examples of the mass ratio
distribution for PCRP from a single power-law mass distribution.
The curves are for the values of $c/d$ of $0.01$, $0.1$, $0.3$, and $0.5$.

For the restricted primary mass range Eq.~(\ref{eq:fqpcrpcuM_app}) has to be
worked out. For $c/x_2\leq q< c/x_1$ the expression for $F_q$ is:
\begin{eqnarray*}
  F_q(q) \propto \int_{c/q}^{x_2}\int_c^{qx}
  \frac{\gamma c^\gamma x^{-\alpha}}{1-(c/x)^\gamma}y^{-\alpha}dydx\,,
\end{eqnarray*}
which can be worked out to:
\begin{eqnarray*}
  F_q(q) \propto \int_{c/q}^{x_2}
  \frac{c^\gamma q^{-\gamma}x^{-2\gamma-1} -
  x^{-\gamma-1}}{(c/x)^\gamma-1}dx\,.
\end{eqnarray*}
This integral can be worked out in the same way as for the full primary mass range.
For $c/x_1\leq q \leq 1$ the expression for $F_q$ becomes:
\begin{eqnarray*}
  F_q(q) \propto \int_{x_1}^{x_2}
  \frac{c^\gamma q^{-\gamma}x^{-2\gamma-1} -
  x^{-\gamma-1}}{(c/x)^\gamma-1}dx\,,
\end{eqnarray*}

The expression for $f_q(q)$ for $\alpha>1$ is:
\begin{equation}
  f_q(q) =\left\{
  \begin{array}{ll}
    0 & q < \frac{c}{x_2} \\[5pt]
    \frac{k}{q^{\gamma+1}}\left(
    \ln\left(\frac{1-(c/x_2)^\gamma}{1-q^\gamma}\right)+\left(\frac{c}{x_2}\right)^\gamma -
    q^\gamma\right) & \frac{c}{x_2}\leq q < \frac{c}{x_1} \\[5pt]
    \frac{k}{q^{\gamma+1}}\left(
    \ln\left(\frac{1-(c/x_2)^\gamma}{1-(c/x_1)^\gamma}\right) + 
    \left(\frac{c}{x_2}\right)^\gamma -
    \left(\frac{c}{x_1}\right)^\gamma\right) & \frac{c}{x_1}\leq q\leq 1
  \end{array}\right.\,.
  \label{eq:fqselpcrpAlt1}
\end{equation}
For $\alpha<1$ the expression for $f_q$ is:
\begin{equation}
  f_q(q) =\left\{
  \begin{array}{ll}
    0 & q < \frac{c}{x_2} \\[5pt]
    \frac{k}{q^{\gamma+1}}\left(
    \ln\left(\frac{(c/x_2)^\gamma-1}{q^\gamma-1}\right)+\left(\frac{c}{x_2}\right)^\gamma -
    q^\gamma\right) & \frac{c}{x_2}\leq q < \frac{c}{x_1} \\[5pt]
    \frac{k}{q^{\gamma+1}}\left(
    \ln\left(\frac{1-(c/x_2)^\gamma}{1-(c/x_1)^\gamma}\right) +
    \left(\frac{c}{x_2}\right)^\gamma -
    \left(\frac{c}{x_1}\right)^\gamma\right) & \frac{c}{x_1}\leq q\leq 1
  \end{array}\right.\,.
  \label{eq:fqselpcrpAlt2}
\end{equation}
The normalisation constant $k$ is:
\begin{equation}
  k=\gamma\left( (c/x_1)^\gamma-(c/x_2)^\gamma\right)^{-1}\,.
\end{equation}

Again, if $x_1=x_2$, $f_q(q)$ will be proportional to $q^{-\alpha}$ for
$c/x_1\leq q\leq 1$ and zero otherwise and the expression is the same as
Eq.~(\ref{eq:fqselrponemass}).

Examples of what happens in the case of PCRP when a
selection is done on primary mass are shown in Fig.~\ref{fig:fqselpcrpexample}. 
The behaviour is qualitatively the same as for the
random pairing case.

\subsection{Primary-constrained pairing} \label{appendix:pcp}

For the pairing mechanism discussed now the assumption is that there is a
physical process which sets the primary mass and the \textit{mass ratio} of the
binary, rather than setting the masses of primary and secondary. For the primary
constrained pairing mechanism (PCP) the assumption is that $M_2$ is determined
from $M_1$ through the mass ratio $q$. The probability densities for $M_1$ and
$q$ are specified in this case and they are assumed to be independent. That is:
\begin{equation}
  f_\mathrm{pcp}(M_1,q)=f_{M_1}(M_1)h_q(q)\,,
  \label{eq:pcpjointdist}
\end{equation}
where the generating mass ratio distribution is written as $h_q(q)$ in order to
distinguish it from the observed mass ratio distribution $f_q(q)$. The latter
can be obtained by integrating $f_\mathrm{pcp}(M_1,q)$ over $M_1$:
\begin{equation}
  f_q(q)=\int f_\mathrm{pcp}(M_1,q)dM_1\,.
  \label{eq:defhq}
\end{equation}

There are a number of choices one can make in generating a binary population
from $f_\mathrm{pcp}$. The generating mass ratio distribution is assumed to be
specified for the interval $0<q\leq 1$ and $c\leq M_1\leq d$, which leads to
the following three possibilities:
\begin{description}
  \item{PCP-I} All values of $q$ are allowed which means that for a given
    primary mass $M_1$, $0<M_2\leq M_1$. Thus binary systems with `sub-stellar'
    secondary components are also allowed.
  \item{PCP-II} All values of $q$ are allowed but only binary systems for which
    the secondary is `stellar' (i.e.\ $M_2\geq c$) are retained. This amounts to
    integrating $f_\mathrm{pcp}$ over the range $c/q\leq M_1\leq d$ and
    re-normalising the resulting distribution of $q$ to 1.
  \item{PCP-III} Only values of $q$ for which $M_2\geq c$ are allowed, i.e.\
    $c/M_1\leq q\leq 1$. This is equivalent to re-normalising the generating
    distribution $h_q(q)$ to the interval $[c/M_1,1]$.
\end{description}

\begin{figure*}[t!]
  \begin{center}
    \includegraphics[width=0.8\textwidth]{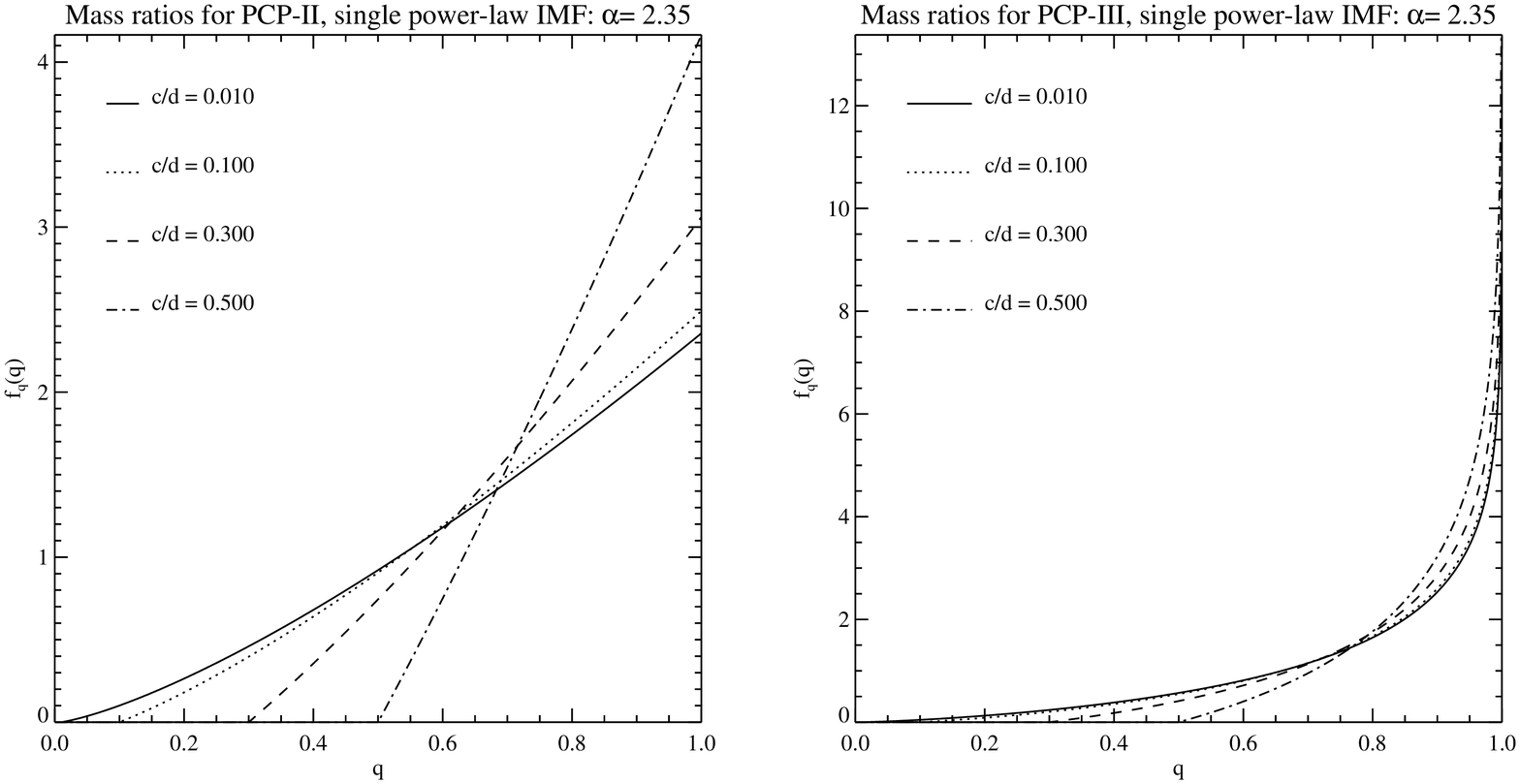}
    \includegraphics[width=0.8\textwidth]{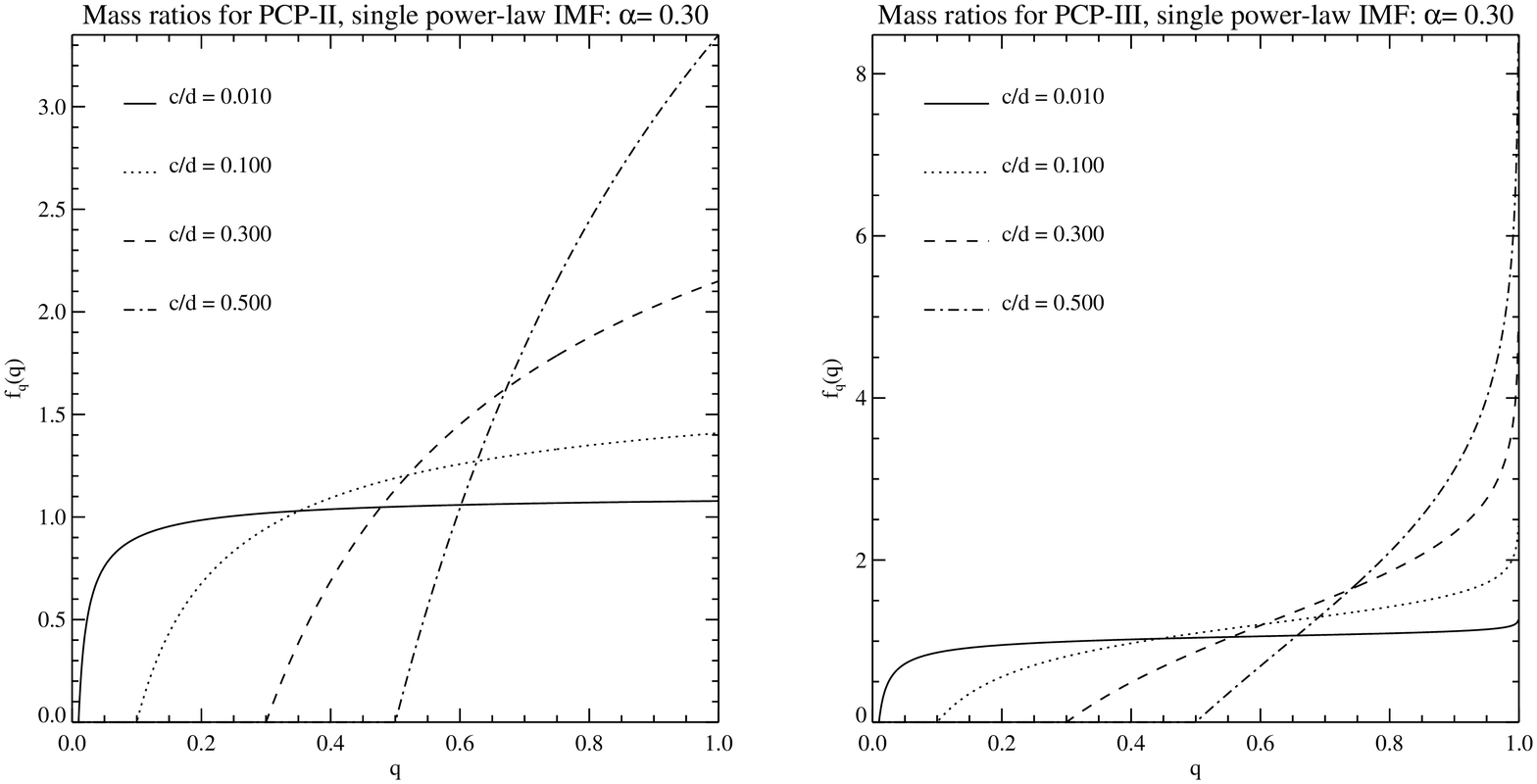}
    \includegraphics[width=0.8\textwidth]{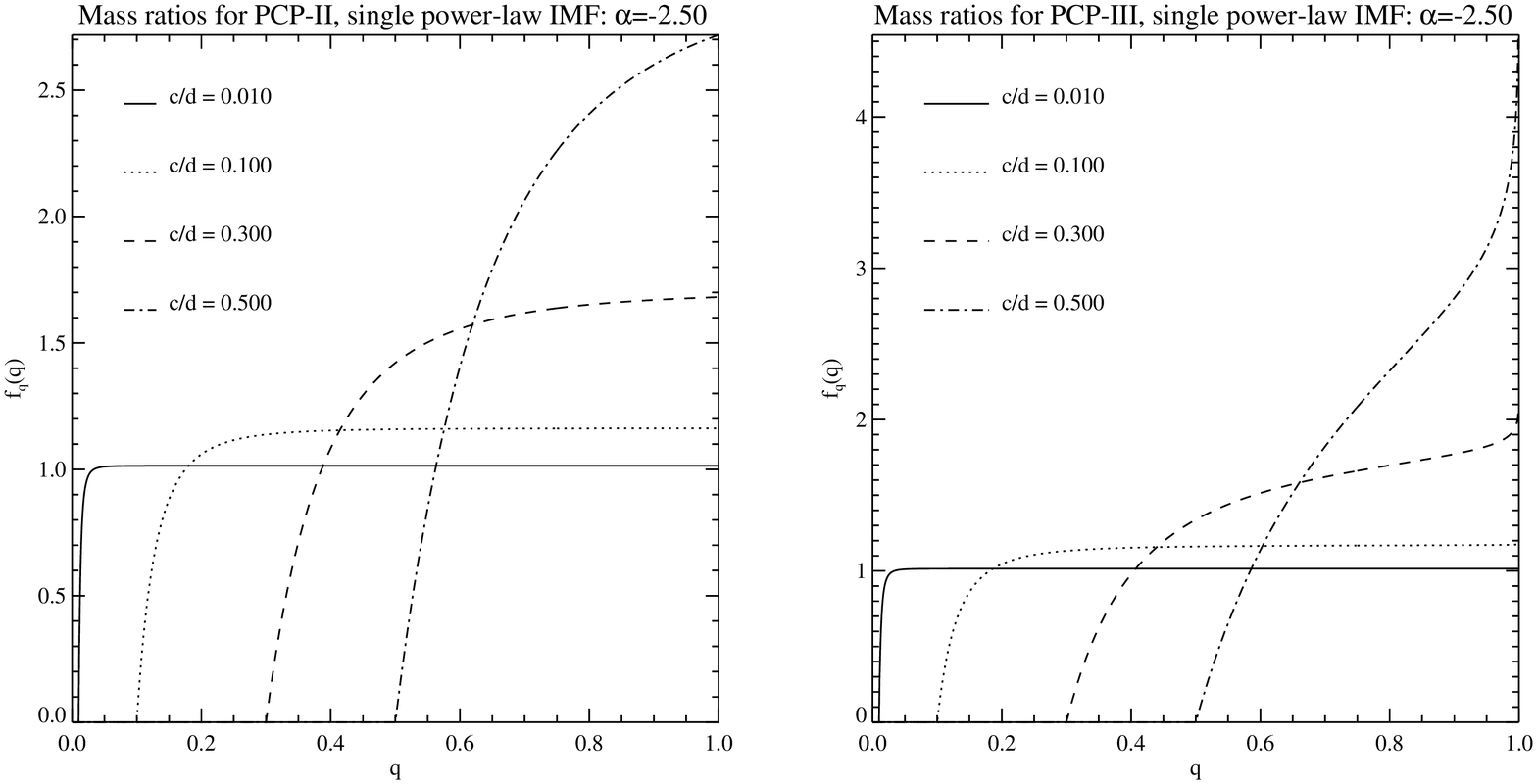}
  \end{center}
  \caption{Mass ratio distributions $f_q(q)$ for PCP-II
  and III for a uniform generating mass ratio distribution and primary masses
  from the single power-law mass distribution for $\alpha=2.35,\,0.30,\,-2.50$. The curves
  are shown for three ratios of the lower to the upper mass limit of the mass distribution:
  $c/d=0.1,\,0.3,\,0.5$.\label{fig:fqpcpcase1}}
\end{figure*}

\subsubsection{Mass ratio distributions for PCP}

Again we first derive the general expressions for $f_q(q)$ before discussing
specific examples.

\paragraph{PCP-I}

In this case one always obtains $f_q(q)=h_q(q)$ because the distributions of $q$
and $M_1$ are independent. Also when restricting the primary mass range the
observed mass ratio distribution is equal to the generating distribution.

\paragraph{PCP-II}

Here the systems with $M_2<c$ are discarded and then:
\begin{equation}
  f_q(q)=k\int_{c/q}^d f_\mathrm{pcp}(M_1,q)dM_1 = k\int_{c/q}^d
  h_q(q)f_{M_1}(M_1) dM_1\,,
  \label{eq:fqpcp2expr_app}
\end{equation}
where $k$ is a normalisation constant which ensures that $\int f_q(q)dq=1$. Of
course $f_q(q)=0$ for $0<q<c/d$. For a restricted primary mass range, $c\leq
x_1\leq M_1\leq x_2\leq d$, the expression for $f_q$ is:
\begin{equation}
  f_q(q) =\left\{
  \begin{array}{ll}
    0 & 0< q<\frac{c}{x_2} \\[5pt]
    k\int_{c/q}^{x_2} h_q(q)f_{M_1}(M_1) dM_1 & \frac{c}{x_2}\leq q<\frac{c}{x_1} \\[5pt]
    k\int_{x_1}^{x_2} h_q(q)f_{M_1}(M_1) dM_1 & \frac{c}{x_1}\leq q\leq 1
  \end{array}\right.\,,
  \label{eq:fqselpcp2expr}
\end{equation}
where $k$ is again a normalisation constant. 
For $q<c/x_2$ the value of $c/q$ is larger than $x_2$ so the lower
integration limit of the integral in Eq.~(\ref{eq:fqpcp2expr_app}) becomes
$x_2$, i.e., all companions with $q<c/x_2$ are rejected, as their mass
is smaller than the minimum mass $c$. 
For $q>c/x_1$ we have $c/q<x_1$ so the lower
integration limit of the integral should be fixed at $x_1$.

\paragraph{PCP-III}

In this case the generating mass ratio distribution is re-normalised to the
interval $[c/M_1,1]$, resulting in a generating distribution ${h'}_q(q)$. The
expressions for $f_q$ are then derived as for the PCP-II case:
\begin{equation}
  f_q(q)=\int_{c/q}^d f_\mathrm{pcp}(M_1,q)dM_1 = \int_{c/q}^d
  {h'}_q(q)f_{M_1}(M_1) dM_1\,,
  \label{eq:fqpcp3expr}
\end{equation}
where a normalisation constant is now not needed (${h'}_q(q)$ is normalised).
For the restricted primary mass range the expression is:
\begin{equation}
  f_q(q) = \left\{ 
  \begin{array}{ll}
    0 & 0< q<\frac{c}{x_2} \\[5pt]
    k\int_{c/q}^{x_2} {h'}_q(q)f_{M_1}(M_1) dM_1 & \frac{c}{x_2}\leq q<\frac{c}{x_1} \\[5pt]
    k\int_{x_1}^{x_2} {h'}_q(q)f_{M_1}(M_1) dM_1 & \frac{c}{x_1}\leq q\leq 1
  \end{array}\right.\,.
  \label{eq:fqselpcp3expr_threesegments}
\end{equation}

\begin{figure*}[t!]
  \begin{center}
    \includegraphics[width=0.8\textwidth]{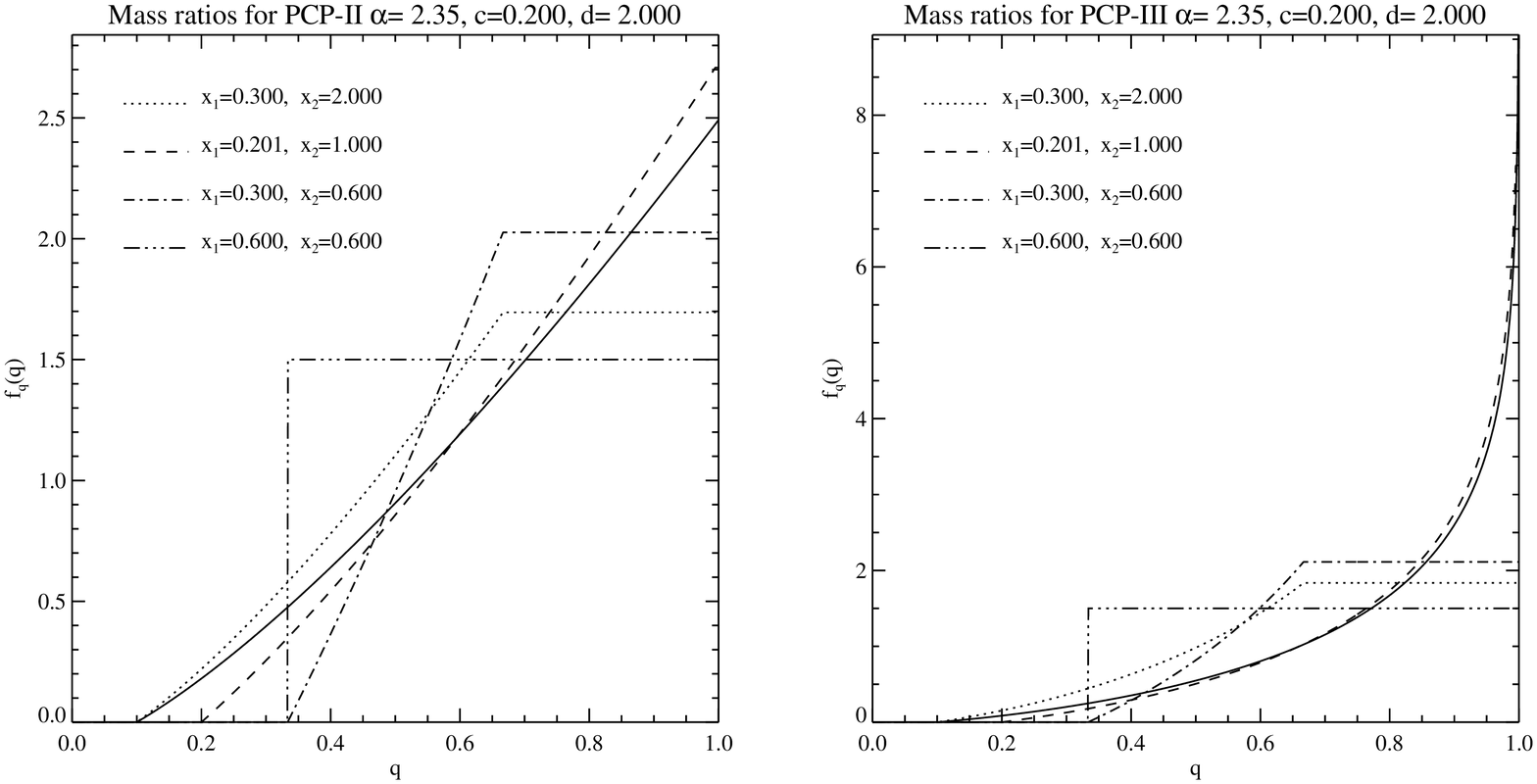}
    \includegraphics[width=0.8\textwidth]{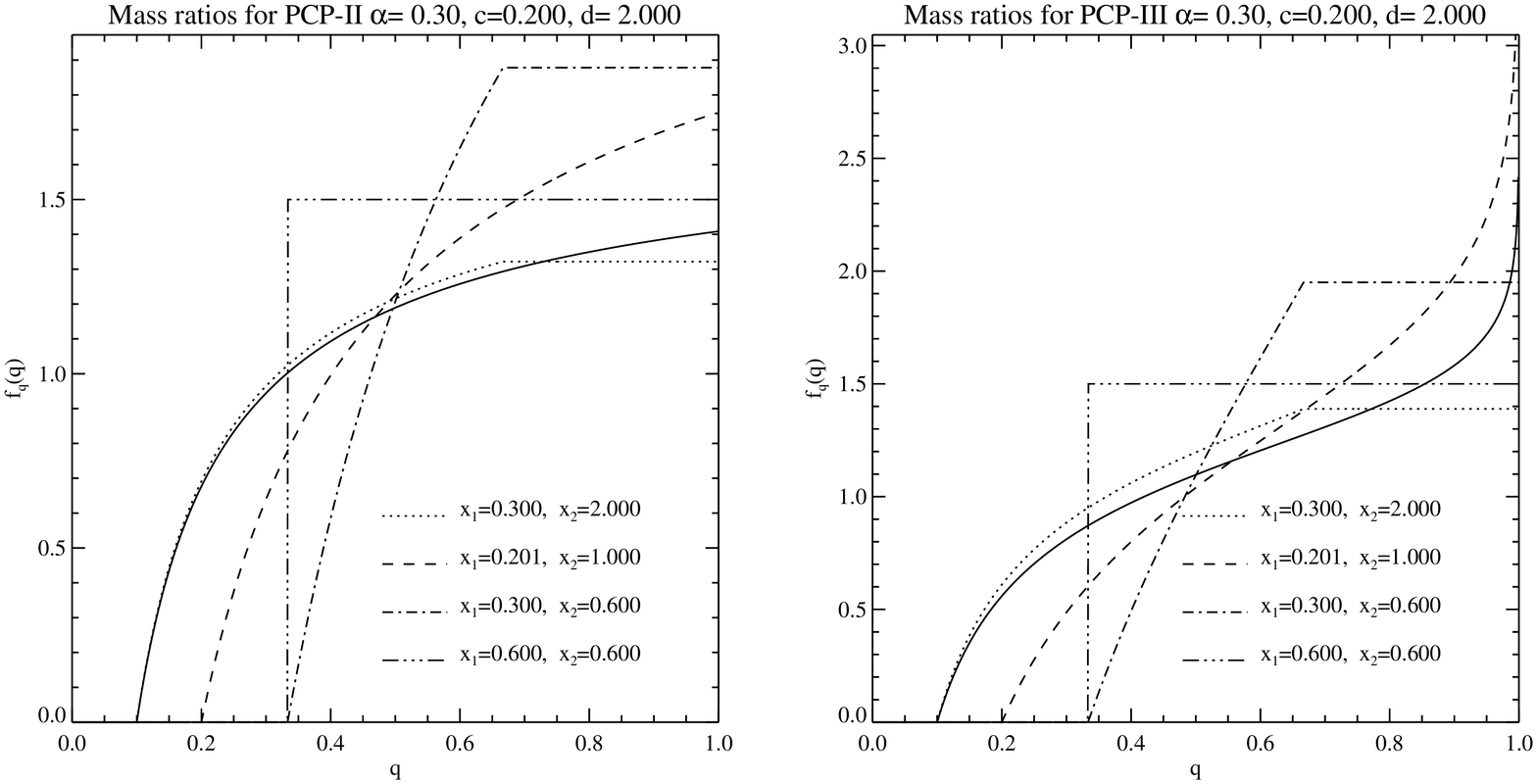}
    \includegraphics[width=0.8\textwidth]{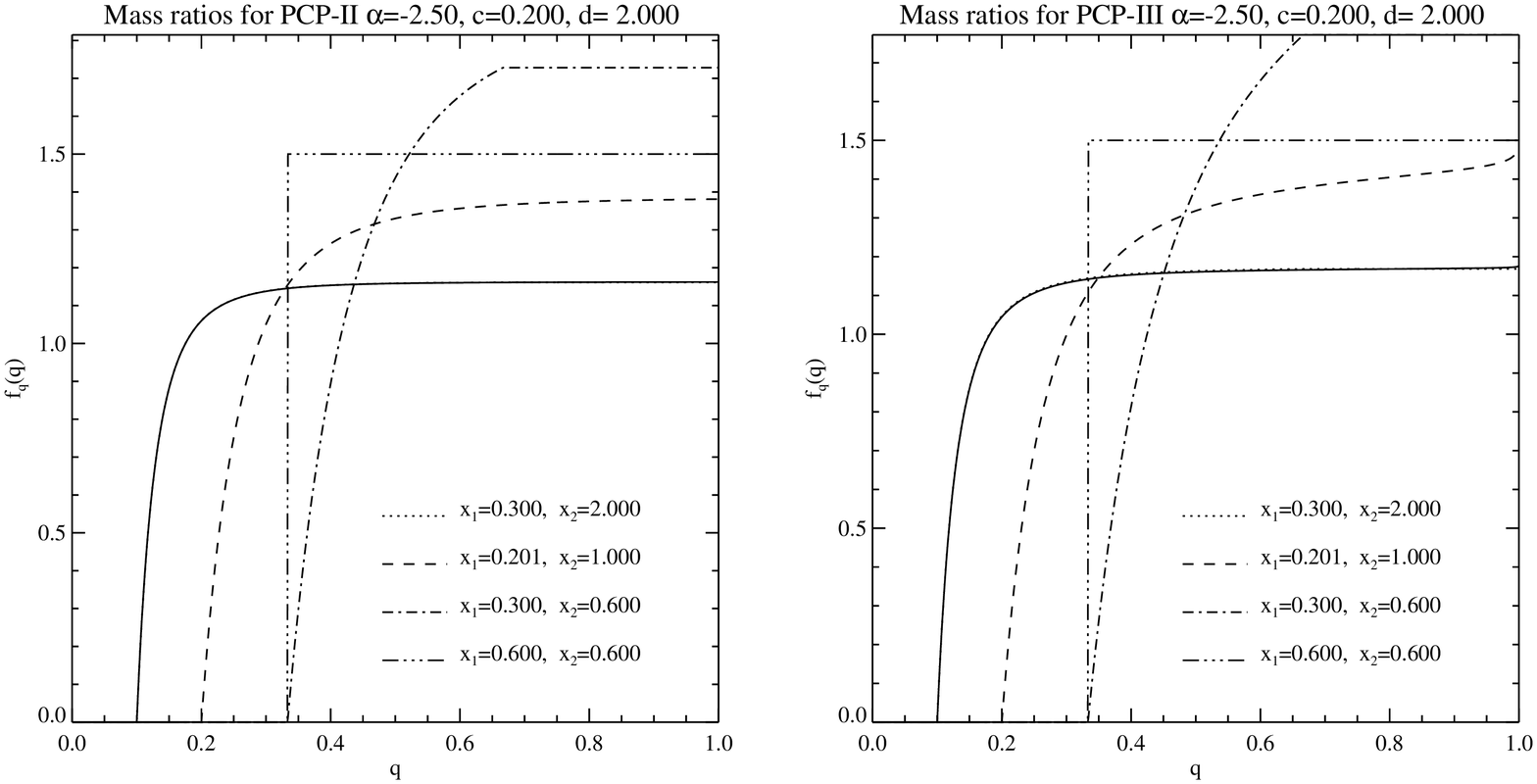}
  \end{center}
  \caption{Mass ratio distributions $f_q(q)$ for PCP-II
  and III for a uniform generating mass ratio distribution and primary masses
  from a single power-law mass distribution. The solid curve shows the complete mass ratio
  distribution (for all binaries in the population) for $c=0.2$ and $d=2.0$. The
  other curves show what happens to the observed $f_q(q)$ if the primary mass
  $M_1$ is restricted to $x_1\leq M_1\leq x_2$.\label{fig:fqselpcpcase1}}
\end{figure*}

\subsubsection{Binary fractions for PCP} \label{appendix:pcp_bf}

\paragraph{PCP-I} The overall and specific binary fractions are always equal to the generating binary fraction $\binf$ for PCP-I

\paragraph{PCP-II} Due to the rejection of low-mass companions with $M_2 < c$, i.e., with $q < \qmin(M_1) = c/M_1$, the binary fraction varies with primary mass:
\begin{equation} \label{equation:overall_bf_pcp}
  \binfm = \binf \int^1_{\qmin(M_1)} h_q(q) \,dq < \binf \,,
\end{equation}
where $h_q(q)$ is the generating mass ratio distribution. Note that $\binfm$ is independent of the generating mass distribution. An example of the mass-dependent binary fraction resulting from PCP-II is shown in Figure~\ref{figure:binaryfraction_versus_spectraltype}. For high-mass binaries with $M_1 \approx d$, very few companions are rejected as $\qmin(M_1) \ll 1$, and hence $\binfm \approx \binf$. For the lowest mass binaries in the sample, on the other hand, $\qmin(M_1) \approx 1$, and therefore $\binfm \approx 0$.
The overall binary fraction can be found by integrating over primary mass
\begin{equation} \label{eq:binfall_for_pcp_and_scp}
  \binfall = \int_{c}^{d} \binfm \, f_{M_1}(M_1) \, dM_1 \,,
\end{equation}
and is always smaller than $\binf$.

\paragraph{PCP-III} The overall and specific binary fractions are always equal to the generating binary fraction $\binf$ for PCP-III

\subsubsection{Uniform mass ratio distribution and single power-law mass distribution}

The following is assumed for $h_q$ and $f_{M_1}(M_1)=f(x)$ (using $x=M_1$ for
ease of notation):
\begin{eqnarray*}
  h_q(q)=1 \quad\quad 0<q \leq 1 \\
  f(x)=ax^{-\alpha} \quad\quad  c\leq x\leq d
  \label{eq:pcpexdef}\,,
\end{eqnarray*}
where $a=\gamma c^\gamma/(1-(c/d)^\gamma)$ and $\alpha\neq1$.

\paragraph{The PCP-II case}

For the full primary mass range we have from Eq.~(\ref{eq:fqpcp2expr_app}):
\begin{equation}
  f_q(q)=k\int_{c/q}^d ax^{-\alpha}dx=
  k\frac{q^\gamma-(c/d)^\gamma}{1-(c/d)^\gamma}\,.
\end{equation}
The expression for $k$ can be found by solving $\int_{c/d}^1 f_q(q)dq=1$ for $k$
which leads to the final expression for the observed mass ratio distribution:
\begin{equation}
  f_q(q)= \left\{
  \begin{array}{ll}
    0 & 0<q<\frac{c}{d} \\
    \frac{(\gamma+1)(q^\gamma-(c/d)^\gamma)}{1-(\gamma+1)(c/d)^\gamma+\gamma(c/d)^{\gamma+1}}
    & \frac{c}{d}\leq q\leq 1
  \end{array}\right.\,.
  \label{eq:fqpcp2case1}
\end{equation}
If the primary mass range is restricted to $x_1\leq q\leq x_2$ the expression
for $f_q$ can be derived using Eq.~(\ref{eq:fqselpcp2expr}):
\begin{equation}
  f_q(q)=\left\{
  \begin{array}{ll}
    0 & 0< q<\frac{c}{x_2} \\[5pt]
    k(q^\gamma-(c/x_2)^\gamma) & \frac{c}{x_2}\leq q<\frac{c}{x_1} \\[5pt]
    k((c/x_1)^\gamma-(c/x_2)^\gamma) & \frac{c}{x_1}\leq q\leq 1
  \end{array}\right.\,,
  \label{eq:fqselpcp2case1}
\end{equation}
where the normalisation constant is:
\begin{equation}
  k=\frac{\gamma+1}{(\gamma+1)\left[(c/x_1)^\gamma-(c/x_2)^\gamma\right]-
  \gamma\left[(c/x_1)^{\gamma+1}-(c/x_2)^{\gamma+1}\right]}\,.
  \label{eq:kprimepcp2case1}
\end{equation}
Note that if $x_1=x_2$ the distribution of $q$ will be uniform on the interval
$[c/x_1,1]$, i.e.\ $f_q(q)=1/(1-c/x_1)$.
The equations above do not hold for $\alpha=0$. This case
is easily evaluated and the result for $\alpha=0$ is:
\begin{equation}
  f_q(q) = \frac{1}{1-c/d+(c/d)\ln(c/d)}\left(1-\frac{c/d}{q}\right)\,.
  \label{eq:fqpcp2case1alphazero}
\end{equation}

\paragraph{The PCP-III case}

Now the mass ratio distribution is re-normalised to the interval $[c/x,1]$ which
gives:
\begin{equation}
  {h'}_q(q)=\frac{1}{1-c/x}\,.
\end{equation}
For the full primary mass range $f_q$ is obtained from Eq.~(\ref{eq:fqpcp3expr}):
\begin{equation}
  f_q(q) = \left\{
  \begin{array}{ll}
    0 & 0<q<\frac{c}{d} \\[5pt]
    a\int_{c/q}^d\frac{x^{1-\alpha}}{x-c}dx & \frac{c}{d}\leq q\leq 1
  \end{array}\right.\,.
  \label{eq:fqpcp3case1}
\end{equation}
Restricting the primary mass range leads to:
\begin{equation}
  f_q(q)=\left\{
  \begin{array}{ll}
    0 & 0< q<\frac{c}{x_2} \\[5pt]
    k\int_{c/q}^{x_2}\frac{x^{1-\alpha}}{x-c}dx & \frac{c}{x_2}\leq q<\frac{c}{x_1} \\[5pt]
    k\int_{x_1}^{x_2}\frac{x^{1-\alpha}}{x-c}dx & \frac{c}{x_1}\leq q\leq 1 \\[5pt]
  \end{array}\right.\,,
  \label{eq:fqselpcp3case1}
\end{equation}
The expressions for the PCP-III case can be written in a slightly more
convenient form when using the substitution $z=c/x$. For the full primary mass
range:
\begin{equation}
  f_q(q) =\left\{
  \begin{array}{ll}
    0 & 0<q<\frac{c}{d} \\[5pt]
    \frac{\gamma}{1-(c/d)^\gamma}\int_{c/d}^q\frac{z^{\gamma-1}}{1-z}dz &
    \frac{c}{d}\leq q\leq 1
  \end{array}\right.\,,
  \label{eq:fqpcp3case1conv}
\end{equation}
and for the restricted range:
\begin{equation}
  f_q(q)=\left\{
  \begin{array}{ll}
    0 & 0< q<\frac{c}{x_2} \\[5pt]
    k\frac{\gamma}{1-(c/d)^\gamma}\int_{c/x_2}^{q}\frac{z^{\gamma-1}}{1-z}dx &
    \frac{c}{x_2}\leq q<\frac{c}{x_1} \\[5pt]
    k\frac{\gamma}{1-(c/d)^\gamma}\int_{c/x_2}^{c/x_1}\frac{z^{\gamma-1}}{1-z}dx
    & \frac{c}{x_1}\leq q\leq 1 \\[5pt]
  \end{array}\right.\,,
  \label{eq:fqselpcp3case1conv}
\end{equation}
The integrals involving the term $z^{\gamma-1}/(1-z)$ can be solved and the
general expression involves the hypergeometric function $_2F_1$. The expression
is:
\begin{equation}
  \int\frac{z^{\gamma-1}}{1-z}dz = \frac{z^\gamma}{\gamma} + 
  \frac{_2F_1(\gamma+1,1;\gamma+2;z) z^{\gamma+1}}{\gamma+1}\,.
  \label{fqpcp3case1convhyper}
\end{equation}
From the properties of the hypergeometric function \citep[see][Chapt.\ 9]{TISP7}
it follows that because $(\gamma+1)+1-(\gamma+2)=0$ the expression above
converges throughout the unit circle in the complex plane except at $|z|=1$. So
there will be a singularity at $q=1$ for the full primary mass range case.

For integer values of $\gamma$ (or $\alpha$) special care should be taken. From
the expression above it is clear that $\gamma=0$ or $\gamma=-1$ should be
treated separately. Furthermore the hypergeometric series for $_2F_1(a,b;c;z)$
is indeterminate for $c=-n$ where $n=0,1,2,\cdots$ if neither $a$ nor $b$ is
equal to $-m$ (where $m<n$ and $m$ is a natural number). Here this means that
all cases $\gamma+2=-n$ should be treated separately which combined with
the condition $\gamma\neq0,-1$ implies that whenever $\gamma=-n$ (i.e.\
$\alpha=-n+1$) the expression above will not apply. The case $\alpha=1$,
$\gamma=0$, is excluded, so the special cases are $\gamma=-n-1$, $\alpha=-n$,
and then the expression for the integrals becomes:
\begin{equation}
  \int\frac{dz}{z^{n+1}(1-z)} = \sum_{k=1}^n \frac{-1}{(n+1-k)z^{n+1-k}}
  -\ln\frac{1-z}{z}\,,
  \label{fqpcp3case1convintgamma}
\end{equation}
where the solution can be found from formula 2.117(4) in \cite{TISP7}. In the
integrals above the integration constants were left out.

For restricted primary mass ranges the results above can be used to evaluate the
integrals listed in Eq.~(\ref{eq:fqselpcp3case1conv}). The normalisation
constant $k$ can be obtained from numerical integration of $f_q(q)$. Again, if
$x_1=x_2$ one obtains $f_q(q)=1/(1-c/x_1)$ for $q$ in the interval $[c/x_1,1]$.
Figs.~\ref{fig:fqpcpcase1} and~\ref{fig:fqselpcpcase1} show examples of the
resulting mass ratio distributions for $h_q(q)=1$ 
and various single power-law mass distributions for the primaries. The
hypergeometric function was calculated using the routine from \citet[][\S\,6.13]{NR3}.

Finally, we have listed the expressions for the binary fraction as a function of primary mass in Table~\ref{table:binfracm_powerlaw}.

\subsection{Split-core pairing} \label{appendix:scp}

The mechanism of split-core pairing (SCP) works on the assumption that binaries
are formed by the splitting of star-forming cores into two components. The
component masses are specified through their mass ratio. The distribution
function for the core masses $M_c$ is given by the core
mass function $f_\mathrm{c}(M_\mathrm{c})$ and the mass ratio distribution $f_q(q)$
is specified independently. The masses of the primary and secondary are then
given by:
\begin{equation}
  M_1=\frac{1}{1+q}M_\mathrm{c} \quad\mathrm{and}\quad
  M_2=\frac{1}{1+q^{-1}}M_\mathrm{c}\,,
  \label{eq:scpm1m2}
\end{equation}
where for simplicity the star formation efficiency is assumed to be 100\% once
the core mass is set. For a constant efficiency as a function of core mass this
assumption has no influence on the results. Furthermore it is assumed that the
minimum core mass is large enough to ensure that the primary is always of
`stellar' mass. That is the minimum core mass has to be at least twice the
minimum stellar mass.

The joint probability density $f_\mathrm{scp}$ for $M_\mathrm{c}$ and $q$ is
written as:
\begin{equation}
  f_\mathrm{scp}(M_\mathrm{c},q)=f_\mathrm{c}(M_\mathrm{c})h_q(q)\,,
  \label{eq:scpjointdist}
\end{equation}
where again the generating mass ratio distribution is written as $h_q(q)$ in
order to distinguish it from the observed mass ratio distribution $f_q(q)$. The
latter can be obtained by integrating $f_\mathrm{scp}(M_\mathrm{c},q)$ over
$M_\mathrm{c}$:
\begin{equation}
  f_q(q)=\int f_\mathrm{scp}(M_\mathrm{c},q)dM_\mathrm{c}\,.
  \label{eq:defhqscp}
\end{equation}

As for PCP there are three cases: The generating mass
ratio distribution is assumed to be specified for the interval $0<q\leq 1$ and
we assume $2c\leq M_\mathrm{c}\leq 2d$ (with $c$ and $2d$ being the minimum and
maximum stellar mass, respectively), which leads to the following three
possibilities:
\begin{description}
  \item{SCP-I} All values of $q$ are allowed, so that for a given core
    mass $M_\mathrm{c}$, $M_1=M_\mathrm{c}/(1+q)$ and $0<M_2\leq M_1$. Thus
    binary systems with `sub-stellar' companions are allowed.
  \item{SCP-II} All values of $q$ are allowed, but only binaries for which
    the secondary is `stellar' (i.e.\ $M_2\geq c$) are retained. This amounts to
    integrating $f_\mathrm{scp}$ over the range $c(1+1/q)\leq M_\mathrm{c}\leq
    2d$ and then re-normalising the resulting distribution of $q$ to 1.
  \item{SCP-III} Only values of $q$ for which $M_2\geq c$ are allowed, i.e.\
    $c/(M_\mathrm{c}-c)\leq q\leq 1$. This is equivalent to re-normalising the
    generating distribution $h_q(q)$ to the interval $[c/(M_\mathrm{c}-c),1]$.
\end{description}
If we would have chosen our core mass distribution in the interval $c \leq M_c \leq 2d$ rather than $2c \leq M_c \leq 2d$, we would have encountered a further complication. For pairing function SCP-II the resulting single star mass is then occasionally smaller than $c$, and for SCP-III the splitting up is not possible if the binary system mass is smaller than $2c$ (see \S\,\ref{section:pairingfunction} for details). For simplicity we avoid this issue in our analysis below, and simply draw masses from the core mass distribution in the range $2c \leq M_c \leq 2d$. 

\subsubsection{Mass ratio distributions for SCP} \label{sec:scp}

We give the general expressions for $f_q(q)$ before discussing specific
examples. For ease of notation we use $x=M_\mathrm{c}$ and
$f(x)=f_\mathrm{c}(M_\mathrm{c})$.

\paragraph{SCP-I}

If no selection on primary mass is made one always obtains $f_q(q)=h_q(q)$
because the distributions of $q$ and $M_\mathrm{c}$ are independent. Unlike
PCP-I, restricting the primary mass range now does cause changes of the observed
mass ratio distribution with respect to the generating one. Several cases have
to be distinguished on the basis of the value of the primary mass selection
limits $x_1$ and $x_2$ compared to the values of $c$, $d$, $2c$, and $2d$. Note
that $x_1$ and $x_2$ are limits on $M_1$, not $M_\mathrm{c}$.

In all cases discussed below a selection $x_1\leq M_1\leq x_2$ in principle
translates to $x_1(1+q)\leq x\leq x_2(1+q)$ but the upper and/or lower limits on
$x$ and $q$ used for integrating $f_\mathrm{scp}$ and normalising $f_q(q)$ are
different for each case. We introduce the following variables
to distinguish the different cases:

\vskip0.15cm
\begin{tabular}{p{0.05cm}llp{0.05cm}l}
  $q_0$ & $= 2c/x_2-1$ & $\quad$ & $q_2$ & $= 2d/x_2-1$ \\
  $q_1$ & $= 2c/x_1-1$ & $\quad$ & $q_3$ & $= 2d/x_1-1$ \\ 
\end{tabular}
\vskip0.15cm

\noindent
The value of $x$ (core-mass) is restricted to $[2c,2d]$, hence for $q<q_0$ or
$q>q_3$ the probability distribution for $q$ vanishes (as these conditions imply
$M_1>x_2$ or $M_1<x_1$, both of which are not allowed). For $q<q_1$ the lower
integration limit for $x$ when determining $f_q$ is fixed at $2c$ and for
$q>q_2$ the upper limit is fixed at $2d$.

\textbf{Case 1:} $x_1\leq c<x_2\leq 2c$\quad In this case the value of $q_0$ is
between 0 and 1 and all other values of $q_i$ are larger than 1. This means that
two cases should be distinguished for $f_q$, $q<q_0$ and $q>q_0$, and that the
lower integration limit for $x$ is always fixed at $2c$. The expression for
$f_q$ is then given by:
\begin{equation}
  f_q(q)=\left\{
  \begin{array}{ll}
    0 & 0< q< q_0 \\[5pt]
    k\int_{2c}^{x_2(1+q)}h_q(q)f(x)dx & q_0\leq q\leq 1
  \end{array}\right.\,,
  \label{eq:fqscp1case1}
\end{equation}
where $k$ is a normalisation constant which follows from:
\begin{equation}
  \int_{q_0}^1 f_q(q)dq=1\,.
  \label{eq:fqscp1case1norm}
\end{equation}

\textbf{Case 2:} $x_1\leq c \,\wedge\, 2c<x_2\leq d$\quad This is the same as case 1
except that now $q_0<0$ so that there is only one part to the expression for
$f_q$. The lower integration limit for $x$ is still $2c$:
\begin{equation}
  f_q(q)= k\int_{2c}^{x_2(1+q)}h_q(q)f(x)dx \quad 0< q\leq 1\,,
  \label{eq:fqscp1case2}
\end{equation}
where the normalisation constant $k$ is obtained by integrating $f_q$ over $[0,1]$.

\textbf{Case 3:} $x_1\leq c \,\wedge\, d<x_2\leq 2d$\quad Now the value of $q_2$ is
in the interval $[0,1]$ which means that for the case $q>q_2$ the upper
integration limit for $x$ is fixed at $2d$ which leads to a plateau in the
probability density $f_q$:
\begin{equation}
  f_q(q)=\left\{
  \begin{array}{ll}
    k\int_{2c}^{x_2(1+q)}h_q(q)f(x)dx & 0< q\leq q_2 \\[5pt]
    k\int_{2c}^{2d}h_q(q)f(x)dx & q_2< q \leq 1
  \end{array}\right.\,,
  \label{eq:fqscp1case3}
\end{equation}
where the normalisation constant $k$ is obtained through integration over the
intervals $[0,q_2]$ and $[q_2,1]$.

\textbf{Case 4:} $c<x_1<x_2\leq 2c$\quad Now the values of $q_0$ and $q_1$ are in the
interval $[0,1]$ which means that the lower integration limit for $x$ depends on
$q$. The probability density for $q$ now consists of three parts:
\begin{equation}
  f_q(q)=\left\{
  \begin{array}{ll}
    0 & 0< q< q_0 \\[5pt]
    k\int_{2c}^{x_2(1+q)}h_q(q)f(x)dx & q_0\leq q< q_1\\[5pt]
    k\int_{x_1(1+q)}^{x_2(1+q)}h_q(q)f(x)dx & q_1\leq q\leq 1\\
  \end{array}\right.\,,
  \label{eq:fqscp1case4}
\end{equation}
where $k$ now has to be obtained from condition (\ref{eq:fqscp1case1norm})
by integrating over the intervals $[q_0,q_1]$ and $[q_1,1]$.

\textbf{Case 5:} $c<x_1\leq 2c \,\wedge\, 2c<x_2\leq d$\quad The value of $q_0$
becomes less than zero and we have:
\begin{equation}
  f_q(q)=\left\{
  \begin{array}{ll}
    k\int_{2c}^{x_2(1+q)}h_q(q)f(x)dx & 0< q< q_1\\[5pt]
    k\int_{x_1(1+q)}^{x_2(1+q)}h_q(q)f(x)dx & q_1\leq q\leq 1\\
  \end{array}\right.\,,
  \label{eq:fqscp1case5}
\end{equation}
where $k$ is now has to be obtained from condition (\ref{eq:fqscp1case1norm})
by integrating over the intervals $[0,q_1]$ and $[q_1,1]$.

\textbf{Case 6:} $c<x_1\leq 2c \,\wedge\, d<x_2\leq 2d$\quad Now the values of $q_1$
and $q_2$ are in the interval $[0,1]$. For $q<q_1$ the lower integration limit
is fixed to $2c$ and for $q>q_2$ the upper integration limit is fixed to $2d$.
In addition it can happen that $q_1<q_2$ or $q_1\geq q_2$. The latter case will
cause a plateau of constant probability density $f_q$ for $q_1\leq q \leq q_2$.
So now there are two `sub-cases'.

\textbf{Case 6a:} $q_1<q_2$
\begin{equation}
  f_q(q)=\left\{
  \begin{array}{ll}
    k\int_{2c}^{x_2(1+q)} h_q(q)f(x)dx & 0< q< q_1 \\[5pt]
    k\int_{x_1(1+q)}^{x_2(1+q)} h_q(q)f(x)dx & q_1\leq q<q_2 \\[5pt]
    k\int_{x_1(1+q)}^{2d} h_q(q)f(x)dx & q_2\leq q \leq 1
  \end{array}\right.\,,
  \label{fqscp1case6a}
\end{equation}
where the normalisation constant $k$ is obtained through integration over three
intervals $[0,q_1]$, $[q_1,q_2]$, and $[q_2,1]$.

\textbf{Case 6b:} $q_1\geq q_2$
\begin{equation}
  f_q(q)=\left\{
  \begin{array}{ll}
    k\int_{2c}^{x_2(1+q)} h_q(q)f(x)dx & 0< q< q_2 \\[5pt]
    k\int_{2c}^{2d} h_q(q)f(x)dx & q_2\leq q<q_1 \\[5pt]
    k\int_{x_1(1+q)}^{2d} h_q(q)f(x)dx & q_2\leq q \leq 1
  \end{array}\right.\,,
  \label{fqscp1case6b}
\end{equation}
where the normalisation constant $k$ is obtained through integration over three
intervals $[0,q_2]$, $[q_2,q_1]$, and $[q_1,1]$.

\textbf{Case 7:} $2c<x_1<x_2\leq d$\quad Now the values of $q_0$ and $q_1$ are less
than zero and both $q_2$ and $q_3$ are larger than 1. Hence the probability
density consists of one part only:
\begin{equation}
  f_q(q)=k\int_{x_1(1+q)}^{x_2(1+q)} h_q(q)f(x)dx \quad 0< q\leq 1\,,
  \label{eq:fqscp1case7}
\end{equation}
where $k$ is obtained by integrating $f_q$ over $[0,1]$.

\textbf{Case 8:} $2c<x_1\leq d \,\wedge\, d<x_2\leq 2d$\quad The value of $q_2$ is
now in $[0,1]$ so $f_q$ will consist of two parts:
\begin{equation}
  f_q(q)=\left\{
  \begin{array}{ll}
    k\int_{x_1(1+q)}^{x_2(1+q)} h_q(q)f(x)dx & 0<q<q_2 \\[5pt]
    k\int_{x_1(1+q)}^{2d} h_q(q)f(x)dx & q_2\leq q\leq 1
  \end{array}\right.\,,
  \label{eq:fqscp1case8}
\end{equation}
where $k$ is obtained by integrating $f_q$ over $[0,q_2]$ and $[q_2,1]$.

\textbf{Case 9:} $d<x_1<x_2\leq d$\quad Now the values of both $q_2$ and $q_3$
are in $[0,1]$ and the probability density will vanish if $q>q_3$:
\begin{equation}
  f_q(q)=\left\{
  \begin{array}{ll}
    k\int_{x_1(1+q)}^{x_2(1+q)} h_q(q)f(x)dx & 0<q<q_2 \\[5pt]
    k\int_{x_1(1+q)}^{2d} h_q(q)f(x)dx & q_2\leq q< q_3 \\[5pt]
    0 & q_3\leq q\leq 1
  \end{array}\right.\,,
  \label{eq:fqscp1case9}
\end{equation}
where $k$ is obtained by integrating $f_q$ over $[0,q_2]$ and $[q_2,q_3]$.

\paragraph{SCP-II}

The expression for $f_q$ is obtained by integrating the joint distribution
$f_\mathrm{scp}$ over $x$ over the range $c(1+1/q)\leq x\leq 2d$ and normalising
the resulting expression for $f_q$ to 1:
\begin{equation}
  f_q(q)=k\int_{c(1+1/q)}^{2d} h_q(q)f(x)dx\,,
  \label{eq:fscp2exprgen}
\end{equation}
where $k$ is obtained from the condition:
\begin{equation}
  \int_{c/(2d-c)}^1 f_q(q)dq=1\,,
  \label{eq:fscp2gennorm}
\end{equation}
with $c/(2d-c)$ being the minimum possible value for $q$.

When the mass range is restricted there are again a number of cases to consider,
depending on the values of $x_1$ and $x_2$. However the situation is less
complicated than for SCP-I. First of all the value of $c(1+1/q)$ is always
larger than $2c$ for $0<q\leq 1$ which means that the value of $q_1$ plays no
role. Secondly the minimum possible value of $q$ is $c/x_2$ and this quantity is
always larger than $q_0$ for $x_2\geq c$ (which is mandatory) and therefore also
the value of $q_0$ plays no role. For $q<c/x_2$ $f_q(q)$ is always zero. The
values of $q_2$ and $q_3$ do matter as discussed below.

\textbf{Case 1:} $x_1<x_2\leq d$\quad In this case the integration limits for $x$,
$x_1(1+q)$ and $x_2(1+q)$ are guaranteed to be less than $2d$. For $q<c/x_1$ the
value of $x_1(1+q)$ is less than $c(1+1/q)$ so the lower limit for integral over
$x$ is then fixed at $c(1+1/q)$. The expression for $f_q(q)$ becomes:
\begin{equation}
  f_q(q)=\left\{
  \begin{array}{ll}
    0 & 0< q<\frac{c}{x_2} \\[5pt]
    k\int_{c(1+1/q)}^{x_2(1+q)} h_q(q)f(x) dx & \frac{c}{x_2}\leq q<\frac{c}{x_1} \\[5pt]
    k\int_{x_1(1+q)}^{x_2(1+q)} h_q(q)f(x) dx & \frac{c}{x_1}\leq q\leq 1
  \end{array}\right.\,,
  \label{eq:fqscp2case1}
\end{equation}
where $k$ is obtained from the condition:
\begin{equation}
  \int_{c/x_2}^1 f_q(q)dq=1\,.
  \label{eq:fscp2case1norm}
\end{equation}

\textbf{Case 2:} $x_1\leq d \,\wedge\, x_2>d$\quad Now the value of $q_2$ is less
than 1 and for $q>q_2$ the upper limit of the integral over $x$ is fixed at $2d$
(the value of $x_2(1+q)$ being larger than $2d$). Now, $q_2\geq c/x_2$ but $q_2$
may be larger or smaller than $c/x_1$. So there are two sub-cases:

\textbf{Case 2a} $q_2\leq c/x_1$
\begin{equation}
  f_q(q)=\left\{
  \begin{array}{ll}
    0 & 0< q<\frac{c}{x_2} \\[5pt]
    k\int_{c(1+1/q)}^{x_2(1+q)} h_q(q)f(x) dx & \frac{c}{x_2}\leq q <q_2 \\[5pt]
    k\int_{c(1+1/q)}^{2d} h_q(q)f(x) dx & q_2\leq q< \frac{c}{x_1} \\[5pt]
    k\int_{x_1(1+q)}^{2d} h_q(q)f(x) dx & \frac{c}{x_1}\leq q\leq 1
  \end{array}\right.\,,
  \label{eq:fqscp2case2a}
\end{equation}
where $k$ is obtained by integrating $f_q$ over $[c/x_2,q_2]$,
$[q_2,c/x_1]$, and $[c/x_1,1]$, and applying condition
(\ref{eq:fscp2case1norm}).

\textbf{Case 2b} $q_2>\frac{c}{x_1}$
\begin{equation}
  f_q(q) =\left\{
  \begin{array}{ll}
    0 & 0< q<\frac{c}{x_2} \\[5pt]
    k\int_{c(1+1/q)}^{x_2(1+q)} h_q(q)f(x) dx & \frac{c}{x_2}\leq q <\frac{c}{x_1} \\[5pt]
    k\int_{x_1(1+q)}^{x_2(1+q)} h_q(q)f(x) dx & \frac{c}{x_1}\leq q< q_2 \\[5pt]
    k\int_{x_1(1+q)}^{2d} h_q(q)f(x) dx & q_2\leq q\leq 1
  \end{array}\right.\,,
  \label{eq:fqscp2case2b}
\end{equation}
where $k$ is obtained by integrating $f_q$ over  $[c/x_2,c/x_1]$,
$[c/x_1,q_2]$, and $[q_2,1]$, and applying condition (\ref{eq:fscp2case1norm}).

\textbf{Case 3:} $x_1>d$\quad This is the same as case 2 except that $q_3<1$ which
means that $f_q(q)=0$ for $q>q_3$ (this is due to the value of $x_1(1+q)$
becoming larger than $2d$). There are the same two sub-cases:

\textbf{Case 3a:} $q_2\leq c/x_1$
\begin{equation}
  f_q(q) =\left\{
  \begin{array}{ll}
    0 & 0< q<\frac{c}{x_2} \\[5pt]
    k\int_{c(1+1/q)}^{x_2(1+q)} h_q(q)f(x) dx & \frac{c}{x_2}\leq q <q_2 \\[5pt]
    k\int_{c(1+1/q)}^{2d} h_q(q)f(x) dx & q_2\leq q< \frac{c}{x_1} \\[5pt]
    k\int_{x_1(1+q)}^{2d} h_q(q)f(x) dx & \frac{c}{x_1}\leq q<q_3 \\[5pt]
    0 & q_3\leq q\leq 1
  \end{array}\right.\,,
  \label{eq:fqscp2case3a}
\end{equation}
where $k$ is obtained by integrating $f_q$ over $[c/x_2,q_2]$,
$[q_2,c/x_1]$, and $[c/x_1,q_3]$, and applying condition
(\ref{eq:fscp2case1norm}).

\textbf{Case 3b} $q_2>c/x_1$
\begin{equation}
  f_q(q) =\left\{
  \begin{array}{ll}
    0 & 0< q<\frac{c}{x_2} \\[5pt]
    k\int_{c(1+1/q)}^{x_2(1+q)} h_q(q)f(x) dx & \frac{c}{x_2}\leq q <\frac{c}{x_1} \\[5pt]
    k\int_{x_1(1+q)}^{x_2(1+q)} h_q(q)f(x) dx & \frac{c}{x_1}\leq q< q_2 \\[5pt]
    k\int_{x_1(1+q)}^{2d} h_q(q)f(x) dx & q_2\leq q<q_3 \\[5pt]
    0 & q_3\leq q\leq 1
  \end{array}\right.\,,
  \label{eq:fqscp2case3b}
\end{equation}
where $k$ is obtained by integrating $f_q$ over $[c/x_2,c/x_1]$,
$[c/x_1,q_2]$, and $[q_2,q_3]$, and applying condition (\ref{eq:fscp2case1norm}).

\paragraph{SCP-III}

Now the generating mass ratio distribution $h_q$ is restricted to the range
$c/(M_\mathrm{c}-c)\leq q\leq 1$ and re-normalised. The corresponding
distribution is ${h'}_q(q)$ and the expression for $f_q$ becomes:
\begin{equation}
  f_q(q)=\int_{c(1+1/q)}^{2d} {h'}_q(q)f(x)dx\,,
  \label{eq:fscp3exprgen}
\end{equation}
where the lower integration limit is set by the condition $M_2\geq c$ and the
distribution is normalised.

When the primary mass range is restricted to $x_1\leq x\leq x_2$ the cases and
expressions for $f_q(q)$ are the same as for the SCP-II case, except that
${h'}_q(q)$ replaces $h_q(q)$ everywhere.

\begin{figure*}[t!]
  \includegraphics[width=0.9\textwidth]{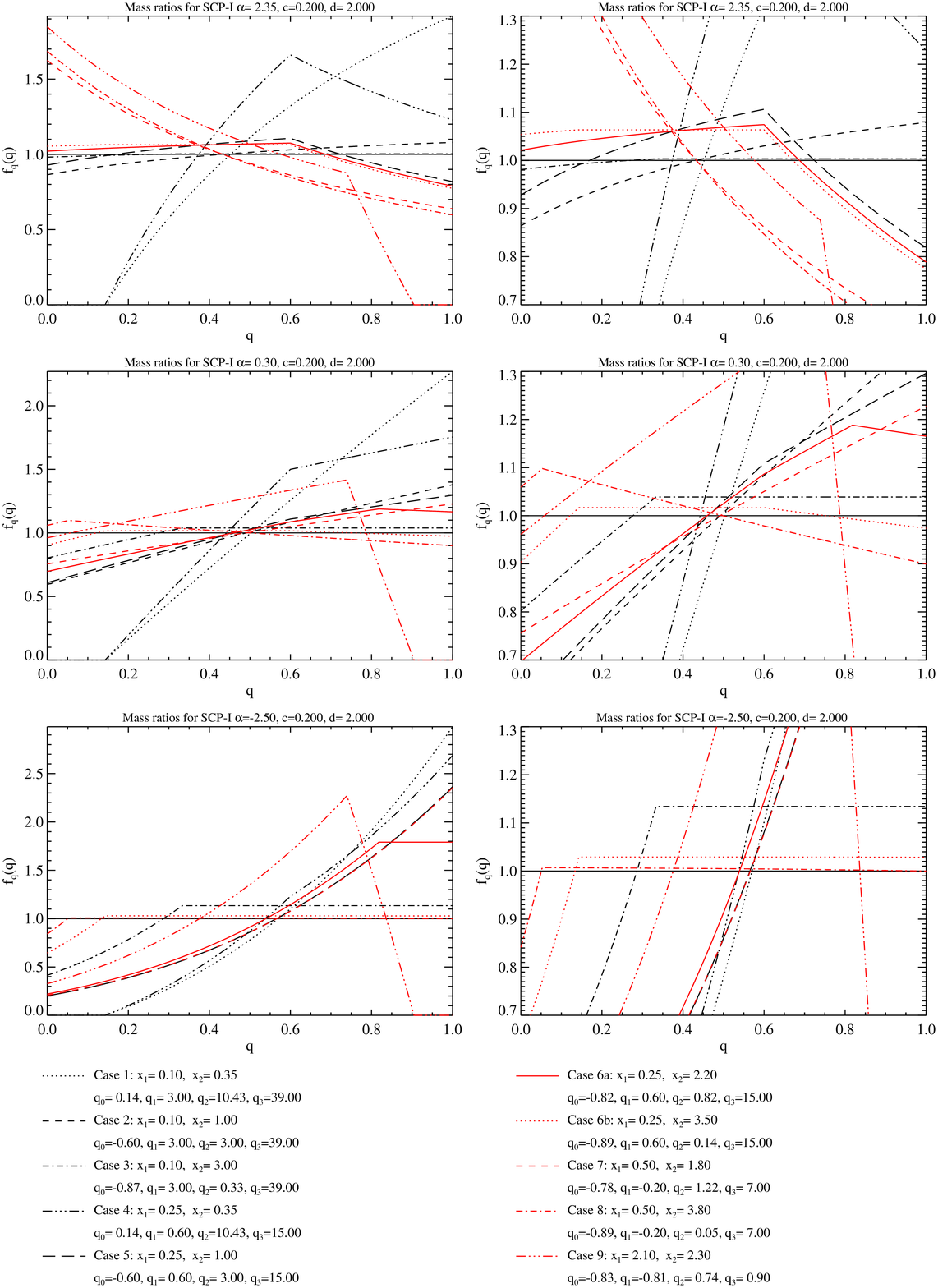}
  \caption{{\em Left:} Mass ratio distributions $f_q(q)$ for split-core pairing I for a
  uniform generating mass ratio distribution and core masses from single
  power-law mass distribution, where $2c\leq M_\mathrm{c}\leq 2d$. The values of $\alpha$
  are $2.35$, $0.30$, and $-2.50$. The black solid curve shows the complete mass
  ratio distribution (for all binaries in the population) for $c=0.2$ and
  $d=2.0$. The other curves show what happens to the observed $f_q(q)$ if the
  primary mass $M_1$ is restricted to $x_1\leq M_1\leq x_2$. The ten cases for
  SCP-I from section~\ref{sec:scp} are listed in the legend. 
  {\em Right:} Same, but with the vertical scale changed to bring out some 
  of the details in the cases close to the line $f_q(q)=1$.
  \label{fig:fqselscp1_paper}}
\end{figure*}

\begin{figure}[tbp]
  \includegraphics[width=0.5\textwidth,height=!]{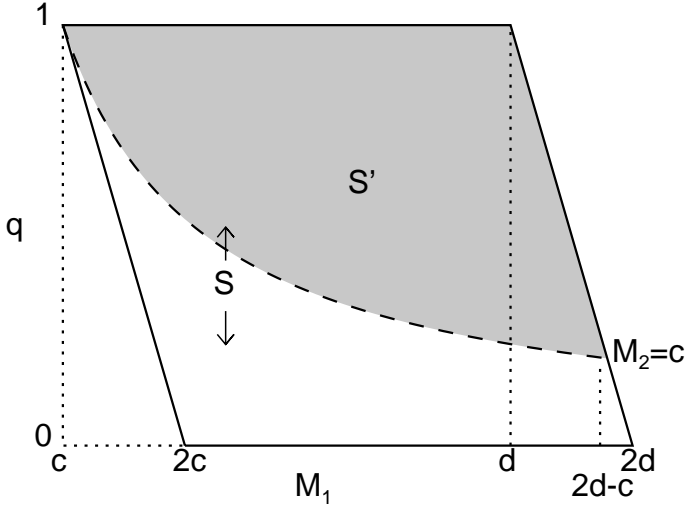} 
  \caption{The integration domains for obtaining the primary mass distribution and specific binary fraction for pairing functions SCP-I, SCP-II and SCP-III. The domain $S$ for SCP-I is the parallelogram enclosed by the solid lines. The domain $S'$ for SCP-II and SCP-III is indicated with the shaded region.
    \label{fig:fqseldomainscp} }
\end{figure}

\subsubsection{Binary fractions for SCP} \label{appendix:scp_bf}

In order to find the specific binary fraction as a function of primary mass, we first need to find the primary mass distribution $f_{M_1}(M_1)$, which can be calculated as:
\begin{equation}
  f_{M_1}(M_1) = \int_{\qmin(M_1)}^{\qmax(M_1)} f_{\rm M_1,q}(M_1,q) \, dq \,,
\end{equation}
where $\qmin(M_1)$ and $\qmax(M_1)$ are the minimum and maximum mass ratios for a given primary mass $M_1$, and $f_{\rm M_1,q}(M_1,q)$ is the joint probability density function for $M_1$ and $q$. The latter can be derived from the generating (core) mass distribution $f_{M_C}(M_C)$ and the generating mass distribution $h_q(q)$:
\begin{equation} \label{eq:joint_probablity_scp}
  f_{M_1,q}(M_1,q) = f_{M_C}(M_C(M_1,q)) \hq (1+q) \,,
\end{equation}
where the factor $(1+q)$ is the Jacobian of the transformation $M_1 = M_C(q+1)^{-1}$ and $q=q$.

\paragraph{SCP-I}

The number of binary systems with a mass $M_1$ is given by $N \binf f_{M_1}(M_1) dM_1$, where $N$ is the total number of systems (singles plus binaries) in the population.  The number of single stars with a mass $M_1$ is given by $N (1-\binf) f_{M_C}(M_1) dM_1$.  The specific binary fraction is thus:
\begin{equation}
  \binf_{M_1}(M_1) = 
   \frac{B_{M_1}}{S_{M_1}+B_{M_1}} =
  \frac{    \binf f_{M_1}(M_1)  }{    \binf f_{M_1}(M_1) + (1-\binf) f_{M_C}(M_1)
  }
\end{equation}
The integration limits for $f_{M_1}(M_1)$ are indicated with domain $S$ in Fig.~\ref{fig:fqseldomainscp}, and are given by:
\begin{equation} \label{eq:qmin_for_scp1}
  \qmin =
  \begin{cases}
    2-M_1/c & c\leq M_1 < 2c \\[3pt]
    0       & 2c \leq M_1 \leq 2d
  \end{cases}\,,
\end{equation}
and
\begin{equation} \label{eq:qmax_for_scp1}
  \qmax =
  \begin{cases}
    1           & c\leq M_1 <d \\[3pt]
    2-M_1/d     & d \leq M_1 \leq 2d
  \end{cases}\,.
\end{equation}
An example for the specific binary fraction resulting from SCP-I is shown in Fig.~\ref{figure:binaryfraction_versus_spectraltype}. The specific binary fraction for $c \leq M_1 \leq 2c$ equals unity, as $f_{M_C}(M_1) = 0$ in this mass range. In the mass range $2c<M_1\leq d$ the specific binary fraction is practically independent of $M_1$. Beyond $M_1=d$, the specific binary fraction rapidly drops to zero at $M_1=2d$. The overall binary fraction $\binfall$ is always equal to $\binf$ for SCP-I.

\paragraph{SCP-II} The integration domain for SCP-II to obtain $f_{M_1}(M_1)$ is indicated with region $S'$ in Fig.~\ref{fig:fqseldomainscp}, i.e., for $c \leq M_1 \leq 2d-c$. The integration limits $\qmin$ and $\qmax$ are given by:
\begin{equation} \label{eq:qmin_for_scp23}
  \qmin =  c/M_1 \quad  c\leq M_1 < 2d-c   \,,
\end{equation}
and
\begin{equation}\label{eq:qmax_for_scp23}
  \qmax =
  \begin{cases}
    1           & c\leq M_1 <d \\[3pt]
    2-M_1/d     & d \leq M_1 \leq 2d-c
  \end{cases}\,.
\end{equation}
The specific binary fraction for a population with a generating binary fraction of unity, i.e., $\binf = 100\%$), is given by:
\begin{equation}
  \binf_{100}(M_1) =
  \frac{
    \int_{q_{\rm min,2}(M_1)}^{q_{\rm max,2}(M_1)} f_{M_1,q}(M_1,q) \, dq 
  }{
    \int_{q_{\rm min,1}(M_1)}^{q_{\rm max,1}(M_1)} f_{M_1,q}(M_1,q) \, dq 
  }
  \,,
\end{equation}
where the joint probability distribution is given by Eq.~(\ref{eq:joint_probablity_scp}). For a given primary mass $M_1$, the value of $\binf_{100}(M_1)$ given by the ratio between the (weighted) lengths of the horizontal line segments of domain $S'$ and $S$ in Fig.~\ref{fig:fqseldomainscp}. The integration limits $q_{\rm min,2}$ and $q_{\rm max,2}$ are thus given by Eqs.~(\ref{eq:qmin_for_scp23}) and~(\ref{eq:qmax_for_scp23}), and the limits $q_{\rm min,1}$ and $q_{\rm max,1}$ are given by Eqs.~(\ref{eq:qmin_for_scp1}) and~(\ref{eq:qmax_for_scp1}), respectively. For systems with an arbitrary value of $\binf$, the number of binary systems is given by 
\begin{equation}
B_{M_1} = N\binf \binf_{100}(M_1) f_{M_1}(M_1) dM_1 \,,
\end{equation}
and the number of single stars by 
\begin{equation}
S_{M_1} = N(1-\binf) f_{M_C}(M_1) dM_1 + N\binf (1-\binf_{100}(M_1)) f_{M_1}(M_1) dM_1 \,,
\end{equation}
where the first term refers to the fraction of clumps that were assigned to be single, and the second term refers to the primaries that have become single due to the rejection of their low-mass companions. The specific binary fraction is then given by:
\begin{equation}
  \binfm = 
  \frac{B_{M_1}}{S_{M_1}+B_{M_1}}
  =
  \frac{
    \binf_{100}(M_1)\binf f_{M_1}(M_1)
  }{
    \binf f_{M_1}(M_1) + (1-\binf) f_{M_C}(M_1)
  }
  \,,
\end{equation}
An example of the specific binary fraction resulting from SCP-II is shown in Fig.~\ref{figure:binaryfraction_versus_spectraltype}. For $c\leq M_1 \leq 2c$, the above equation reduces to $\binf(M_1) =\binf_{100}(M_1)$ as $f_{M_C}(M_1)=0$. The binary fraction among the lowest mass stars ($M_1=c$) equals unity, after which it drops until $M_1=2c$. Beyond that minimum mass it rises again and reaches its maximum at $M_1=d$, after which it drops again go zero at $M_1=2d$, for the same reasons as for the SCP-I case.
The overall binary fraction can be calculated by integrating $\binf(M_1)$ over primary mass, weighed by the primary mass distribution $f_{M_1}(M_1)$, see Eq.~(\ref{eq:binfall_for_pcp_and_scp}).

\paragraph{SCP-III}

The values for $\qmin$ and $\qmax$ for SCP-III are identical to those for SCP-II, and are given by Eqs.~(\ref{eq:qmin_for_scp23}) and~(\ref{eq:qmax_for_scp23}). The expression for the specific binary fraction for SCP-III is the same as in (the one for SCP-I). Note, however, that the primary mass distribution $f_{M_1}(M_1)$ is different due to the different limits $\qmin$ and $\qmax$. The right-hand panel in Fig.~\ref{figure:binaryfraction_versus_spectraltype} shows an example of $\binfm$ resulting from SCP-III. The term $f_{M_C}(M_1)$ vanishes for $c\leq M_1 \leq 2c$, so that $\binfm = 1$ in this mass range. A discontinuity appears at $M_1=2c$, beyond which $\binfm$ reaches its lowest point, after which it steadily rises until $M_1=d$. Beyond $M_1=d$, the specific binary fraction decreases again to $\binfm=0$ at $M_1=2d$, for the same reasons as for SCP-I and SCP-II. The overall binary fraction $\binfall$ is always equal to $\binf$ for SCP-III.

\subsubsection{Uniform mass ratio distribution and single power-law mass distribution} \label{appendix:scp_powerlaw}

In the expressions given in this section the term $1-(c/d)^\gamma$ has often
been absorbed in the normalisation constants (which are therefore not strictly
consistent with the expressions above).
The following is assumed for $h_q$ and $f_\mathrm{c}(M_\mathrm{c})=f(x)$ (using
$x=M_\mathrm{c}$ for ease of notation):
\begin{eqnarray*}
  h_q(q)=1 \quad\quad 0<q \leq 1 \\
  f(x)=ax^{-\alpha} \quad\quad 2c\leq x\leq 2d
  \label{eq:scpexdef}\,,
\end{eqnarray*}
where $a=\gamma (2c)^\gamma/(1-(c/d)^\gamma)$ and $\gamma=\alpha-1$, and
$\alpha\neq1$.

\paragraph{The SCP-I case} Without restricting the primary mass range the
expression for the mass ratio distribution is simply $f_q(q)=1$. When the
primary mass range is restricted the cases listed above have to be worked out.
The corresponding expressions for $\fq$ are listed in 
Table~\ref{table:bigequations_scpi_powerlaw}.
Fig.~\ref{fig:fqselscp1_paper} shows an example of what
$f_q(q)$ looks like for all the SCP-I cases discussed above.

\paragraph{The SCP-II case} Without restrictions on the primary mass range the
expression for the mass ratio distribution follows from
Eq.~(\ref{eq:fscp2exprgen}) and is:
\begin{equation}
  f_q(q)= \frac{k}{1-(c/d)^\gamma}\left[
  \left(\frac{2q}{1+q}\right)^\gamma - \left(\frac{c}{d}\right)^\gamma\right]\,,
  \label{eq:fqscp2ex1all}
\end{equation}
where the normalisation constant follows from:
\begin{eqnarray}
  1&=& \int_{q_\mathrm{min}}^1 f_q(q)dq \nonumber\\[5pt]
   &=& \frac{k}{1-(c/d)^\gamma}\left( 
  \int_{q_\mathrm{min}}^1 \left(\frac{2q}{1+q}\right)^\gamma dq
  -\left(\frac{c}{d}\right)^\gamma(1-q_\mathrm{min}) \right)\,,
  \label{eq:fqscp2ex1allnorm}
\end{eqnarray}
where $q_\mathrm{min}=(c/d)/(2-c/d)$. The integral in this expression will be
dealt with below.
With a restricted primary mass range the expressions for $f_q$ are
obtained for the three cases discussed above. These expressions are listed in 
Table~\ref{table:bigequations_scpii_powerlaw}. Again the term $1-(c/d)^\gamma$
is absorbed in the normalisation constants $k$. 
All the normalisation constants for the SCP-II case contain the following
integral:
\begin{equation}
  \int\left(\frac{2q}{1+q}\right)^\gamma dq
\end{equation}
For $\gamma\neq 0,-1,-2,-3,\cdots$ the integral evaluates to an expression
involving a hypergeometric function:
\begin{equation}
  \frac{2^\gamma q^{1+\gamma} {_2F_1}(\gamma+1, \gamma; \gamma+2;
  -q)}{\gamma+1}\,.
\end{equation}
where the result was obtained by using the website
\texttt{integrals.wolfram.com}. The value $\gamma=0$ is not allowed as
$\alpha=1$ was excluded. For $\gamma=-1$ ($\alpha=0$) the integral is:
\begin{equation}
  \tfrac{1}{2}(q+\ln q)\,.
\end{equation}
From the properties of the hypergeometric function (see PCP-III case) it follows
that for $\gamma+2=-n$ the expression above does not converge. This means that
for $\gamma=-p$, where $p=2,3,4,\cdots$ the integral has to be evaluated
separately. In this case we can write:
\begin{eqnarray*}
  \int\left(\frac{2q}{1+q}\right)^\gamma dq & = 2^{-p}\int
  \left(1+\frac{1}{q}\right)^p dq \\[5pt]
  & = 2^{-p}\int \sum_{r=0}^p \left(\begin{array}{c}p\\ r\end{array}\right) q^{-r} dq\,,
\end{eqnarray*}
which evaluates to:
\begin{equation}
  2^{-p}\left(1+p\ln q + \sum_{r=2}^p \left(\begin{array}{c}p\\ r\end{array}\right)
    \frac{1}{1-r}q^{1-r} \right)\,.
\end{equation}

\paragraph{The SCP-III case} Without restrictions on the primary mass range the
expression for the mass ratio distribution is:
\begin{equation}
  f_q(q)=\int_{c(1+1/q)}^{2d} \left(\frac{x-c}{x-2c}\right)ax^{-\alpha}dx\,,
  \label{eq:fqscp3ex1all}
\end{equation}
where the term for ${h'}_q$ follows from ${h'}_q(q)=1/(1-c/(x-c))$. Note that
the distribution is normalised. To bring out better the dependence on $c/d$ one
can also write (substituting $z=x/d$):
\begin{equation}
  f_q(q)=\frac{\gamma(2c/d)^\gamma}{1-(c/d)^\gamma} \int_{(c/d)(1+1/q)}^2
  \left( \frac{z-(c/d)}{z-(2c/d)}\right)z^{-\alpha} dz
  \label{eq:fqscp3ex1allalt}
\end{equation}
The integral over $z$:
\begin{equation}
  \int\left( \frac{z-(c/d)}{z-(2c/d)}\right)z^{-\alpha} dz\nonumber
\end{equation}
can again be evaluated using the hypergeometric function
for $\alpha\neq0,1,2,3,\cdots$, and the result is:
\begin{equation}
  \frac{z^{1-\alpha}\left(_2F_1(1-\alpha,1;2-\alpha;\frac{z}{2c/d})-2\right)}{2(\alpha-1)}\,.
  \label{eq:fqscp3ex1allalt:hyper}
\end{equation}
The excluded values of $\alpha$ again follow from the convergence properties of
the hypergeometric function. For $\alpha=0$ the integral evaluates to:
\begin{equation}
  z-(c/d)\ln(z-2c/d)\,.
  \label{eq:fqscp3ex1allalt:alphazero}
\end{equation}
The case $\alpha=1$ is excluded and for $\alpha=p=2,3,4,\cdots$ the integral can
be written as:
\begin{equation}
  \int \left(\frac{1}{z^{p-1}(z-2c/d)} - \frac{c/d}{z^p(z-2c/d}\right) dz
\end{equation}
The solution can be found again from formula 2.117(4) in \cite{TISP7}:
\begin{equation}
  \begin{array}{l}
    \sum_{r=1}^{p-2}\frac{1}{(p-1-r)(2c/d)^r z^{p-1-r}} + 
    \left(\frac{2c}{d}\right)^{1-p}\ln\left(\frac{z-2c/d}{z}\right) - \\[5pt]
    \left(\frac{c}{d}\right) \left( \sum_{r=1}^{p-1}\frac{1}{(p-r)(2c/d)^r z^{p-r}} +
    \left(\frac{2c}{d}\right)^{-p}\ln\left(\frac{z-2c/d}{z}\right) \right)\,.
  \end{array}
  \label{eq:fqscp3ex1allalt:alphaint}
\end{equation}

\begin{figure*}[t!]
  \begin{center}
    \includegraphics[width=0.8\textwidth]{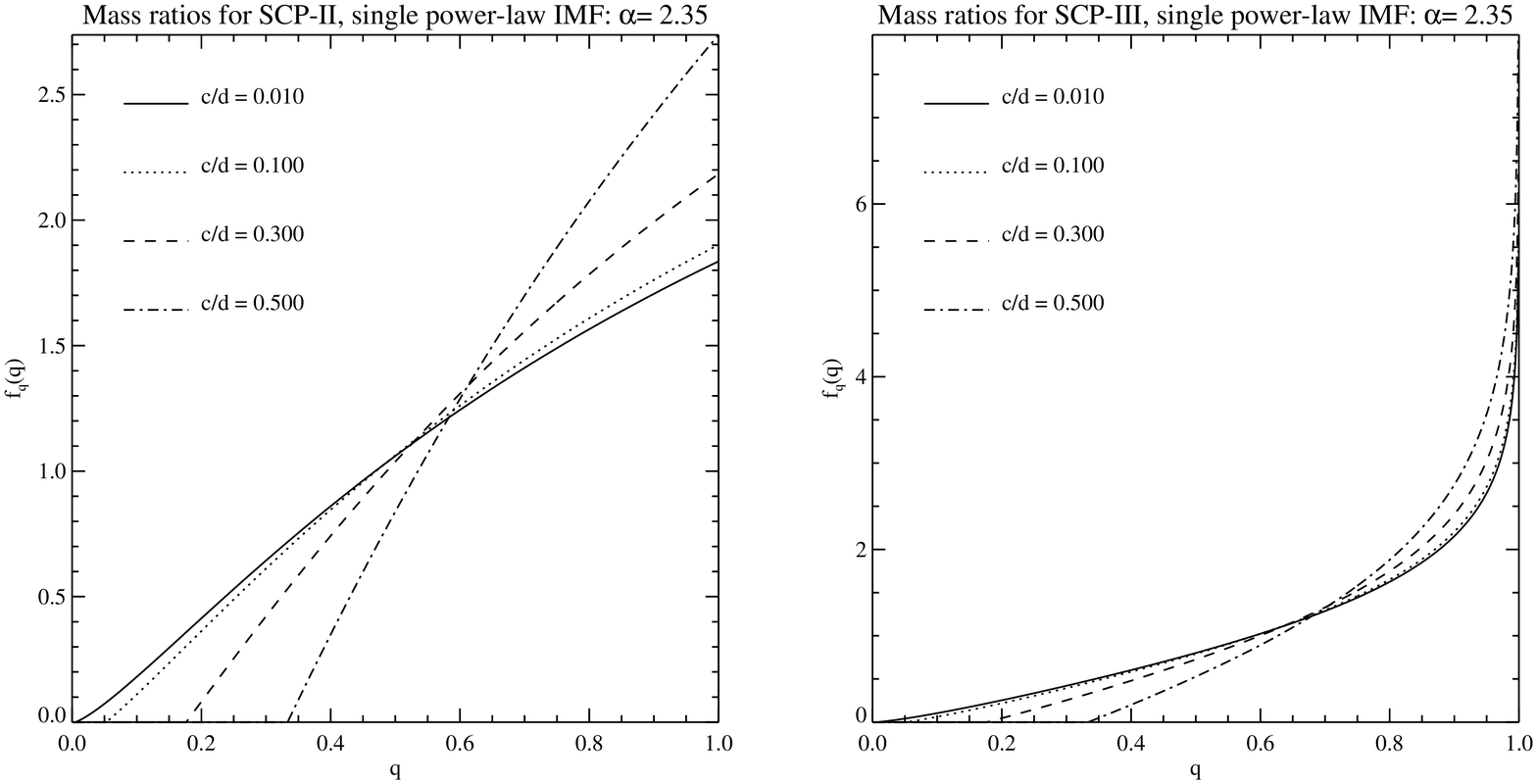}
    \includegraphics[width=0.8\textwidth]{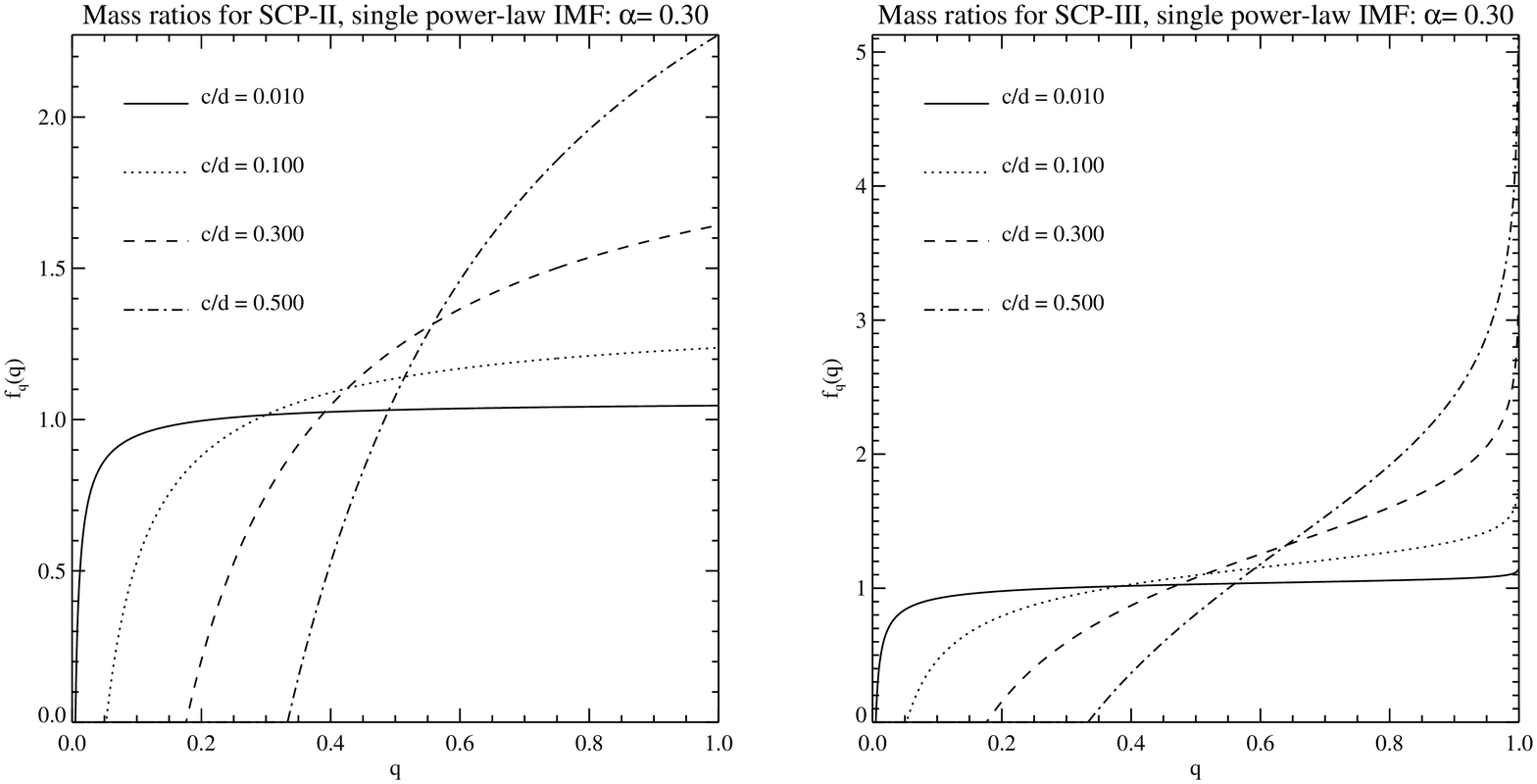}
    \includegraphics[width=0.8\textwidth]{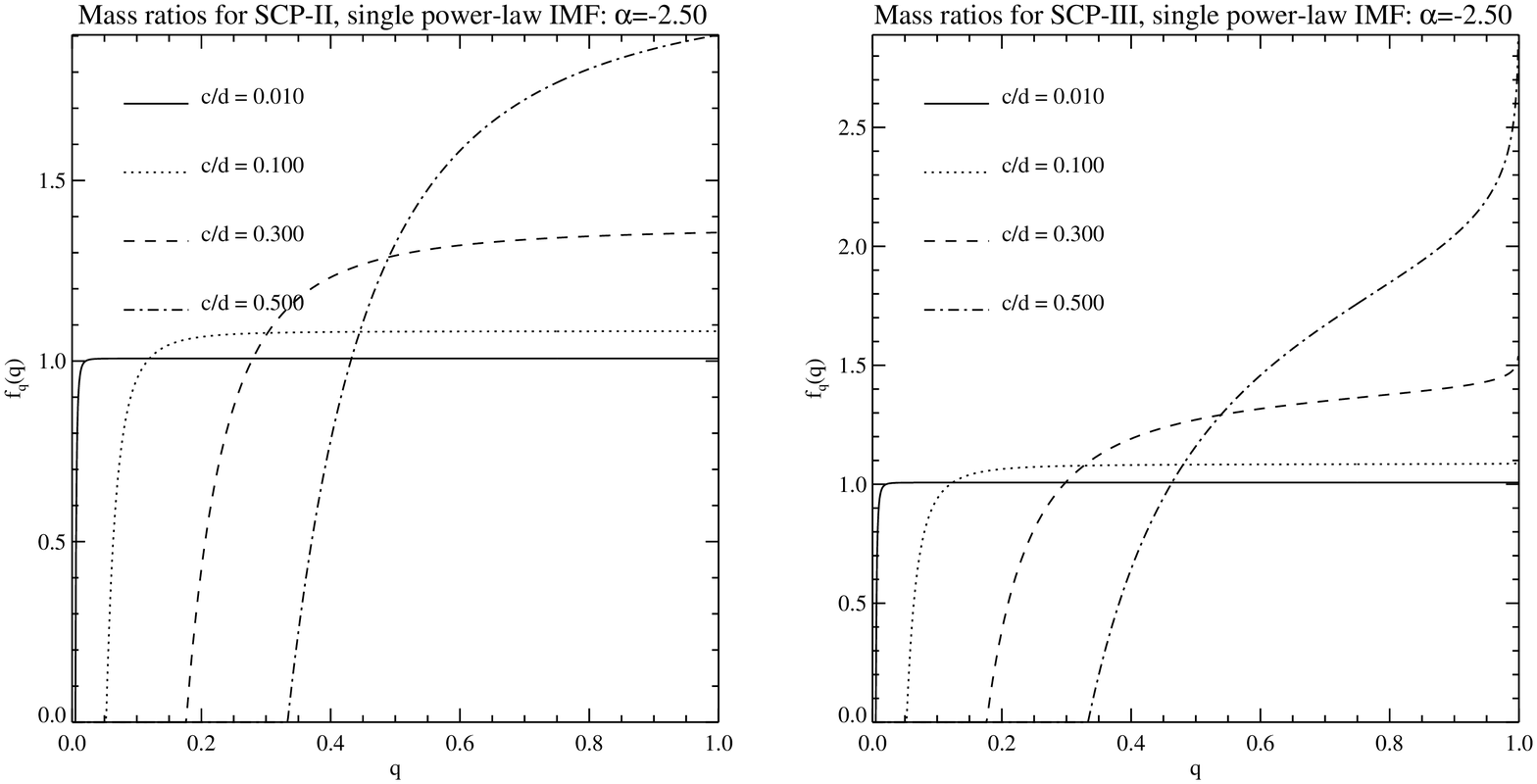}
  \end{center}
  \caption{Mass ratio distributions $f_q(q)$ for split-core pairing II and III
  for a uniform generating mass ratio distribution and core masses from single
  power-law mass distribution for $\alpha=2.35,\,0.30,\,-2.50$, where $2c\leq M_\mathrm{c}\leq
  2d$. The curves are shown for
  $c/d=0.01,\,0.1,\,0.3,\,0.5$.\label{fig:fqscp23ex1}}
\end{figure*}

\begin{figure*}[t!]
  \begin{center}
    \includegraphics[width=0.9\textwidth]{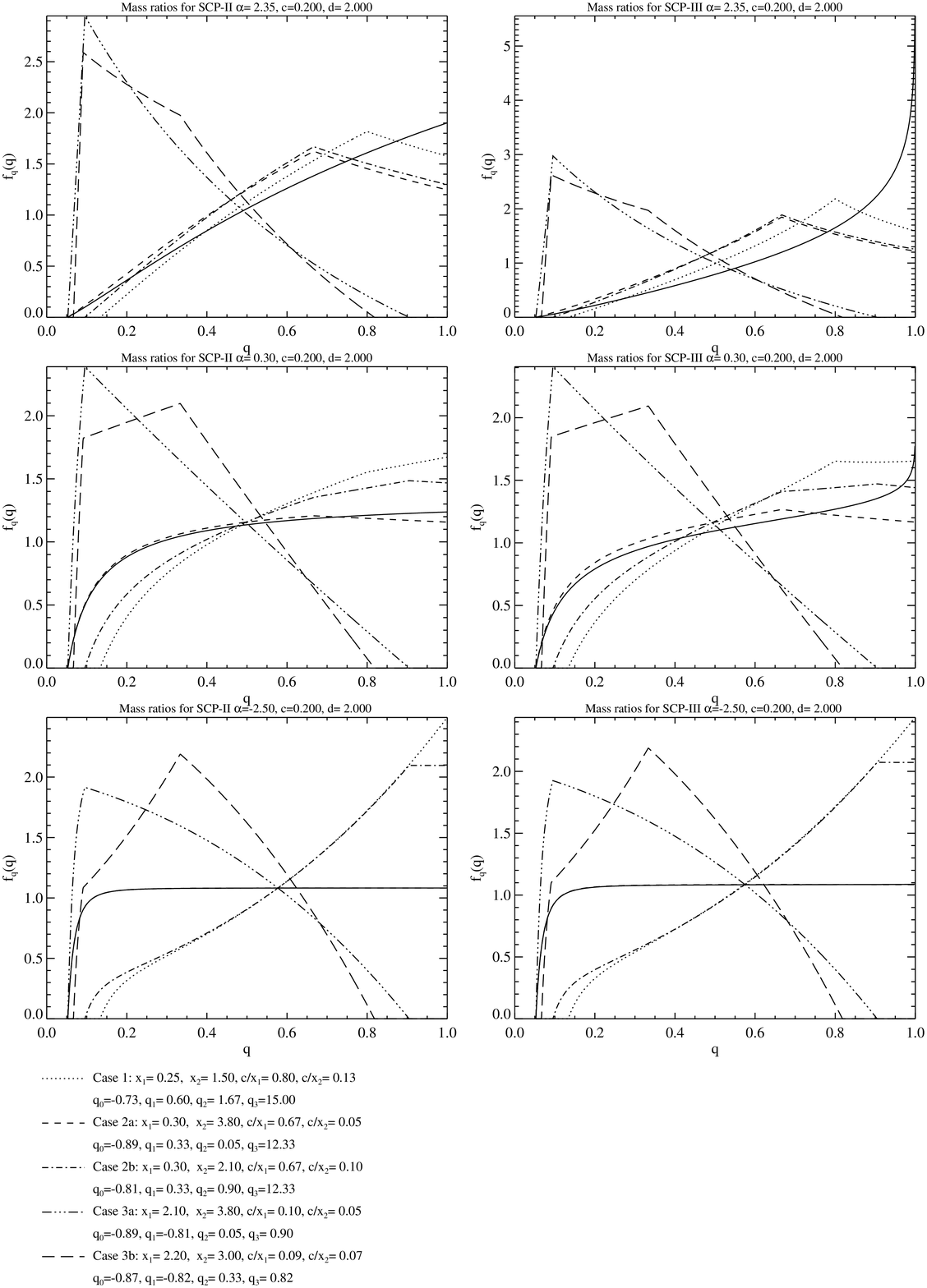}
  \end{center}
  \caption{Mass ratio distributions $f_q(q)$ for split-core pairing
  II and III for a uniform generating mass ratio distribution and core masses
  from a single power-law mass distribution with $\alpha=2.35$, where 
  $2c\leq M_\mathrm{c}\leq 2d$. 
  The black solid curve shows the complete mass ratio distribution (for all
  binaries in the population) for $c=0.2$ and $d=2.0$. The other curves show
  what happens to the observed $f_q(q)$ if the primary mass $M_1$ is restricted
  to $x_1\leq M_1\leq x_2$. Results are shown for $\alpha=2.35$ ({\em top}), $\alpha=0.3$
  ({\em middle}) and $\alpha=-2.5$ ({\em bottom}). The examples cases are listed in the
  legend. \label{fig:fqselscp23paper}}
\end{figure*}

With restrictions on the primary mass range the integral discussed above has to
be evaluated for the various integration intervals corresponding to the cases
discussed for SCP-II. The normalisation constants $k$ can be obtained from a
numerical integration of the functions $f_q(q)$. For $x_1=x_2$ the mass ratio
distribution becomes $f_q(q)=1/(1-c/(x_1-c))=(x_1-c)/(x_1-2c)$.
Figs.~\ref{fig:fqscp23ex1} and~\ref{fig:fqselscp23paper} show examples of what
$f_q(q)$ looks like for all the SCP-II and SCP-III cases discussed above.
Finally, we have listed the expressions for the binary fraction as a function of 
primary mass in Table~\ref{table:binfracm_powerlaw}.

\begin{table*}

\begin{tabularx}{\textwidth}{|p{0.01cm}XX|}

  \hline

  \mbox{SCP-I} & \multicolumn{1}{c}{Mass ratio distribution} & \multicolumn{1}{c|}{Normalization constant} \\

  \hline

  \mbox{Case 1} &

\begin{equation*}
  f_q(q)=\left\{
  \begin{array}{ll}
    0 & 0< q< q_0 \\[5pt]
    k \left[1-(2c/x_2)^\gamma(1+q)^{-\gamma}\right] & q_0\leq q\leq 1\\
  \end{array}\right.
  \label{eq:fqscp1ex1ca1}
\end{equation*}

&

\begin{equation*}
  k=\frac{1-\gamma}{(1-q_0)(1-\gamma)-2^{1-\gamma}(2c/x_2)^\gamma+2c/x_2}
  \label{eq:fscp1ex1ca1norm}
\end{equation*}

\\

  \hline

  \mbox{Case 2} &

\begin{equation*}
  f_q(q)= k\left[1-(2c/x_2)^\gamma(1+q)^{-\gamma}\right] \quad 0< q\leq 1
  \label{eq:fqscp1ex1ca2}
\end{equation*}

&

\begin{equation*}
  k=\frac{1-\gamma}{(1-\gamma)-(2c/x_2)^\gamma(2^{1-\gamma}-1)}
  \label{fqscp1ex1ca2norm}
\end{equation*}

\\

  \hline

  \mbox{Case 3} &

\begin{equation*}
  f_q(q)=\left\{
  \begin{array}{ll}
    k\frac{1-(2c/x_2)^\gamma(1+q)^{-\gamma}}{1-(c/d)^\gamma} & 0< q\leq q_2\\[5pt]
    k & q_2< q\leq 1
  \end{array}\right.
  \label{eq:fqscp1ex1ca3}
\end{equation*}

&

\begin{eqnarray*}
  k^{-1}=\frac{1}{1-(c/d)^\gamma}\left[q_2-\frac{(2c/x_2)^\gamma(
  (1+q_2)^{1-\gamma}-1)}{1-\gamma}\right]+1-q_2
  \label{fqscp1ex1ca3norm}
\end{eqnarray*}

\\

  \hline

  \mbox{Case 4} &

\begin{equation*}
  f_q(q)=\left\{
  \begin{array}{ll}
    0 & 0< q< q_0 \\[5pt]
    k\left[1-\left(\frac{2c}{x_2}\right)^\gamma(1+q)^{-\gamma}\right] & q_0\leq q< q_1\\[5pt]
    k\left[ \left(\frac{2c}{x_1}\right)^\gamma-\left(\frac{2c}{x_2}\right)^\gamma\right]
    (1+q)^{-\gamma} & q_1\leq q\leq 1\\
  \end{array}\right.
  \label{eq:fqscp1ex1ca4}
\end{equation*}

&

\begin{equation*}
  k=\frac{1-\gamma}{2^{1-\gamma}[(2c/x_1)^\gamma-(2c/x_2)^\gamma]-\gamma(2c/x_1-2c/x_2)}
  \label{eq:fscp1ex1ca4norm}
\end{equation*}

\\

  \hline

  \mbox{Case 5} &

\begin{equation*}
  f_q(q)=\left\{
  \begin{array}{ll}
    k\left[1-\left(\frac{2c}{x_2}\right)^\gamma(1+q)^{-\gamma}\right] & 0< q< q_1\\[5pt]
    k\left[ \left(\frac{2c}{x_1}\right)^\gamma-\left(\frac{2c}{x_2}\right)^\gamma\right]
    (1+q)^{-\gamma} & q_1\leq q\leq 1\\
  \end{array}\right.
  \label{eq:fqscp1ex1ca5}
\end{equation*}

&

\begin{eqnarray*}
  k= &
  \frac{1-\gamma}{q_1(1-\gamma)+[(2c/x_1)^\gamma-(2c/x_2)^\gamma]2^{1-\gamma}+(2c/x_2)^\gamma-2c/x_1}
  \label{eq:fscp1ex1ca5norm}
\end{eqnarray*}

\\

  \hline

  \mbox{Case 6a} &

\begin{equation*}
  f_q(q) =\left\{
  \begin{array}{ll}
    k\left[1-\left(\frac{2c}{x_2}\right)^\gamma(1+q)^{-\gamma}\right] & 0< q<q_1 \\[5pt]
    k\left[ \left(\frac{2c}{x_1}\right)^\gamma-\left(\frac{2c}{x_2}\right)^\gamma\right]
    (1+q)^{-\gamma} & q_1\leq q<q_2 \\[5pt]
    k\left[\left(\frac{2c}{x_1}\right)^\gamma(1+q)^{1-\gamma}-\left(\frac{c}{d}\right)^\gamma\right]
    & q_2\leq q\leq 1
  \end{array}\right.
  \label{eq:fqscp1ex1ca6a}
\end{equation*}

&

\begin{eqnarray*}
  k = &\frac{1-\gamma}{q_1(1-\gamma)-(2c/x_2)^\gamma\left[
  (1+q_1)^{1-\gamma}-1\right]} + \\[5pt]
  & \frac{1-\gamma}{\left[(2c/x_1)^\gamma-(2c/x_2)^\gamma\right]
  \left[(1+q_2)^{1-\gamma}-(1+q_1)^{1-\gamma}\right]} + \\[5pt]
  & \frac{1-\gamma}{(2c/x_1)^\gamma(2^{1-\gamma}-(1+q_2)^{1-\gamma}) -
  (c/d)^\gamma(1-q_2)(1-\gamma)}
  \label{eq:fscp1ex1ca6anorm}
\end{eqnarray*}

\\

  \hline

  \mbox{Case 6b} &

\begin{equation*}
  f_q(q) =\left\{
  \begin{array}{ll}
    k\left[1-\left(\frac{2c}{x_2}\right)^\gamma(1+q)^{-\gamma}\right] & 0< q<q_2 \\[5pt]
    k[1-\left(\frac{c}{d}\right)^\gamma] & q_2\leq q<q_1 \\[5pt]
    k\left[\left(\frac{2c}{x_1}\right)^\gamma(1+q)^{1-\gamma}-\left(\frac{c}{d}\right)^\gamma\right]
    & q_1\leq q\leq 1
  \end{array}\right.
  \label{eq:fqscp1ex1ca6b}
\end{equation*}

&

\begin{eqnarray*}
    k= & \frac{1-\gamma}{q_2(1-\gamma)-(2c/x_2)^\gamma\left[
    (1+q_2)^{1-\gamma}-1\right]} + \\[5pt]
    & \frac{1-\gamma}{(1-(c/d)^\gamma)(q_1-q_2)(1-\gamma)} +\\[5pt]
    & \frac{1-\gamma}{(2c/x_1)^\gamma(2^{1-\gamma}-(1+q_1)^{1-\gamma}) -
    (c/d)^\gamma(1-q_1)(1-\gamma)}
  \label{eq:fscp1ex1ca6bnorm}
\end{eqnarray*}

\\

  \hline

  \mbox{Case 7} &

\begin{equation*}
  f_q(q)=k(1+q)^{-\gamma}
  \label{eq:fqscp1ex1ca7}
\end{equation*}

&

\begin{equation*}
  k=\frac{1-\gamma}{2^{1-\gamma}-1} \quad \quad  \quad \quad \quad\quad\quad \quad  \quad \quad \quad\quad
\end{equation*}

\\

  \hline

  \mbox{Case 8} &

\begin{equation*}
  f_q(q) =\left\{
  \begin{array}{ll}
    k\left[ \left(\frac{2c}{x_1}\right)^\gamma-\left(\frac{2c}{x_2}\right)^\gamma\right]
    (1+q)^{-\gamma} & 0<q<q_2 \\[5pt]
    k\left[\left(\frac{2c}{x_1}\right)^\gamma(1+q)^{1-\gamma}-\left(\frac{c}{d}\right)^\gamma\right]
    & q_2\leq q\leq 1
  \end{array}\right.
  \label{eq:fqscp1ex1ca8}
\end{equation*}

&

\begin{eqnarray*}
    k= & \frac{1-\gamma}{\left[(2c/x_1)^\gamma-(2c/x_2)^\gamma\right]
    \left[(1+q_2)^{1-\gamma}-1\right]} + \\[5pt]
    & \frac{1-\gamma}{(2c/x_1)^\gamma(2^{1-\gamma}-(1+q_2)^{1-\gamma})} -
    \\[5pt]
    & \frac{1-\gamma}{(c/d)^\gamma(1-q_2)(1-\gamma)}
  \label{eq:fscp1ex1ca8norm}
\end{eqnarray*}

\\

  \hline

  \mbox{Case 9} &

\begin{equation*}
  f_q(q) =\left\{
  \begin{array}{ll}
    k\left[ \left(\frac{2c}{x_1}\right)^\gamma-\left(\frac{2c}{x_2}\right)^\gamma\right]
    (1+q)^{-\gamma} & 0<q<q_2 \\[5pt]
    k\left[\left(\frac{2c}{x_1}\right)^\gamma(1+q)^{1-\gamma}-\left(\frac{c}{d}\right)^\gamma\right]
    & q_2\leq q<q_3 \\[5pt]
    0 & q_3\leq q\leq 1
  \end{array}\right.
  \label{eq:fqscp1ex1ca9}
\end{equation*}

&

\begin{eqnarray*}
  k= & \frac{1-\gamma}{\left[(2c/x_1)^\gamma-(2c/x_2)^\gamma\right]
  \left[(1+q_2)^{1-\gamma}-1\right]} + \\[5pt]
  & \frac{1-\gamma}{(2c/x_1)^\gamma( (1+q_3)^{1-\gamma}-(1+q_2)^{1-\gamma})} -
  \\[5pt]
  & \frac{1-\gamma}{(c/d)^\gamma(q_3-q_2)(1-\gamma)}
  \label{eq:fscp1ex1ca9norm}
\end{eqnarray*}

\\

  \hline

\end{tabularx}

\caption{The specific mass ratio distribution $\fq$ for pairing function SCP-I, resulting from a power-law generating mass distribution $f_m(m)$ and a uniform generating mass ratio distribution $h_q(q)$. See Appendix~\ref{appendix:scp_powerlaw} for details. }

\label{table:bigequations_scpi_powerlaw}

\end{table*}

\begin{table*}

\begin{tabularx}{\textwidth}{|p{0.01cm}XX|}

  \hline

  \mbox{SCP-II} & \multicolumn{1}{c}{Mass ratio distribution} & \multicolumn{1}{c|}{Normalization constant} \\

  \hline

  \mbox{Case 1} &

\begin{equation*}
  f_q(q) =\left\{
  \begin{array}{ll}
    0 & 0< q<\frac{c}{x_2} \\[5pt]
    k\left[ \left(\frac{2q}{1+q}\right)^\gamma - \left(\frac{2c/x_2}{1+q}\right)^{\gamma}
    \right]& \frac{c}{x_2}\leq q<\frac{c}{x_1} \\[5pt]
    k\left[ \left(\frac{2c}{x_1}\right)^\gamma-
    \left(\frac{2c}{x_2}\right)^\gamma\right](1+q)^{-\gamma} & \frac{c}{x_1}\leq q\leq 1
  \end{array}\right.
  \label{eq:fqscp2ex1ca1}
\end{equation*}

&

\begin{eqnarray*}
  k^{-1} & = \int_{c/x_2}^{c/x_1}\left(\frac{2q}{1+q}\right)^\gamma dq -
  \frac{(2c/x_2)^\gamma}{1-\gamma}\left[
  \left(1+\frac{c}{x_1}\right)^{1-\gamma} -
  \left(1+\frac{c}{x_2}\right)^{1-\gamma}\right] +\\[5pt]
  & \frac{(2c/x_1)^\gamma-(2c/x_2)^\gamma}{1-\gamma}
  \left[2^{1-\gamma}-\left(1+\frac{c}{x_1}\right)^{1-\gamma}\right]
  \label{eq:fqscp2ex1ca1norm}
\end{eqnarray*}

\\

  \hline

  \mbox{Case 2a} &

\begin{equation*}
  f_q(q) =\left\{
  \begin{array}{ll}
    0 & 0< q<\frac{c}{x_2} \\[5pt]
    k\left[ \left(\frac{2q}{1+q}\right)^\gamma -
    \left(\frac{2c/x_2}{1+q}\right)^\gamma\right] & \frac{c}{x_2}\leq q <q_2 \\[5pt]
    k\left[ \left(\frac{2q}{1+q}\right)^\gamma - \left(\frac{c}{d}\right)^\gamma\right]
    & q_2\leq q< \frac{c}{x_1} \\[5pt]
    k\left[ \left(\frac{2c/x_1}{1+q}\right)^\gamma
    -\left(\frac{c}{d}\right)^\gamma\right] & \frac{c}{x_1}\leq q\leq 1
  \end{array}\right.\,.
  \label{eq:fqscp2ex1ca2a}
\end{equation*}

&

\begin{equation*}
  \begin{array}{ll}
    k^{-1} & = \int_{c/x_2}^{c/x_1}\left(\frac{2q}{1+q}\right)^\gamma dq -
    \\[5pt]
    & \frac{(2c/x_2)^\gamma}{1-\gamma}\left[
    (1+q_2)^{1-\gamma}-\left(1+\frac{c}{x_2}\right)^{1-\gamma}\right] +
    \\[5pt]
    & \frac{(2c/x_1)^\gamma}{1-\gamma}\left[2^{1-\gamma}-\left(1+\frac{c}{x_1}\right)^{1-\gamma}\right]
    - \left(\frac{c}{d}\right)^\gamma(1-q_2)
  \end{array}
  \label{eq:fqscp2ex1ca2anorm}
\end{equation*}

\\

  \hline

  \mbox{Case 2b} &

\begin{equation*}
  f_q(q) =\left\{
  \begin{array}{ll}
    0 & 0< q<\frac{c}{x_2} \\[5pt]
    k\left[ \left(\frac{2q}{1+q}\right)^\gamma -
    \left(\frac{2c/x_2}{1+q}\right)^\gamma\right]  & \frac{c}{x_2}\leq q <\frac{c}{x_1} \\[5pt]
    k\left[ \left(\frac{2c}{x_1}\right)^\gamma-\left(\frac{2c}{x_2}\right)^\gamma\right](1+q)^{-\gamma} & \frac{c}{x_1}\leq q< q_2 \\[5pt]
    k\left[ \left(\frac{2c/x_1}{1+q}\right)^\gamma -\left(\frac{c}{d}\right)^\gamma\right] & q_2\leq q\leq 1
  \end{array}\right.\,.
  \label{eq:fqscp2ex1ca2b}
\end{equation*}

&

\begin{equation*}
  \begin{array}{ll}
    k^{-1} & = \int_{c/x_2}^{c/x_1}\left(\frac{2q}{1+q}\right)^\gamma dq -\\[5pt]
    & \frac{(2c/x_2)^\gamma}{1-\gamma}\left[
    \left(1+\frac{c}{x_1}\right)^{1-\gamma}-\left(1+\frac{c}{x_2}\right)^{1-\gamma}\right] + \\[5pt]
    & \frac{(2c/x_1)^\gamma-(2c/x_2)^\gamma}{1-\gamma}
    \left[ (1+q_2)^{1-\gamma}-\left(1+\frac{c}{x_1}\right)^{1-\gamma}\right] + \\[5pt]
    & \frac{(2c/x_1)^\gamma}{1-\gamma}\left[2^{1-\gamma}-(1+q_2)^{1-\gamma}\right]
    - \left(\frac{c}{d}\right)^\gamma(1-q_2)
  \end{array}
  \label{eq:fqscp2ex1ca2bnorm}
\end{equation*}

\\

  \hline

  \mbox{Case 3a} &
  
\begin{equation*}
  f_q(q) =\left\{
  \begin{array}{ll}
    0 & 0< q<\frac{c}{x_2} \\[5pt]
    k\left[ \left(\frac{2q}{1+q}\right)^\gamma -
    \left(\frac{2c/x_2}{1+q}\right)^\gamma\right] & \frac{c}{x_2}\leq q <q_2 \\[5pt]
    k\left[ \left(\frac{2q}{1+q}\right)^\gamma - \left(\frac{c}{d}\right)^\gamma\right] & q_2\leq q< \frac{c}{x_1} \\[5pt]
    k\left[ \left(\frac{2c/x_1}{1+q}\right)^\gamma
    -\left(\frac{c}{d}\right)^\gamma\right] & \frac{c}{x_1}\leq q< q_3\\[5pt]
    0 & q_3\leq q\leq 1
  \end{array}\right.
  \label{eq:fqscp2ex1ca3a}
\end{equation*}

&

\begin{equation*}
  \begin{array}{ll}
    k^{-1} & = \int_{c/x_2}^{c/x_1}\left(\frac{2q}{1+q}\right)^\gamma dq -\\[5pt]
    & \frac{(2c/x_2)^\gamma}{1-\gamma}\left[
    (1+q_2)^{1-\gamma}-\left(1+\frac{c}{x_2}\right)^{1-\gamma}\right] + \\[5pt]
    & \frac{(2c/x_1)^\gamma}{1-\gamma}
    \left[(1+q_3)^{1-\gamma}-\left(1+\frac{c}{x_1}\right)^{1-\gamma}\right]
    - \left(\frac{c}{d}\right)^\gamma(q_3-q_2)
  \end{array}
  \label{eq:fqscp2ex1ca3anorm}
\end{equation*}

\\

  \hline

  \mbox{Case 3b} &

\begin{equation*}
  f_q(q) =\left\{
  \begin{array}{ll}
    0 & 0< q<\frac{c}{x_2} \\[5pt]
    k\left[ \left(\frac{2q}{1+q}\right)^\gamma -
    \left(\frac{2c/x_2}{1+q}\right)^\gamma\right]  & \frac{c}{x_2}\leq q <\frac{c}{x_1} \\[5pt]
    k\left[ \left(\frac{2c/x_1}{1+q}\right)^\gamma-\left(\frac{2c/x_2}{1+q}\right)^\gamma\right] &
    \frac{c}{x_1}\leq q< q_2 \\[5pt]
    k\left[ \left(\frac{2c/x_1}{1+q}\right)^\gamma -\left(\frac{c}{d}\right)^\gamma\right] & q_2\leq q<q_3 \\[5pt]
    0 & q_3\leq q\leq 1
  \end{array}\right.
  \label{eq:fqscp2ex1ca3b}
\end{equation*}

&

\begin{equation*}
  \begin{array}{ll}
    k^{-1} & = \int_{c/x_2}^{c/x_1}\left(\frac{2q}{1+q}\right)^\gamma dq -\\[5pt]
    & \frac{(2c/x_2)^\gamma}{1-\gamma}\left[
    \left(1+\frac{c}{x_1}\right)^{1-\gamma}-\left(1+\frac{c}{x_2}\right)^{1-\gamma}\right] + \\[5pt]
    & \frac{(2c/x_1)^\gamma-(2c/x_2)^\gamma}{1-\gamma}
    \left[ (1+q_2)^{1-\gamma}-\left(1+\frac{c}{x_1}\right)^{1-\gamma}\right] +\\[5pt]
    & \frac{(2c/x_1)^\gamma}{1-\gamma}\left[(1+q_3)^{1-\gamma}-(1+q_2)^{1-\gamma}\right]
    - \left(\frac{c}{d}\right)^\gamma(q_3-q_2)
  \end{array}
  \label{eq:fqscp2ex1ca3bnorm}
\end{equation*}

\\

  \hline

\end{tabularx}

\caption{The specific mass ratio distribution $\fq$ for pairing function SCP-II, resulting from a power-law generating mass distribution $f_m(m)$ and a uniform generating mass ratio distribution $h_q(q)$. See Appendix~\ref{appendix:scp_powerlaw} for details. }

\label{table:bigequations_scpii_powerlaw}

\end{table*}

\begin{table*}

\begin{tabularx}{\textwidth}{|p{0.01cm}XX|}

  \hline

  \mbox{Pairing function} & \multicolumn{1}{c}{Specific binary fraction} & \multicolumn{1}{c|}{Remarks} \\

  \hline

  \mbox{RP} &

\begin{equation*}  
\binfm = \left( \frac{\binf^{-1}-1}{2 F_M(M_1)} + 1 \right)^{-1}
\quad \quad
c \leq M_1 \leq d
\end{equation*}

&

\begin{equation*}  
F_M(M_1) = \frac{ M_1^{1-\alpha} - c^{1-\alpha} }{ d^{1-\alpha} - c^{1-\alpha} }
\end{equation*}

\\

  \hline

  \mbox{PCRP} &

\begin{equation*}  
\binfm = \binf
\quad \quad
c \leq M_1 \leq d
\end{equation*}
&

\\

  \hline

  \mbox{PCP-I} &

\begin{equation*}  
\binfm = \binf
\quad \quad
c \leq M_1 \leq d
\end{equation*}
&

\\

  \hline

  \mbox{PCP-II} &

\begin{equation*}  
\binfm = \binf \left( 1 - \frac{c}{M_1} \right)
\quad \quad
c \leq M_1 \leq d
\end{equation*}
&

\\

  \hline

  \mbox{PCP-III} &

\begin{equation*}  
\binfm = \binf
\quad \quad
c \leq M_1 \leq d
\end{equation*}
&

\\

  \hline

  \mbox{SCP-I} &

  \begin{equation*}  
    \binfm = \left\{
    \begin{array}{ll}
      1                                                    & c \leq M_1 \leq 2c \\
      \left( 1 + \frac{\alpha+2}{\binf \cdot R(M_1)} \right)^{-1} & 2c \leq M_1 \leq d \\
      \left( 1 + \frac{\alpha+2}{\binf \cdot S(M_1)} \right)^{-1} & d \leq M_1 \leq 2d \\
    \end{array}
    \right.
  \end{equation*}

  &
  
  \begin{equation*}  
    R(M_1) = 2^{\alpha+2} - \left( 3-\frac{M_1}{c} \right)^{\alpha+2}
  \end{equation*}
  
  \begin{equation*}  
    S(M_1) = \left( 3 - \frac{M_1}{d} \right)^{\alpha+2} - 1
  \end{equation*}

\\

  \hline

  \mbox{SCP-II} &

  \begin{equation*}  
    \binfm = \left\{
    \begin{array}{ll}
      \binf_{100,R}(M_1)                                                    & c \leq M_1 \leq 2c \\
      \frac{\binf_{100,S}(M_1)(\alpha+2)}{1+S(M_1)(\binf^{-1}-1)}           & 2c \leq M_1 \leq d \\
      \frac{\binf_{100,T}(M_1)(\alpha+2)}{1+T(M_1)(\binf^{-1}-1)}           & d \leq M_1 \leq 2d-c \\
      0                                                                     & 2d-c \leq M_1 \leq 2d \\
    \end{array}
    \right.
  \end{equation*}

&

  \begin{equation*}  
    \binf_{100,R}(M_1) = 
    \frac{
      2^{\alpha+2} - \left( 1+c/M_1\right)^{\alpha+2}
    }{
      2^{\alpha+2} - \left( 3-M_1/c\right)^{\alpha+2}
    } 
  \end{equation*}
  
  \begin{equation*}  
    \binf_{100,S}(M_1) = 
    \frac{
      2^{\alpha+2} - \left( 1+c/M_1\right)^{\alpha+2}
    }{
      2^{\alpha+2} - 1
    } 
  \end{equation*}
  
  \begin{equation*}  
    \binf_{100,T}(M_1) = 
    \frac{
      \left( 3-M_1/d\right)^{\alpha+2} - \left( 1+c/M_1\right)^{\alpha+2}
    }{
      \left( 3-M_1/c\right)^{\alpha+2} - 1
    } 
  \end{equation*}

  \begin{equation*}  
    S(M_1) = 2^{\alpha+2} - \left( 1+\frac{c}{M_1} \right)^{\alpha+2} 
  \end{equation*}
  
  \begin{equation*}  
    T(M_1) = \left( 3 - \frac{M_1}{d} \right)^{\alpha+2} - \left( 1+\frac{c}{M_1} \right)^{\alpha+2} 
  \end{equation*}

\\

  \hline

  \mbox{SCP-III} &

  \begin{equation*}  
    \binfm = \left\{
    \begin{array}{ll}
      1                                                           & c \leq M_1 \leq 2c \\
      \left( 1 + \frac{\alpha+2}{\binf \cdot R(M_1)} \right)^{-1} & 2c \leq M_1 \leq d \\
      \left( 1 + \frac{\alpha+2}{\binf \cdot S(M_1)} \right)^{-1} & d \leq M_1 \leq 2d-c \\
      0                                                           & 2d-c \leq M_1 \leq 2d \\
    \end{array}
    \right.
  \end{equation*}

&

  \begin{equation*}  
    R(M_1) = \left( 1 - \frac{c}M_1{}  \right)^{-1}
    \left[ 2^{\alpha+2} - \left( 1+\frac{c}{M_1} \right)^{\alpha+2} \right]
  \end{equation*}
  
  \begin{equation*}  
    S(M_1) = \left( 1 - \frac{c}M_1{}  \right)^{-1}
    \left[ \left( 3 - \frac{M_1}{d} \right)^{\alpha+2} - \left( 1+\frac{c}{M_1} \right)^{\alpha+2} \right]
  \end{equation*}

\\

  \hline

\end{tabularx}

\caption{The specific binary fraction $\binfm$ for the nine pairing functions described in this paper, resulting from a power-law mass distribution $f_M(M)$ and a uniform mass ratio distribution $h_q(q)$. See Appendices~\ref{appendix:rp}--\ref{appendix:scp} for a detailed description, and Fig.~\ref{figure:binaryfraction_versus_spectraltype} for a visualisation. }

\label{table:binfracm_powerlaw}

\end{table*}

\end{appendix}

\end{document}